\documentclass[aps,twocolumn,showpacs,floatfix,groupedaddress,amsmath,amssymb,prb]{revtex4}
\usepackage{graphicx}
\usepackage{mathrsfs}
\usepackage{slashbox}
\usepackage{array}


\newcommand{\ket} [1] {| #1 \rangle}
\newcommand{\bra} [1] {\langle #1 |}
\newcommand{\braket}[2]{\langle #1 | #2 \rangle}

\newcommand{\fuser}{\Upsilon^{\mbox{\tiny \,fuse}}}
\newcommand{\spliter}{\Upsilon^{\mbox{\tiny \,split}}}

\newcommand{\splitspin}[3]{\Upsilon^{\mbox{\tiny \,split}}_{j_{#1}t_{j_{#1}}m_{j_{#1}}\rightarrow j_{#2}t_{j_{#2}}m_{j_{#2}},j_{#3}t_{j_{#3}}m_{j_{#3}}}}
\newcommand{\fusespin}[3]{\Upsilon^{\mbox{\tiny \,fuse}}_{j_{#1}t_{j_{#1}}m_{j_{#1}},j_{#2}t_{j_{#2}}m_{j_{#2}}\rightarrow j_{#3}t_{j_{#3}}m_{j_{#3}}}}

\newcommand{\tfuser}{X^{\mbox{\tiny \,fuse}}}
\newcommand{\tspliter}{X^{\mbox{\tiny \,split}}}

\newcommand{\tfuse}[3]{X^{\mbox{\tiny \,fuse}}_{#1, #2 \rightarrow #3}}

\newcommand{\tsplitspin}[3]{X^{\mbox{\tiny \,split}}_{j_{#1}t_{j_{#1}}\rightarrow j_{#2}t_{j_{#2}},j_{#3}t_{j_{#3}}}}
\newcommand{\tfusespin}[3]{X^{\mbox{\tiny \,fuse}}_{j_{#1}t_{j_{#1}},j_{#2}t_{j_{#2}}\rightarrow j_{#3}t_{j_{#3}}}}

\newcommand{\csplitt}[3]{C^{\mbox{\tiny \,split}}_{#1\rightarrow #2,#3}}
\newcommand{\cfuse}[3]{C^{\mbox{\tiny \,fuse}}_{#1, #2 \rightarrow #3}}
\newcommand{\cfuser}{C^{\mbox{\tiny \,fuse}}}
\newcommand{\cspliter}{C^{\mbox{\tiny \,split}}}

\newcommand{\csplitspin}[3]{C^{\mbox{\tiny \,split}}_{j_{#1}m_{j_{#1}}\rightarrow j_{#2}m_{j_{#2}},j_{#3}m_{j_{#3}}}}
\newcommand{\cfusespin}[3]{C^{\mbox{\tiny \,fuse}}_{j_{#1}m_{j_{#1}},j_{#2}m_{j_{#2}}\rightarrow j_{#3}m_{j_{#3}}}}

\makeatletter

\newcommand{\Rmnum}[1]{\expandafter\@slowromancap\romannumeral #1@}
\newcommand{\half}{\frac{1}{2}}
\newcommand{\mhalf}{\frac{-1}{2}}

\newcommand{\rmi}{\mathrm{i}}

\newcommand{\eref}[1]{Eq.(\ref{#1})}

\newcommand{\fref}[1]{Fig.\ref{#1}}
\newcommand{\tree}{\boldsymbol{\tau}}

\newcommand{\markend}{~\rule{0.4em}{1.8ex}} 

\newcommand{\mycup}{\Omega^{\mbox{\tiny \,cup}}}
\newcommand{\mycap}{\Omega^{\mbox{\tiny \,cap}}}

\newcommand{\cupf}[3]{\mu^{\mbox{\tiny \,cup}}_{j_{#1}j_{#2}j_{#3}}}
\newcommand{\capf}[3]{\mu^{\mbox{\tiny \,cap}}_{j_{#1}j_{#2}j_{#3}}}

\newcommand{\braid}[3]{R^{\mbox{\tiny \,swap}}_{j_{#1},j_{#2} \rightarrow j_{#3}}}
\newcommand{\braider}{R^{\mbox{\tiny \,swap}}}

\newcommand{\specialcell}[2][c]{\begin{tabular}[#1]{@{}c@{}}#2\end{tabular}}

\newcolumntype{C}[1]{%
 >{\vbox to 1ex\bgroup\vfill\centering}%
 p{#1}%
 <{\egroup}}

\begin{document}
\title{Tensor network states and algorithms in the presence of a global SU(2) symmetry}
\author{Sukhwinder Singh}
\affiliation{Center for Engineered Quantum Systems, Dept. of Physics \& Astronomy, Macquarie University, 2109 NSW, Australia}
\author{Guifre Vidal}
\affiliation{Perimeter Institute for Theoretical Physics, Waterloo, Ontario, N2L 2Y5, Canada}

\begin{abstract}
The benefits of exploiting the presence of symmetries in tensor network algorithms have been extensively demonstrated in the context of matrix product states (MPSs). These include the ability to select a specific symmetry sector (e.g. with a given particle number or spin), to ensure the exact preservation of total charge, and to significantly reduce computational costs. Compared to the case of a generic tensor network, the practical implementation of symmetries in the MPS is simplified by the fact that tensors only have three indices (they are trivalent, just as the Clebsch-Gordan coefficients of the symmetry group) and are organized as a one-dimensional array of tensors, without closed loops. Instead, a more complex tensor network, one where tensors have a larger number of indices and/or a more elaborate network structure, requires a more general treatment. In two recent papers, namely $(i)$ [Phys. Rev. A \textbf{82}, 050301 (2010)] and $(ii)$ [Phys. Rev. B \textbf{83}, 115125 (2011)], we described how to incorporate a global internal symmetry into a generic tensor network algorithm based on decomposing and manipulating tensors that are invariant under the symmetry. In $(i)$ we considered a generic symmetry group $\mathcal{G}$ that is compact, completely reducible and multiplicity free, acting as a global internal symmetry. Then in $(ii)$ we described the implementation of Abelian group symmetries in much more detail, considering a U(1) symmetry (\textit{e.g.}, conservation of global particle number) as a concrete example. In this paper we describe the implementation of non-Abelian group symmetries in great detail. For concreteness we consider an SU(2) symmetry (\textit{e.g.}, conservation of global quantum spin).
Our formalism can be readily extended to more exotic symmetries associated with conservation of total fermionic or anyonic charge.
As a practical demonstration, we describe the SU(2)-invariant version of the multi-scale entanglement renormalization ansatz and apply it to study the low energy spectrum of a quantum spin chain with a global SU(2) symmetry.
\end{abstract}


\maketitle

\section {Introduction \label{sec:intro}}

In recent years, tensor network states
\cite{
Fannes92, Ostlund95, Vidal03, PerezGarcia07, 
White92, White93, Schollwoeck05, McCulloch08, 
Ramasesha96, Sierra97, Tatsuaki00, McCulloch02, Bergkvist06, Pittel06, McCulloch07, PerezGarcia08, Sanz09, 
Vidal04, Daley04, White04, Schollwoeck05b, Daley05, Vidal07, Danshita07, Muth09, Mishmash09, Singh10, Cai10, 
Shi06, 
Vidal07b, Vidal08, Evenbly09, Giovannetti08, Pfeifer09, Vidal10, 
Yan11, Jiang12, Depenbrock12, 
Tagliacozzo09, Murg10, 
Verstraete04, Sierra98, Nishino98, Nishio04, Murg07, Jordan08, Gu08, Jiang08, Xie09, Murg09, Wang11, 
Evenbly10f, Evenbly10b, Aguado08, Cincio08, Evenbly09b, Koenig09,Evenbly10, 
Corboz09, Kraus09, Pineda09, Corboz09b, Barthel09, Shi09, Corboz10b, Pizorn10, Gu10, 
Pollmann10, Chen11, Schuch11, Chen11b, Gu12}
have become an important tool to study quantum many-body systems on a lattice. On the theoretical side, they offer a natural framework to investigate and classify the possible phases of quantum matter
\cite{Pollmann10, Chen11, Schuch11, Chen11b, Gu12}. On the numerical side, they are the basis of novel computational approaches capable of addressing non-perturbatively a large range of interacting systems, including two-dimensional systems of frustrated spins\cite{Murg09, Evenbly10, Yan11, Jiang12, Depenbrock12} and of interacting fermions.\cite{Corboz09, Kraus09, Pineda09, Corboz09b, Barthel09, Shi09, Corboz10b, Pizorn10, Gu10}

Tensor network states for one dimensional systems include the matrix product state\cite{Fannes92, Ostlund95, Vidal03, PerezGarcia07} (MPS), which is the basis of the density matrix renormalization group\cite{White92, White93} (DMRG) algorithm for computing ground states and the time-evolving block-decimation\cite{Vidal04} (TEBD) algorithm for simulating time evolution; the tree tensor network\cite{Shi06} (TTN); and the multi-scale entanglement renormalization ansatz\cite{Vidal07b, Vidal08} (MERA) for critical systems. In two (and more) spatial dimensions, one can still use a MPS \cite{Yan11, Jiang12, Depenbrock12} or TTN,\cite{Tagliacozzo09, Murg10} although these tensor networks can only represent small systems since they do not offer an efficient, scalable description. In contrast, a projected entangled pair state\cite{Verstraete04} (PEPS), which is a higher dimensional generalization of MPS, as well as higher dimensional versions of the MERA,\cite{Evenbly09b} offer a scalable description in two and larger dimensions.
Presently, the main limitation of tensor network methods comes from the fact that simulation costs increase rapidly with the amount of entanglement in the system, which introduces a bias towards weakly entangled phases. In addition, in dimensions larger than one, both PEPS and MERA can only efficiently represent ground states that obey a boundary law for entanglement entropy,\cite{Verstraete06, Evenbly11} although the recently proposed branching MERA overcomes this limitation.\cite{Evenbly12,Evenbly12b}

In this paper we are concerned with incorporating symmetries into tensor networks. The presence of symmetries can be a powerful advantage in numerical approaches. A many-body Hamiltonian $\hat H$ may be invariant under certain transformations, which form a group of symmetries.\cite{Cornwell97} The symmetry group divides the Hilbert space of the theory into symmetry sectors labeled by quantum numbers or conserved charges. By targeting a specific symmetry sector during a calculation, computational costs can often be significantly reduced while explicitly preserving the symmetry. Here we will be concerned with \textit{global internal} symmetries of lattice models, where \textit{internal} means that the symmetry acts on the Hilbert space of each site of the lattice, whereas \textit{global} means that the symmetry acts identically on all sites.

In this paper we address, in a pedagogical way, the implementation of (global internal) non-Abelian symmetries in tensor network algorithms. For concreteness we consider the context of SU(2) symmetry which corresponds, for instance, to spin isotropy. Following Ref.~\onlinecite{Singh09} we consider tensors that are invariant under the symmetry and describe how an SU(2)-invariant tensor compactly decomposes into a \textit{degeneracy} part, which contains all degrees of freedom not determined by symmetry, and a \textit{structural} part that corresponds to \textit{intertwiners} or generalized Clebsch-Gordan coefficients) of SU(2). [In contrast to the Abelian case, which we addressed in detail in Ref.~\onlinecite{Singh11}, here the Clebsch-Gordan coefficients of the group are non-trivial.] Consequently, a tensor network made of SU(2)-invariant tensors decomposes as a linear superposition of \textit{spin networks}\cite{spinnetwork} which, here, encode the constraints imposed by the symmetry on the tensor network. We describe how this decomposition of an SU(2)-invariant tensor network can be maintained and exploited for computational gain in a tensor network algorithm. As a practical demonstration we describe the SU(2)-invariant version of the MERA and apply it to study the low energy spectrum of a quantum spin chain with spin isotropy.

\begin{figure}[t]
  \includegraphics[width=8cm]{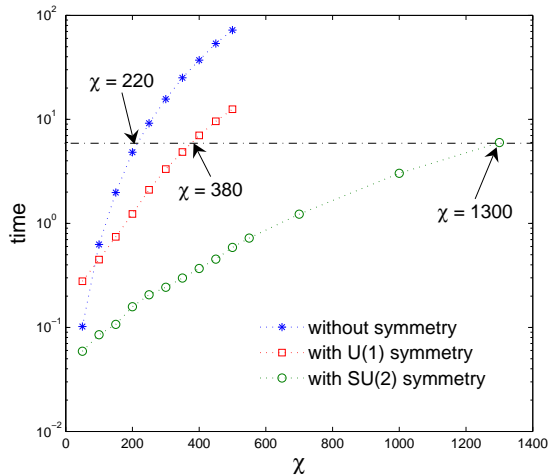}
\caption{(Color online) Computational gain obtained by exploiting the symmetry in an MPS algorithm. Computation time (in seconds) for one iteration of the ``infinite Time Evolving Block Decimation''\cite{Vidal03} (TEBD) algorithm, as a function of the MPS bond dimension $\chi$ is shown. Here $\chi$ is a refinement parameter, a larger $\chi$ leads to a better accuracy of the method. For sufficiently large $\chi$, exploiting symmetry leads to reductions in computation time. The horizontal line on this graph shows that this reduction in computation time equates to the ability to evaluate MPSs with a higher bond dimension $\chi$: For the same cost per iteration incurred when optimizing a regular MPS in MATLAB with bond dimension $\chi=220$, one may choose instead to optimize a U(1)-symmetric MPS with $\chi=380$ or an SU(2)-symmetric MPS with $\chi=1300$. \label{fig:mpscompare}}
\end{figure}

Our approach to incorporate the symmetry into tensor network algorithms takes into account only the total spin $j$ while the Clebsch-Gordan coefficients and the quantum number $m$, corresponding to the spin projection along the z-axis, appear only to develop the formalism and facilitate discussion. Instead, the only data required from the symmetry group are the (i) list of irreps of the group,  (ii) the fusion rules, that is, the decomposition of the tensor product of two irreps into the direct sum of irreps, (iii) the recoupling coefficients (or 6j-symbols) of the group that relate the different ways of fusing three irreps, and (iv) the swap coefficients, \eref{eq:su2swap}. On the one hand, working only with this data allows for  a \textit{manifestly} SU(2)-invariant treatment of tensors. On the other, it leads to reduction in computational costs in symmetric tensor network algorithms, since the Clebsch-Gordan coefficients are eliminated from the description of SU(2)-invariant tensor networks.

Moreover, by not emphasizing the internal degrees of freedom $m$ of the symmetry group our formalism can be readily generalized (see Sec.~\ref{sec:outlook}) to include more exotic symmetry constraints associated, for instance, with the deformed group $SU(2)_k$ that appears in the context of conservation of total anyonic charge in lattice models of anyons (in case of $SU(2)_k$ there is no internal irrep space corresponding to the label $m$ and therefore no Clebsch-Gordan coefficients). Our formalism may be generalized to this case by replacing the data (i)-(iv) for SU(2) by that for $SU(2)_k$. In fact, the present work corresponds to the special case $k\rightarrow \infty$ and as such also serves to illustrate the basic ingredients that are required for the implementation of anyonic constraints albeit in the more familiar context of SU(2) symmetry and spin systems.

\subsection{Related work}

The implementation of symmetries is well understood in the context of the MPS. Both space (e.g. translation invariance) and global internal symmetries (Abelian and non-Abelian) have been thoroughly incorporated into algorithms based on an MPS, namely in DMRG\cite{White92, White93, Schollwoeck05, McCulloch08, Ramasesha96, Sierra97, Tatsuaki00, McCulloch02, Bergkvist06, Pittel06, McCulloch07, PerezGarcia08, Sanz09} and TEBD,\cite{Vidal03, Vidal04, Daley04, White04, Schollwoeck05b, Daley05, Vidal07, Danshita07, Muth09, Mishmash09, Singh10, Cai10} where it has been exploited to obtain computational gains. Figure~\ref{fig:mpscompare} is demonstrative of the colossal computational gain that has been obtained by exploiting the symmetry in the context of the MPS. Figure \ref{fig:meracompare} shows an analogous comparison in the context of the MERA.

The use of symmetric tensors in more complex tensor networks has also been discussed in Refs.~\onlinecite{PerezGarcia10, Zhao10}. In particular, Ref.~\onlinecite{PerezGarcia10} has shown that under convenient conditions (injectivity), a PEPS that represents a symmetric state can be represented with symmetric tensors, generalizing similar results for MPS obtained in Ref.~\onlinecite{PerezGarcia08}. We notice that these studies are not concerned with how to computationally protect or exploit the symmetry, which is the focus of the present paper.

The implementation of non-Abelian symmetries in generic tensor network algorithms was recently addressed in Ref.~\onlinecite{Andreas12}, based again on the use of symmetric tensors.\cite{Singh10} A remarkable aspect of Ref. \onlinecite{Andreas12} is that it deals with non-Abelian groups with multiplicity,\cite{multiplicity} such as SU(3) (also see Sec.~\ref{sec:outlook}). However, we note that in Ref. \onlinecite{Andreas12}, the Clebsch-Gordan coefficients that appear in the decomposition of invariant tensors explicitly participate in the contraction of tensors. This is in contrast with our approach where Clebsch-Gordan are not used (only 6j-symbols are required), which leads to computational gains. In addition, in spite of claiming to address generic tensor networks, Ref. \onlinecite{Andreas12} only demonstrates the approach in the context of an MPS, for which there is already extensive literature. In contrast, here we will describe and address the practical computational issues that arise in using non-Abelian symmetries in complex tensor networks.

The implementation of conservation of total fermionic and anyonic charges was described in e.g. Refs.~\onlinecite{Corboz09}-\onlinecite{Gu10} and Refs.~\onlinecite{Pfeifer10} and \onlinecite{Koenig10}, respectively.

\begin{figure}[t]
  \includegraphics[width=8.5cm]{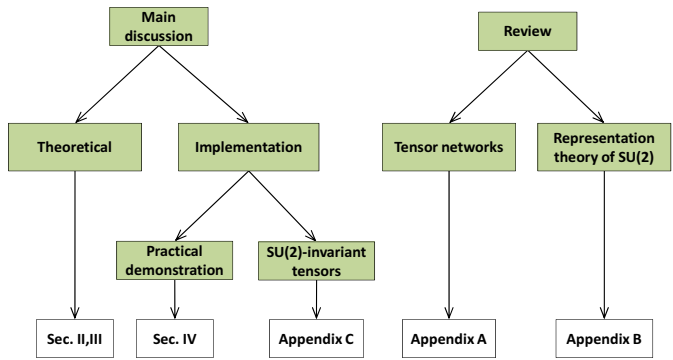}
\caption{(Color online)  The schematic organization of the paper.}
\label{fig:schema}
\end{figure}


\subsection{Organization of the paper}

The schematic organization of this paper is shown in Fig. \ref{fig:schema}. It closely follows Ref.~\onlinecite{Singh11}, where implementation of Abelian symmetries was presented by the authors. Therefore the differences between the Abelian and non-Abelian implementation are highlighted throughout the discussion. We note that our specific implementation of non-Abelian symmetries in tensor networks, which is based on \textit{tree decompositions} of SU(2)-invariant tensors, is presented in App. \ref{sec:tree}.

In Sec.~\ref{sec:su2tensors} we characterize SU(2) invariant tensors and describe their compact canonical decomposition into degeneracy and structural parts. Then in Sec.~\ref{sec:tnsu2} we consider tensor networks made of such tensors. We describe how a set $\mathcal{P}$ of primitive tensor networks manipulations is adapted to the presence of the symmetry. 
Section~\ref{sec:mera} contains a practical demonstration of exploiting SU(2) symmetry in a tensor network algorithm by presenting MERA calculations of the ground state and low energy states of the spin-1/2 anti-ferromagnetic Heisenberg chain.
Section~\ref{sec:outlook} contains a brief summary of the paper and some remarks pertaining to extending the present formalism to more general symmetry constraints.

An important part of the paper is devoted to appendices, where we present review material and specific implementations. App. \ref{sec:tensor} reviews the tensor network formalism and the diagrammatic representation of tensors. It also describes a set $\mathcal{P}$ of primitive operations for manipulating tensor networks, namely, the reversal, permutation, fusion and splitting of the indices of a tensor and matrix operations, namely, matrix multiplication and matrix factorization. App. \ref{sec:symmetry} reviews the relevant results from the representation theory of SU(2). In App. \ref{sec:tree} we describe our implementation of the set $\mathcal{P}$ of primitive tensor networks manipulations based on decomposing individual SU(2) invariant tensors as a tree.


\section{SU(2)-invariant tensors\label{sec:su2tensors}}

In this section we consider tensors that are invariant under the action of SU(2) and explain how such tensors decompose into a compact canonical form which exploits their symmetry. 
For a review on the tensor network formalism and its diagrammatic notation, see App. \ref{sec:tensor}.
For a review on the representation theory of SU(2), see App \ref{sec:symmetry}.


\subsection{Symmetry constraints}
Let $\hat{T}$ be a rank-$k$ tensor with components $\hat{T}_{i_1 i_2 \cdots i_k}$ (see App. \ref{sec:tensor}) and $\vec{D}$ denote the directions of its indices $i_1, i_2, \cdots, i_k$. Here, $\vec{D}(l) = $`in' if index $i_l$ is incoming and $\vec{D}(l) = $`out' if it is outgoing. Each index $i_l$ is associated with a vector space $\mathbb{V}^{(l)}$ on which SU(2) acts by means of transformations $\hat{W}_{\textbf{r}}^{(l)}$. Also consider the action of SU(2) on the space
\begin{equation}
\mathbb{V}^{(l)} \otimes \mathbb{V}^{(2)} \otimes \cdots \otimes \mathbb{V}^{(k)} \label{eq:kdecomp}
\end{equation}
given by
\begin{equation}
	\hat{Y}^{(1)}_{\textbf{r}}\otimes \hat{Y}^{(2)}_{\textbf{r}}\otimes \ldots \otimes \hat{Y}^{(k)}_{\textbf{r}},
\label{eq:Xtrans}
\end{equation}
where
\begin{equation}
\hat{Y}^{(l)}_{\textbf{r}} = \left\{
	\begin{array}{cc} \hat{W}^{(l)~*}_{\textbf{r}}& ~~~\mbox{ if } \vec{D}(l) = \mbox{ `in' } ,\\
	 									\hat{W}^{(l)}_{\textbf{r}}& ~~~~\mbox{ if } \vec{D}(l) = \mbox{ `out' }.
	\end{array} \right.
\label{eq:y}
\end{equation}
($\hat{W}_{\textbf{r}}^{(l)~*}$ denotes the complex conjugate of $\hat{W}_{\textbf{r}}^{(l)}$.) That is, $\hat{Y}^{(l)}_{\textbf{r}}$ acts differently depending on whether index $i_l$ is an incoming or outgoing index. 

We say that tensor $\hat{T}$ is SU(2)-invariant if it is invariant under the transformation of \eref{eq:Xtrans}. In components,
\begin{align}
	\sum_{i_1, i_2, \ldots, i_k} 	\left(\hat{Y}^{(1)}_{\textbf{r}}\right)_{i_1'i_1} \left(\hat{Y}^{(2)}_{\textbf{r}} \right)_{i_2' i_2} \ldots \left( \hat{Y}^{(k)}_{\textbf{r}} \right)_{i_k' i_k} \hat{T}_{i_1 i_2 \ldots i_k}& \nonumber \\
	=\hat{T}_{i_1' i_2' \ldots i_k'}&,
	\label{eq:Tinv}
\end{align}
for all $\textbf{r} \in \mathbb{R}^3$.

\begin{figure}[t]
  \includegraphics[width=7.5cm]{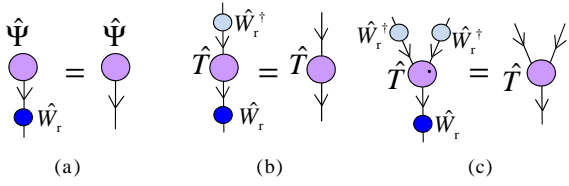}
\caption{(Color online) Constraint fulfilled by (a) an SU(2)-invariant vector, (b) an SU(2)-invariant matrix, and (c) a rank-$3$ SU(2)-invariant tensor with two incoming and one outgoing index.\label{fig:su2inv}}
\end{figure}

\textit{Example 1.} An SU(2)-invariant vector $\ket{\Psi}$ fulfills
\begin{equation}
	(\hat\Psi)_{a'} = \sum_{a} \left(\hat{W}_{\textbf{r}}\right)_{a'a} \hat\Psi_{a} ~~~~~~\forall\ \textbf{r} \in \mathbb{R}^3,
\end{equation}
in accordance with \eref{eq:invPsi1}, see \fref{fig:su2inv}(a).\markend

\textit{Example 2.} An SU(2)-invariant matrix $\hat{T}$ fulfills
\begin{eqnarray}
	\hat{T}_{a'b'} &=& \sum_{a,b} \left(\hat{W}^{(1)}_{\textbf{r}}\right)_{a'a} \left( \hat{W}^{(2)~*}_{\textbf{r}} \right)_{b'b} \hat{T}_{ab} \\
	&=& \sum_{a,b} \left(\hat{W}^{(1)}_{\textbf{r}}\right)_{a'a} \hat{T}_{ab} \left(\hat{W}^{(2)~\dagger}_{\textbf{r}}\right)_{bb'},
\end{eqnarray}
for all $\textbf{r} \in \mathbb{R}^3$, in accordance with \eref{eq:invOp2}, see \fref{fig:su2inv}(b).\markend

\textit{Example 3.} Tensor $\hat{T}$ with components $\hat{T}_{abc}$ where $a$ and $b$ are incoming indices and $c$ is an outgoing index is SU(2)-invariant iff
\begin{eqnarray}
	\hat{T}_{a' b' c'} &=& \sum_{a,b,c} \left(\hat{W}^{(1)~*}_{\textbf{r}}\right)_{a'a}\left(\hat{W}^{(2)~*}_{\textbf{r}} \right)_{b'b} \left( \hat{W}^{(3)}_{\textbf{r}} \right)_{c'c} \hat{T}_{abc}~~~~~~~\label{eq:threeinveg0}\\
&=& \sum_{a,b,c} \left(\hat{W}^{(1)}_{\textbf{r}}\right)_{c'c}\hat{T}_{abc} \left(\hat{W}^{(3)~\dagger}_{\textbf{r}}\right)_{aa'}\left(\hat{W}^{(3)~\dagger}_{\textbf{r}}\right)_{bb'} \label{eq:threeinveg}
\end{eqnarray}
for all $\textbf{r} \in \mathbb{R}^3$, see \fref{fig:su2inv}(c).\markend


\subsection{Block structure\label{sec:su2tensors:block}}
Let us now write tensor $\hat{T}$ that fulfills \eref{eq:Tinv} in the spin basis for each factor space in \eref{eq:kdecomp}, that is, $i_1=(j_1,t_{j_1},m_{j_1}), i_2=(j_2,t_{j_2},m_{j_2}), \ldots, i_k=(j_k,t_{j_k},m_{j_k})$. For fixed value of $j$'s, each index $i_l$ decomposes [\fref{fig:123}(a)] into a \textit{degeneracy index} $(j_l,t_{j_l})$ and a \textit{spin index} $(j_l, m_{j_l})$ in accordance with the decomposition (\ref{eq:decoV}) of the vector space $\mathbb{V}^{(l)}$ associated with it. Tensor $\hat{T}$ decomposes into a \textit{degeneracy tensor} $\hat{P}$ that carries all the degeneracy indices and a \textit{structural tensor} (or an intertwiner) that carries all the spin indices. In particular, this implies that an SU(2)-invariant tensor $\hat{T}$ has a \textit{sparse block structure} - several of its components are identically zero. Next, we describe this block structure for SU(2)-invariant tensors with one, two, three and then an arbitrary number of indices.


\subsubsection{One index\label{sec:su2tensors:block:one}}

\begin{figure}[t]
  \includegraphics[width=8.5cm]{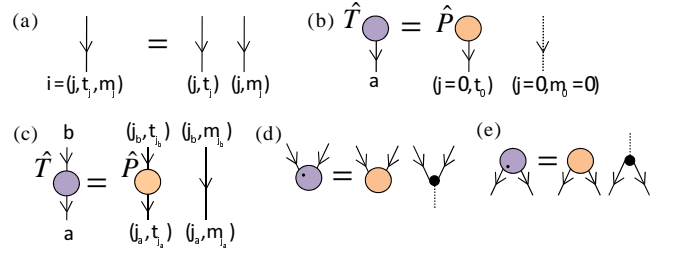}
\caption{(Color online) (a) Each index $i=(j,t_j,m_j)$ of an SU(2)-invariant tensor that is written in the spin basis decomposes into a degeneracy index $(j,t_{j})$ and a spin index $(j,m_{j})$ in accordance with the decomposition (\ref{eq:decoV}) of the vector space associated with it. (b) Only $j=0$ is relevant for an SU(2)-invariant tensor with one index. (c)-(e) For fixed value of $j_a$ and $j_b$ a rank-2 SU(2)-invariant tensor $\hat{T}_{ab}$ decomposes into a degeneracy tensor $\hat{P}_{j_aj_b}$ and a structural term that depending on the directions of $a$ and $b$ is the Identity $(c)$, tensor $\cfuse{j_am_{j_a}}{j_b m_{j_b}}{00}$ $(d)$, or tensor $\csplitt{00}{j_am_{j_a}}{j_b m_{j_b}}$ $(e)$.
\label{fig:123}}
\end{figure}

Consider an SU(2)-invariant tensor $\hat{T}$ with an index $a=(j,t_{j},m_{j})=(0,t_0,0)$ (that is, only irrep $j=0$ is relevant on the one index). We have [\fref{fig:123}(b)]
\begin{equation}
\hat{T}_a = (\hat{P})_{t_0},
\label{eq:canonone}
\end{equation}
where $(\hat{P})_{t_0}$, shorthand for $(\hat{T}_{j=0})_{t_0, m_{0}=0}$, encodes the non-trivial components of $\hat{T}$.

\subsubsection{Two indices\label{sec:su2tensors:block:two}}
An SU(2)-invariant tensor $\hat{T}$ with an outgoing index $a=(j_a,m_{j_a},t_{j_a})$ and an incoming index $b=(j_b,m_{j_b},t_{j_b})$
decomposes as (Schur's lemma)
\begin{align}
(\hat{T})_{ab} = (\hat{P}_{j_aj_b})_{t_{j_a} t_{j_b}} \delta_{j_aj_b} \delta_{m_{j_a} m_{j_b}}, \label{eq:struct2}
\end{align}
where (degeneracy) tensors $\hat{P}_{j_aj_b}$ with components $(\hat{P}_{j_aj_b})_{t_{j_a} t_{j_b}}$ carry the two degeneracy indices $(j_a,t_{j_a})$ and $(j_b,t_{j_b})$, see \fref{fig:123}(c). The term $\delta_{j_aj_b} \delta_{m_{j_a} m_{j_b}}$ corresponds to the constraints imposed by the symmetry on the tensor $\hat{T}$, namely, only components $(\hat{T})_{ab}$ with $j_a=j_b$ and $m_{j_a}=m_{j_b}$ are non-trivial.

The constraints, and therefore the decomposition of $\hat{T}$, depend on the arrangement of arrows on the indices (since different arrangement of arrows correspond to a different action of the group). When both $a$ and $b$ are incoming indices, tensor $\hat{T}$ decomposes as
\begin{equation}
(\hat{T})_{ab} = (\hat{P}_{j_aj_b})_{t_{j_a} t_{j_b}} \cfuse{j_am_{j_a}}{j_b m_{j_b}}{00},
\end{equation}
where $\cfuse{j_am_{j_a}}{j_b m_{j_b}}{00}$ corresponds to the constraint that spins $j_a$ and $j_b$ fuse into a total spin $j=0$, \fref{fig:123}(d). That is, components $(\hat{T})_{ab}$ are identically zero unless $j_a=j_b$ and $m_{j_a}+m_{j_b}=0$.

Analogously, when $a$ and $b$ are both outgoing indices we have [\fref{fig:123}(e)]
\begin{equation}
(\hat{T})_{ab} = (\hat{P}_{j_aj_b})_{t_{j_a} t_{j_b}} \csplitt{00}{j_am_{j_a}}{j_b m_{j_b}}.
\end{equation}

The decomposition of an SU(2)-invariant tensor $\hat{T}$ with two indices $a$ and $b$ can generally be written as
\begin{align}
(\hat{T})_{ab} = (\hat{P}_{j_aj_b})_{t_{j_a} t_{j_b}} (\hat{Q}_{j_aj_b})_{m_{j_a} m_{j_b}},
\label{eq:canon:two1}
\end{align}
which can also be recast in a block-diagonal form,
\begin{align}
\hat{T} &= \bigoplus_j \hat{T}_j, \nonumber \\
&=\bigoplus_j (\hat{P}_j \otimes \hat{Q}_j),~~~j = j_a = j_b.\label{eq:canon:two2}
\end{align}

\textit{Example 4.} Let us estimate the sparseness of an SU(2)-invariant tensor when it is decomposed into degeneracy and structural parts. Consider an SU(2)-invariant tensor $\hat{T}$ with two incoming indices each of which is associated with the vector space $\mathbb{V}$,
\begin{equation}
\mathbb{V}~\equiv~(\mathbb{D}_0 \otimes \mathbb{V}_0)\oplus(\mathbb{D}_1 \otimes \mathbb{V}_1)\oplus(\mathbb{D}_2 \otimes \mathbb{V}_2),\nonumber
\end{equation}
where the dimensions of the degeneracy spaces $\mathbb{D}_0, \mathbb{D}_1$ and $\mathbb{D}_2$ are $d_0 = 1, d_1=3$ and $d_2=1$ respectively. Tensor $\hat{T}$ decomposes as
\begin{equation}
\hat{T} \equiv (\hat{P}_0 \otimes \hat{Q}_0) \oplus (\hat{P}_1 \otimes \hat{Q}_1) \oplus (\hat{P}_2 \otimes \hat{Q}_2), \nonumber
\end{equation}
where $\hat{P}_j$ is a (degeneracy) matrix that acts on the space $\mathbb{D}_j$ and $\hat{Q}_j$ is a matrix made of Clebsch-Gordan coefficients, $(\hat{Q}_j)_{m_jm'_j}\equiv\cfuse{jm_{j}}{j m'_{j}}{00}$, that acts on the space $\mathbb{V}_j$. The dimension of the vector space $\mathbb{V}$ is $\sum_j d_j\Delta_j = 15$. Thus, the total number of complex coefficients contained in $\hat{T}$ is $|\hat{T}| = 15 \times 15 = 225$. However, $\hat{T}$ can be stored compactly by only storing the degeneracy tensors $\hat{P}_0, \hat{P}_1$ and $\hat{P}_2$. The total number of complex coefficients to be stored in this case is equal to
\begin{equation}
|\hat{P_0}|+|\hat{P_1}|+|\hat{P_2}| = d_0^2 + d_1^2 + d_2^2 = 11.\nonumber
\end{equation}
That is, by exploiting the symmetry the number of coefficients that need to be stored is about twenty times smaller.\markend

\begin{figure}[t]
  \includegraphics[width=8cm]{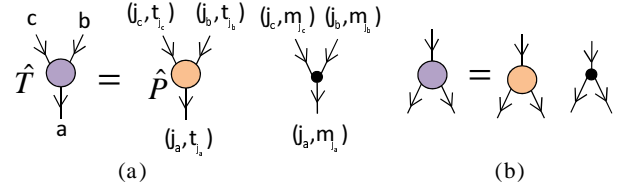}
\caption{(Color online) Examples of the decomposition of rank-$3$ SU(2)-invariant tensors into degeneracy tensors $\hat{P}$ and Clebsch-Gordan tensors.
\label{fig:three}}
\end{figure}


\subsubsection{Three indices\label{sec:su2tensors:block:three}}
The Wigner-Eckart theorem establishes that an SU(2)-invariant tensor $\hat{T}$ with three indices $a=(j_a,m_{j_a},t_{j_a}), b=(j_b,m_{j_b},t_{j_b})$ and $c=(j_c,m_{j_c},t_{j_c})$ decomposes as
\begin{align}
(\hat{T})_{abc} = (\hat{P}_{j_aj_bj_c})_{t_{j_a} t_{j_b} t_{j_c}} (\hat{Q}_{j_aj_bj_c})_{m_{j_a} m_{j_b}m_{j_c}},
\label{eq:three22}
\end{align}
where tensors $\hat{Q}_{j_aj_bj_c}$ depend on the particular choice $\vec{D}$ of incoming and outgoing indices. The components $(\hat{Q}_{j_aj_bj_c})_{m_{j_a} m_{j_b}m_{j_c}}$ are the Clebsch-Gordan coefficients given as (e.g. see \fref{fig:three})
\begin{align}
	  \cfusespin{b}{a}{c}&~~~~\mbox{if }\vec{D} = \{\mbox{`in', `in', `out'}\},\\
	 	\cfusespin{a}{c}{b}&~~~~\mbox{if }\vec{D} = \{\mbox{`in', `out', `in'}\},\\
	 	\cfusespin{c}{b}{a}&~~~~\mbox{if }\vec{D} = \{\mbox{`out', `in', `in'}\}, \\
	 	\csplitspin{b}{c}{a}&~~~~\mbox{if }\vec{D} = \{\mbox{`out', `in', `out'}\},\\
	 	\csplitspin{c}{a}{b}&~~~~\mbox{if }\vec{D} = \{\mbox{`out', `out', `in'}\},\label{eq:forReversal}\\
	 	\csplitspin{a}{b}{c}&~~~~\mbox{if }\vec{D} = \{\mbox{`in', `out', `out'}\}.
\label{eq:threeall}
\end{align}

(The remaining two cases $\vec{D} = \{\mbox{`in', `in', `in'}\}$ and $\vec{D} = \{\mbox{`out', `out', `out'}\}$ correspond to rank-$4$ intertwiners associated with fusing $j_a$, $j_b$ and $j_c$ into a total spin $j=0$, and are covered in the next subsection.) The decomposition \eref{eq:three22} can be recast in a ``block'' form,
\begin{align}
\hat{T} &\equiv \bigoplus_{j_aj_bj_c} \hat{T}_{j_aj_bj_c}, \nonumber \\
&\equiv \bigoplus_{j_aj_bj_c} \left(\hat{P}_{j_aj_bj_c} \otimes \hat{Q}_{j_aj_bj_c}\right),\label{eq:threetensor}
\end{align}
where we use the direct sum symbol $\bigoplus$ to denote that the different tensors (or blocks) $\hat{T}_{j_aj_bj_c}$ are supported on orthonormal subspaces of the tensor product of the spaces associated with indices $a,b$ and $c$; the direct sum is over all compatible values of $j_a, j_b$ and $j_c$.


\subsubsection{$k>3$ indices\label{sec:su2tensors:block:k}}
A rank-$4$ SU(2)-invariant tensor $\hat{T}$ with incoming indices $a=(j_a,t_{j_a},m_{j_a}), b=(j_b,t_{j_b},m_{j_b})$ and $c=(j_c,t_{j_c},m_{j_c})$ and outgoing index $d=(j_d,t_{j_d},m_{j_d})$ can be decomposed as
\begin{align}
(T)_{abcd} = \sum_{j_e}&(\hat{P}^{j_e}_{j_aj_bj_cj_d})_{t_{j_a}t_{j_b}t_{j_c}t_{j_d}} \times\nonumber \\
&~~~~~~~(\hat{Q}^{j_e}_{j_a j_b j_c j_d})_{m_{j_a} m_{j_b} m_{j_c} m_{j_d}},
\label{eq:deco41}
\end{align}
where $\hat{Q}^{j_e}_{j_a j_b j_c j_d}$ is the intertwiner (analogous to \eref{eq:cob1}) that describes the fusion of $j_a, j_b$ and $j_c$ into a total spin $j_d$ by first fusing $j_c$ and $j_b$ into an intermediate spin $j_e$ and then fusing $j_e$ with $j_a$.

\begin{figure}[t]
  \includegraphics[width=8.5cm]{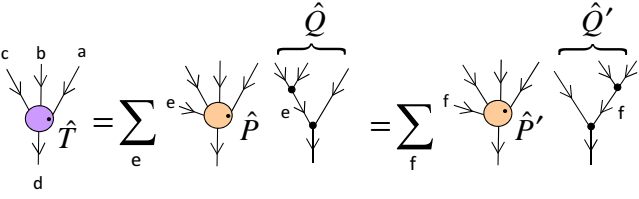}
\caption{(Color online) Two different canonical decompositions of a rank-$4$ SU(2)-invariant tensor $\hat{T}$ into $(\hat{P}, \hat{Q})$ and $(\hat{P}', \hat{Q}')$ tensors corresponding to two different choices of the fusion tree.
\label{fig:four}}
\end{figure}

Alternatively, tensor $\hat{T}$ can be decomposed as
\begin{align}
(T)_{abcd} = \sum_{j_f}&(\hat{P}'^{j_f}_{j_aj_bj_cj_d})_{t_{j_a}t_{j_b}t_{j_c}t_{j_d}} \times\nonumber \\
&~~~~~(\hat{Q}'^{j_f}_{j_a j_b j_c j_d})_{m_{j_a} m_{j_b} m_{j_c} m_{j_d}},
\label{eq:deco42}
\end{align}
where $\hat{Q}'^{j_f}_{j_a j_b j_c j_d}$ (analogous to \eref{eq:cob2}) is the intertwiner associated with fusing $j_a, j_b$ and $j_c$ differently. That is, first fusing $j_b$ and $j_a$ into an intermediate spin $j_f$ and then fusing spin $j_c$ with $j_f$. Since Eqs.~(\ref{eq:deco41}) and (\ref{eq:deco42}) represent the same tensor $\hat{T}$, $\hat{P}$ and $\hat{P}'$ are related by an F-move [\eref{eq:fmove}],
\begin{equation}
\hat{P}'^{j_f}_{j_aj_bj_cj_d} = \sum_{j_e}\hat{F}^{j_ej_f}_{j_aj_bj_cj_d} \hat{P}^{j_e}_{j_aj_bj_cj_d}.
\label{eq:fmove2}
\end{equation}
For a different choice of incoming and outgoing indices, the degeneracy tensors $\hat{P}$ and $\hat{P}'$ are related by a different F-move e.g. \fref{fig:fmove1}(a). (Analogous to the rank-$3$ case, a rank-$4$ SU(2)-invariant tensor with all incoming or all outgoing indices corresponds to rank-$5$ intertwiners associated with fusing the four spins into a total spin $j=0$.)

More generally, an SU(2)-invariant tensor $\hat{T}$ with $k$ indices $i_1 = (j_1, t_{j_1}, m_{j_1}), i_2 = (j_2, t_{j_2}, m_{j_2})$ and $i_k = (j_k, t_{j_k}, m_{j_k})$ decomposes as
\begin{align}
(\hat{T})_{i_1 i_2\ldots i_k} \equiv \sum_{j_{e_1} \ldots j_{e_{k-3}}}&\left(\hat{P}^{j_{e_1}j_{e_2}\ldots j_{e_{k-3}}}_{j_{1}j_2 \ldots j_{k}}\right)_{t_{j_1}t_{j_2} \ldots t_{j_k}}\times \nonumber \\
&\left(\hat{Q}^{j_{e_1} j_{e_2}\ldots j_{e_{k-3}}}_{j_{1} j_2\ldots j_{k}}\right)_{m_{j_1}m_{j_2}\ldots m_{j_k}},
\label{eq:Tcanon}
\end{align}
where $\hat{Q}^{j_{e_1}j_{e_2}\ldots j_{e_{k-3}}}_{j_{1} \ldots j_{k}}$ is a rank-$k$ intertwiners of SU(2). The intertwiner $\hat{Q}^{j_{e_1}j_{e_2}\ldots j_{e_{k-3}}}_{j_{1} \ldots j_{k}}$ can further be decomposed into a trivalent tree tensor network made of $\cfuser$ and $\cspliter$ tensors. This decomposition is completely specified by its underlying graph, the \textit{fusion-splitting tree}\cite{fusetree}, that is additionally decorated by labeling its links with $j$'s. Here, the fusion-splitting tree consists of $k-2$ vertices associated with fusions (two incoming arrows and one outgoing arrow) and/or splittings (one incoming arrow and two outgoing arrows), $k-3$ internal edges that interconnect the vertices and $k$ open edges. The internal edges are labeled by $\{j_{e_1}, j_{e_2},\ldots, j_{e_{k-3}}\}$ while the open edges are labeled by $\{j_1,j_2,\ldots,j_k\}$.

We refer to the decomposition $(\hat{P}, \hat{Q})$ as the \textit{canonical decomposition} or the \textit{canonical form} of tensor $\hat{T}$. The canonical form is the most compact description of an SU(2)-invariant tensor $\hat{T}$ in that $\hat{T}$ can be stored in memory by only storing the degeneracy tensors $\hat{P}$ and the underlying fusion-splitting tree. A different choice of the tree will produce different sets of tensors $\hat{P}'$ and $\hat{Q}'$, related to $P$ and $Q$ by F-moves (and also possibly $\braider$ coefficients in case the underlying fusion-splitting tree has crossings). We also refer the reader to App. C where we describe \textit{tree decompositions}, a canonical form of SU(2)-invariant tensors that is based on \textit{splitting trees}, that is, trees made of only splitting vertices.

\begin{figure}[t]
  \includegraphics[width=7cm]{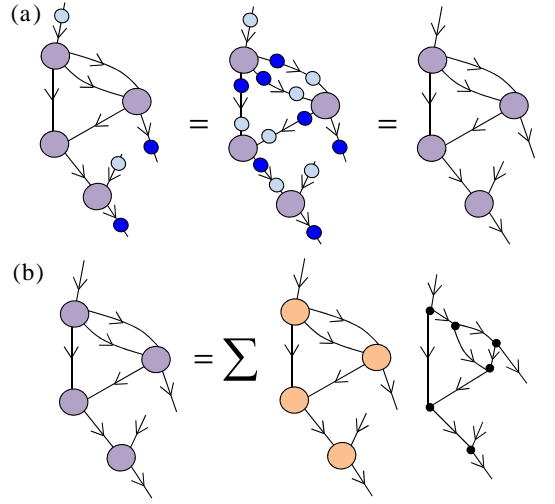}
\caption{(Color online) (a) A tensor network $\mathcal{N}$ made of SU(2)-invariant tensors represents an SU(2)-invariant tensor $\hat{T}$. This is seen by means of two equalities. The first equality is obtained by inserting resolutions of Identity $\hat{I} = \hat{W}_{\textbf{r}}\hat{W}^{\dagger}_{\textbf{r}}$ on each index connecting two tensors in $\mathcal{N}$ (the small dark circles depict the transformation $\hat{W}_{\textbf{r}}$ whereas the small light circles depict the adjoint transformation $\hat{W}^{\dagger}_{\textbf{r}}$). The second equality follows from the fact that each tensor in $\mathcal{N}$ is SU(2)-invariant. (b) For fixed values of $j$'s on all indices, the tensor network $\mathcal{N}$ decomposes into a tensor network made degeneracy tensors and a spin network. The sum is over all $j$'s, and the corresponding $t$'s and $m$'s, that are associated with the contracted indices. \label{fig:symTN}}
\end{figure}


\section{SU(2)-invariant tensor networks\label{sec:tnsu2}}

In this section we describe how to incorporate SU(2) symmetry into tensor networks. We refer to App. \ref{sec:tensor} for a review on the tensor network formalism. 

In particular, in App.~\ref{sec:tensor:linearmap} we review how a tensor network $\mathcal{N}$ can be interpreted as a collection of linear maps composed into a single linear map $\hat{T}$ of which $\mathcal{N}$ is a tensor network decomposition. By introducing a spin operator on the vector space associated to each line of $\mathcal{N}$, we can define a unitary representation of SU(2) on each index of each tensor in $\mathcal{N}$. Then we say that $\mathcal{N}$ is an SU(2)-invariant tensor network if all its tensors are SU(2)-invariant. Notice that, by construction, if $\mathcal{N}$ is an SU(2)-invariant tensor network, then the resulting linear map $\hat{T}$ is also SU(2)-invariant. This is illustrated in \fref{fig:symTN}(a).

\textit{Linear superposition of spin networks.} We can now investigate how the tensor network decomposes if we write each of its tensors $\hat{T}$ in the $(\hat{P},\hat{Q})$ form. For any fixed value of the $j$'s on all the indices, the whole tensor network factorizes into two terms, as illustrated in \fref{fig:symTN}(b). The first one is a tensor network of degeneracy tensors. The second one is a directed graph with edges labeled by spins $j$ and vertices labeled by intertwining operators of SU(2) i.e. the $\hat{Q}$ tensors. This is nothing other than a spin network. Accordingly, an SU(2)-invariant tensor network for the $\ket{\Psi} \in \mathbb{V}^{\otimes L}$ of a lattice $\mathcal{L}$ of $L$ sites can be regarded as a linear superposition of spin networks with $L$ open edges. The number of spin networks in the linear superposition grows exponentially with the size of the tensor network. The expansion coefficients are given by the degeneracy tensors.


\subsection{Tensor network states and algorithms with SU(2) symmetry \label{sec:tnsu2:statesAlgo}}

As reviewed in App.~\ref{sec:tensor:TNstates}, a tensor network $\mathcal{N}$ can be used to describe certain pure states $\ket{\Psi} \in \mathbb{V}^{\otimes L}$ of a lattice $\mathcal{L}$. If $\mathcal{N}$ is an SU(2)-invariant tensor network then it will describe a pure state $\ket{\Psi}$ that has a well-defined total spin $j=0$. That is, an SU(2)-invariant pure state fulfills
\begin{equation}
\hat{J}_{\alpha} \ket{\Psi} = 0, ~~~ \alpha=x,y,z. \nonumber
\end{equation}
In this way we can obtain a more refined version of popular tensor network states such as MPS, TTN, MERA, PEPS, etc. As a variational Ansatz, an SU(2)-invariant tensor network state is more constrained (by the appearance of spin networks in its canonical decomposition) than a regular tensor network state, and consequently it can represent fewer states $\ket{\Psi} \in \mathbb{V}^{\otimes L}$. However, it also depends on fewer parameters (those encoded in the degeneracy tensors). This implies a more economical description, as well as the possibility of reducing computational costs during its manipulation.

In the rest of the section we explain how to reduce computational costs by exploiting the symmetry. This is based on storing and manipulating SU(2)-invariant tensors expressed in the canonical form of \eref{eq:Tcanon} and the consequent decomposition of the tensor network as a linear superposition of spin networks. By decomposing the tensor network in this way, the primitive tensor network manipulations belonging to the set $\mathcal{P}$ reviewed in App.~\ref{sec:tensor:TN}, namely, reversal, permutation and reshaping indices, matrix multiplication and factorizations, can be addressed \textit{separately} for the degeneracy term and the spin network term, see Table \Rmnum{1}.

App.~\ref{sec:tree} describes our implementation of the primitive tensor network manipulations based on tree decompositions of SU(2)-invariant tensors.

\begin{table}
\centering
{
    \begin{tabular}{  c | c | c | c }
    \hline\hline
     Operation & Degeneracy tensors & Intertwiners & Definition\\\hline
     Reversal & \textit{regular} e.g. \eref{eq:bend} & $\mycup, \mycap$ &\specialcell{Eqs.(\ref{eq:cup})-(\ref{eq:cap})\\ \fref{fig:fmove}(c)-(e)}\\ \hline
     Swap & \textit{regular} e.g. \eref{eq:permute} & $\braider$ &\specialcell{Eqs.(\ref{eq:su2swap})-(\ref{eq:braid3}) \\ \fref{fig:cg}(d)}\\ \hline
     Fusion & $\tfuser$ & $\cfuser$ & \specialcell{Eqs.(\ref{eq:cg}),(\ref{eq:fusedecompose}) \\ Figs.\ref{fig:cg},\ref{fig:su2fuse}}\\ \hline
     Splitting & $\tspliter$ & $\cspliter$& \specialcell{Eqs.(\ref{eq:revcg}),(\ref{eq:u1split1})\\ Figs.\ref{fig:cg},\ref{fig:su2fuse}}\\ \hline \hline
    \end{tabular}
    \caption{Transformations that play an instrumental role in the reversal, swap and reshape (fusion and splitting) of indices of SU(2)-invariant tensors when working in the canonical form. These manipulations are addressed separately and differently for the degeneracy tensors and for the structural tensors (intertwiners).
}} \label{table:basictensors}
\end{table}


\subsection{Reversal of indices \label{sec:tnsu2:bend}}
\begin{figure}[t]
  \includegraphics[width=8.65cm]{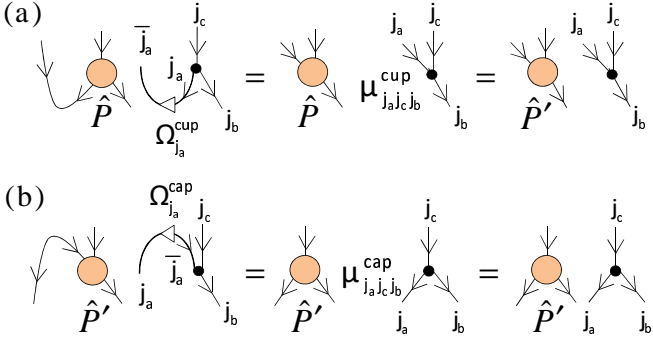}
\caption{(Color online) [\textit{Contrast with \fref{fig:tensorman}(a)}] (a) The leftward bending of an outgoing index $a=(j_a,t_{j_a},m_{j_a})$ of an SU(2)-invariant tensor $\hat{T}$ given in the canonical form $(\hat{P},\cspliter)$ to obtain another SU(2)-invariant tensor $\hat{T}'$, shown in two steps. The first step shows that for fixed values of $j_a, j_b$ and $j_c$, the degeneracy index $(j_a,t_{j_a})$ and the spin index $(j_a,m_{j_a})$ can be bent separately. Degeneracy tensor $\hat{P}_{j_aj_bj_c}$ remains unchanged (like a regular tensor) by bending its index $(j_a,t_{j_a})$. In contrast, bending the index $(j_a,m_{j_a})$ corresponds to ``multiplying'' the structural tensor $\cspliter$ with the cup $\mycup_{j_a}$ such that $(j_a,m_{j_a})$ is contracted. This multiplication corresponds to applying the F-move of \fref{fig:fmove1}(c). The second step corresponds to multiplying the resulting factor $\cupf{a}{c}{b}$ into the degeneracy tensors $\hat{P}$ to obtain  $\hat{T}'$ in the canonical form $(\hat{P}',\cfuser)$, \eref{eq:cup3}. (b) Tensor $\hat{T}$ is recovered from $\hat{T}'$ by bending $\overline{a}$ downward using the cap transformation $\mycap_{j_{\overline{a}}}$ as shown and restoring the canonical form by applying the inverse F-move [\fref{fig:fmove1}(d)], leading to \eref{eq:cap3}.
\label{fig:bend}}
\end{figure}

Unlike regular tensors, bending indices of an SU(2)-invariant tensor $\hat{T}$ produces a different SU(2)-invariant tensor. Moreover, the resulting tensor is generally different when bending an index from the left than when bending it from the right. That is, the ``parity'' of the bend has to be taken into account. For example, consider an SU(2)-invariant tensor $\hat{T}$ with outgoing indices $a = (j_a,t_{j_a},m_{j_a})$ and $b = (j_b,t_{j_b},m_{j_b})$ and incoming index $c = (j_c,t_{j_c},m_{j_c})$ and components [Eqs.(\ref{eq:three22}),(\ref{eq:forReversal})],
\begin{equation}
\hat{T}_{abc} = (\hat{P}_{j_aj_bj_c})_{t_{j_a}t_{j_b}t_{j_c}} \csplitspin{c}{a}{b},\label{eq:simeg}
\end{equation}
and let $\hat{T}'$ denote the SU(2)-invariant tensor that is obtained from $\hat{T}$ by bending the outgoing index $a$ from the \textit{left}. In the canonical form tensor $\hat{T}'$ has components $\hat{T}'_{\overline{a}bc}$,
\begin{equation}
\hat{T}'_{\overline{a}bc} = (\hat{P}'_{\overline{j}_aj_bj_c})_{\overline{t}_{j_a}t_{j_b}t_{j_c}} \cfuser_{\overline{j}_{a}\overline{m}_{j_a},j_{c}m_{j_{c}}\rightarrow j_{b}m_{j_{b}}} \label{eq:temp2}
\end{equation}
where
\begin{equation}
\hat{P}'_{\overline{j}_aj_bj_c} = \cupf{a}{c}{b} \hat{P}_{j_aj_bj_c}.\label{eq:cup3}
\end{equation}
Here $\overline{a} = (\overline{j}_a,\overline{t}_{j_a},\overline{m}_{j_a}) = (j_a, t_{j_a}, -m_{j_a})$ denotes the index obtained by bending $a$ and the factor $\cupf{a}{c}{b}$ is defined according to \eref{eq:cup}.
The derivation of \eref{eq:cup3} is shown in \fref{fig:bend}(a). Note that the components of degeneracy tensor $\hat{P}_{j_aj_bj_c}$ remain unchanged when bending its index $(j_a, t_{j_a})$ (recall that the degeneracy tensors are not constrained by the symmetry). On the other hand, bending of the spin index ($j_a, m_{j_a}$) is subject to the symmetry constraint that $j_a$ and $\overline{j}_a$ must fuse to a total spin $j=0$. This corresponds to multiplying the structural tensor $\cspliter$ with the cup $\mycup_{j_a}$, as shown in the figure.

The SU(2)-invariant tensor $\hat{T}$ may be recovered from $\hat{T}'$ by bending down index $\overline{a}$ from the left. This is achieved by multiplying the structural tensor with the cap $\mycap_{j_a}$, as shown in \fref{fig:bend}(b). We have
\begin{equation}
\hat{P}_{j_aj_bj_c} = \capf{a}{c}{b} \hat{P}'_{\overline{j}_aj_bj_c},\label{eq:cap3}
\end{equation}
where the factor $\capf{a}{c}{b}$ is defined according to \eref{eq:cap}.

Next, bending the same index $a$ of the SU(2)-invariant tensor $\hat{T}$ from the \textit{right} may result in a different SU(2)-invariant tensor $\hat{T}''$. In the canonical form, the SU(2)-invariant tensors $\hat{T}'$ and $\hat{T}''$ are related as (see \fref{fig:fmove1}(e))
\begin{equation}
\hat{P}''_{\overline{j}_aj_bj_c} = \braider_{\overline{j}_a,j_a \rightarrow 0} \hat{P}'_{\overline{j}_aj_bj_c}.
\end{equation}
Notice that in case $j_a \in \{0,1,2,\ldots\}$ the factor $\braider_{\overline{j}_a,j_a \rightarrow 0}$ is equal to one and the leftward or rightward bending of $a$ results in the same final tensor, $\hat{T}'' = \hat{T}'$.

More generally, consider bending an outgoing index $i_r=(j_r,t_{j_r},m_{j_r})$ of a rank-$k$ SU(2)-invariant tensor given in the canonical form of \eref{eq:Tcanon}. When bending $i_r$ from the left \eref{eq:cup3} is generalized to
\begin{align}
\hat{P}'^{j_{e_1}\ldots j_{e_{k-3}}}_{j_{1} \ldots \overline{j}_r \ldots j_{k}} &= \cupf{r}{}{e_s} \hat{P}^{j_{e_1}\ldots j_{e_{k-3}}}_{j_{1} \ldots j_r \ldots j_{k}}, \label{eq:cupk}
\end{align}
where $j_r, j_{e_s}$ and $j$ label three edges that meet at a vertex in the underlying fusion-splitting tree. Depending on the specific choice of the tree, here $j$ may label either an open or an internal edge. Equation (\ref{eq:cap3}) can be generalized in a similar way for the downward bending of an index of a rank-$k$ SU(2)-invariant tensor.

\begin{figure}[t]
  \includegraphics[width=8.5cm]{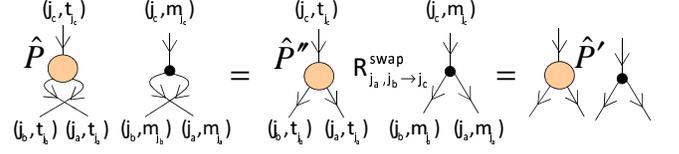}
\caption{(Color online) [\textit{Contrast with \fref{fig:tensorman}(b)}] Swapping indices $a=(j_a,t_{j_a},m_{j_a})$ and $b=(j_b,t_{j_b},m_{j_b})$ of an SU(2)-invariant tensor $\hat{T}$ that is given in the canonical form $(\hat{P},\cspliter)$ to obtain another SU(2)-invariant tensor $\hat{T}'$, shown in two steps. The first step shows that for fixed values of $j_a, j_b$ and $j_c$ the corresponding indices of the degeneracy tensor and the structural tensor can be swapped separately. Swapping indices $(j_a,t_{j_a})$ and $(j_b,t_{j_b})$ of degeneracy tensor $\hat{P}_{j_aj_bj_b}$ is achieved by swapping the labels $j_a$ and $j_b$ and also the indices $t_{j_a}$ and $t_{j_b}$. The analogous swap of the indices $(j_a,m_{j_a})$ and $(j_b,m_{j_b})$ of the structural tensor is straightforward and can be performed algebraically [\eref{eq:su2swap}] - the resulting tensor is simply proportional to another Clebsch-Gordan tensor $\csplitspin{c}{b}{a}$ times the factor $\braid{a}{b}{c}$. The second step corresponds to multiplying the factor $\braid{a}{b}{c}$ into the degeneracy tensors $\hat{P}_{j_aj_bj_c}$ to obtain $\hat{T}'$ in the canonical form, \eref{eq:perm11}.
\label{fig:permutecanon}}
\end{figure}


\subsection{Permutation of indices\label{sec:tnsu2:permute}}

Next consider permuting indices of an SU(2)-invariant tensor that is given in the canonical form. As reviewed in App.\ref{sec:tensor:manipulations}, an arbitrary permutation of indices of a tensor can be decomposed into a sequence of reversals and pairwise swaps. Here we explain how to swap two adjacent indices, both `in' or both `out', of an SU(2)-invariant tensor given in the canonical form. For example, tensor $\hat{T}'$ of \eref{eq:permute} obtained by swapping indices $a$ and $b$ of an SU(2)-invariant tensor $\hat{T}$ [\eref{eq:simeg}] has components
\begin{equation}
\hat{T}'_{bac} = (\hat{P}'_{j_bj_aj_c})_{t_{j_b}t_{j_a}t_{j_c}} \csplitspin{c}{b}{a},
\end{equation}
where
\begin{equation}
(\hat{P}'_{j_bj_aj_c})_{t_{j_b}t_{j_a}t_{j_c}} = \braid{a}{b}{c} (\hat{P}_{j_aj_bj_c})_{t_{j_a}t_{j_b}t_{j_c}},\label{eq:perm11}
\end{equation}
as explained by \fref{fig:permutecanon}.

\begin{figure}[t]
  \includegraphics[width=7cm]{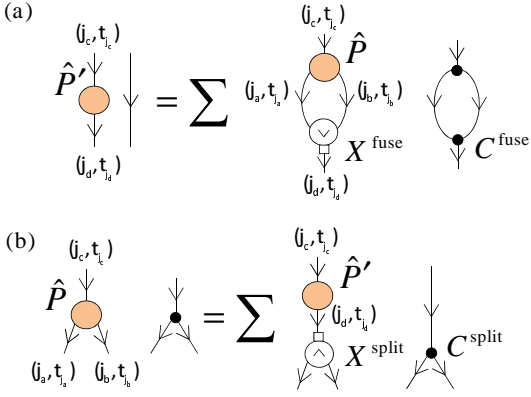}
\caption{(Color online) [\textit{Contrast with \fref{fig:tensorman}(d)-(e)}] (a) Fusion, by means of the transformation $\fuser$, of two indices of an SU(2)-invariant tensor $\hat{T}$ given in the canonical form $(\hat{P},\cspliter)$ to obtain an SU(2)-invariant matrix $\hat{T}'$. For fixed values of $j_a, j_b$ and $j_c$, the fusion is addressed separately for the degeneracy tensors and structural tensors by decomposing the transformation $\fuser$ into $\tfuser$ and $\cfuser$ parts. Degeneracy indices are fused by multiplying tensors $\hat{P}_{j_aj_bj_c}$ with $\tfuser$ as shown. For a fixed value of $j_c$ tensor $\hat{P}'_{j_cj_d}$ ($j_d=j_c$) is the sum of all reshaped degeneracy tensors $\hat{P}_{j_aj_bj_c}$ with compatible labels $j_a, j_b$ and $j_c$. The fusion of the spin indices is trivial since the structural tensor cancels out with the applied $\cfuser$ [\fref{fig:cg}(c)], leading to \eref{eq:su2fusecanon}. (b) Tensor $\hat{T}$ is recovered by analogously splitting back the index $(j_d,t_{j_d})$ of $\hat{T}'$ using the transformation $\spliter$ which decomposes into $\tspliter$ and $\cspliter$ parts.
\label{fig:reshapecanon}}
\end{figure}

More generally, consider the swap of indices $i_{r} = (j_r, t_{j_r}, m_{j_r})$ and $i_{r+1} = (j_{r+1}, t_{j_{r+1}}, m_{j_{r+1}})$ of a rank-$k$ SU(2)-invariant tensor. Let us work in a canonical form in which the two indices $i_r$ and $i_{r+1}$ fuse to an intermediate index $i_{e_s}=(j_{e_s}, t_{j_{e_s}}, m_{j_{e_s}})$ in the underlying fusion-splitting tree. Then \eref{eq:perm11} can be readily generalized as
\begin{equation}
\hat{P}'^{j_{e_1}\ldots j_{e_{k-3}}}_{j_{1} \ldots j_{r+1}, j_r \ldots j_{k}} = \braid{r}{r+1}{e_s} \hat{P}^{j_{e_1}\ldots j_{e_{k-3}}}_{j_{1} \ldots j_{r}, j_{r+1} \ldots j_{k}}.\label{eq:kperm}
\end{equation}

Notice that the canonical form of an SU(2)-invariant tensor facilitates a computational speedup for permutation of indices since computational cost is incurred only by the permutation of indices of the degeneracy tensors. Figure~\ref{fig:permutereshapecompare} illustrates the computational speedup corresponding to a permutation of indices performed using our reference implementation in \small{MATLAB}.


\subsection{Reshape of indices\label{sec:tnsu2:reshape}}

Two adjacent \textit{outgoing} indices of an SU(2)-invariant tensor that is given in the canonical form can be fused together by using the transformation $\fuser$ [\eref{eq:basischange}]. For example, tensor $\hat{T}'$ of \eref{eq:fuse} obtained by fusing indices $a$ and $b$ of the SU(2)-invariant tensor $\hat{T}$ [\eref{eq:simeg}] has components
\begin{equation}
\hat{T}'_{dc} = (\hat{P}'_{j_dj_c})_{t_{j_d}t_{j_c}}\delta_{j_dj_c}\delta_{m_{j_d}m_{j_c}},\label{eq:fuse3}
\end{equation}
where (see \fref{fig:reshapecanon}(a))
\begin{align}
(\hat{P}'_{j_dj_c})_{t_{j_d}t_{j_c}} = \sum_{j_aj_b}\sum_{t_{j_a}t_{j_b}} \tfusespin{a}{b}{d} (\hat{P}_{j_aj_bj_c})_{t_{j_a}t_{j_b}t_{j_c}}.\label{eq:su2fusecanon}
\end{align}
(The sum is over all $j_a$ and $j_b$ that are compatible with $j_c$.)

The original tensor $\hat{T}$ may be recovered from $\hat{T}'$ by splitting index $d$ back into indices $a$ and $b$. This is achieved by using the inverse transformation $\spliter$ [\eref{eq:revbasischange}]. In the canonical form we have (see \fref{fig:reshapecanon}(b)),
\begin{align}
(\hat{P}_{j_aj_bj_c})_{t_{j_a}t_{j_b}t_{j_c}} = \sum_{j_d}\sum_{t_{j_d}} \tsplitspin{d}{a}{b} (\hat{P}'_{j_dj_c})_{t_{j_d}t_{j_c}}.\label{eq:su2splitcanon}
\end{align}

More generally, consider reshaping indices of a rank-$k$ SU(2)-invariant tensor, for instance, by fusing two \textit{outgoing} indices $i_{r} = (j_r, t_{j_r}, m_{j_r})$ and $i_{r+1} = (j_{r+1}, t_{j_{r+1}}, m_{j_{r+1}})$. Let us once again work in a canonical form in which the adjacent indices $i_r$ and $i_{r+1}$ fuse to an intermediate index $i_{e_s}=(j_{e_s}, t_{j_{e_s}}, m_{j_{e_s}})$ in the underlying fusion-splitting tree. In such a canonical form \eref{eq:su2fusecanon} generalizes to
\begin{align}
\hat{P}'^{j_{e_1}\ldots  j_{e_{k-3}}}_{j_{1} \ldots j_{e_s} \ldots j_{k}} = \sum \tfusespin{r}{r+1}{e_s} \hat{P}^{j_{e_1}\ldots j_{e_{k-3}}}_{j_{1} \ldots j_r, j_{r+1} \ldots j_{k}},\label{eq:kfuse}
\end{align}
where the sum is over the two degeneracy indices $(j_r, t_{j_r})$ and $(j_{r+1}, t_{j_{r+1}})$ that are fused.

\begin{figure}[t]
  \includegraphics[width=7cm]{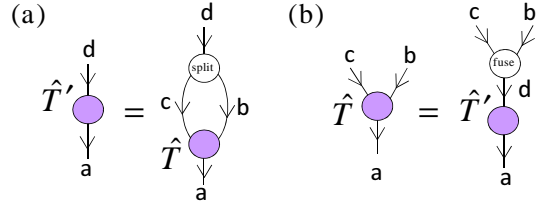}
\caption{(Color online) When reshaping incoming indices transformations $\fuser$ and $\spliter$ reverse roles. (a) Fusion of incoming indices is by means of transformation $\spliter$. (b) Splitting of an incoming index is by means of transformation $\fuser$.\label{fig:reshapein}}
\end{figure}

Finally, when fusing and splitting \textit{incoming} indices, the transformations $\fuser$ and $\spliter$ reverse roles, as illustrated in Fig. \ref{fig:reshapein}.

\begin{figure}[t]
  \includegraphics[width=8cm]{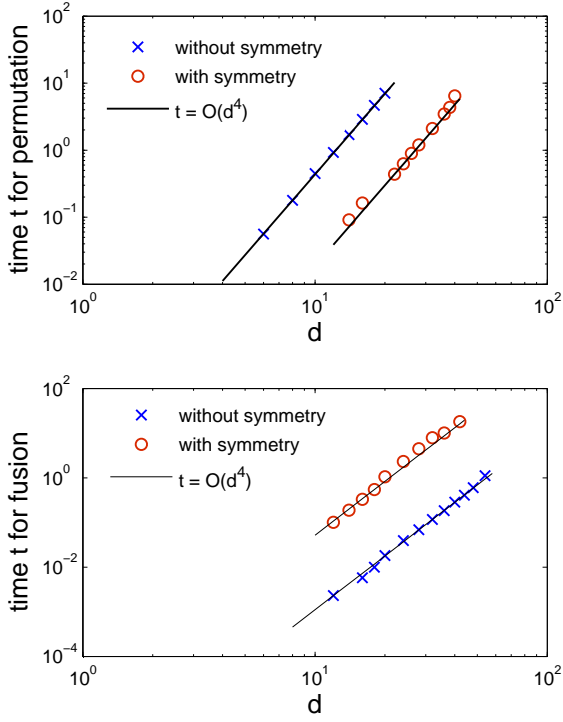}
\caption{(Color online) Computation times (in seconds) required to permute indices of a rank-four tensor $\hat{T}$ as a function of the size of the indices. All four indices of $\hat{T}$ have the same size $9d$, and therefore the tensor contains $|\hat{T}| = 9^4d^4$ coefficients. The figures compare the time required to perform these operations using a regular tensor and an SU(2)-invariant tensor, where in the second case each index contains three different values of spin $j=0,1,2$, each with degeneracy $d$, and the canonical form of \eref{eq:deco41} is used. The upper figure shows the time required to permute two indices: For large $d$ exploiting the symmetry of an SU(2)-invariant tensor by using the canonical form results in shorter computation times. The lower figure shows the time required to fuse two adjacent indices. In this case maintaining the canonical form requires more computation time. Notice that in both figures the asymptotic cost scales as $O(d^4)$, or the size of $\hat{T}$, since this is the number of coefficients which need to be rearranged. We note that the fixed-cost overheads associated with symmetric manipulations could potentially vary substantially with choice of programming language, compiler, and machine architecture. The results given here show the performance of our MATLAB implementation of SU(2) symmetry.}
\label{fig:permutereshapecompare}
\end{figure}

Notice in \eref{eq:kfuse} that reshape of indices in the canonical form corresponds to rearranging components of the degeneracy tensors and then taking a linear combination of them. This requires more work, and can therefore incur additional computational cost, than reshaping indices of regular tensors which is a simple rearrangement of the tensor components. For instance, \fref{fig:permutereshapecompare} shows that fusing indices of SU(2)-invariant tensors can be more expensive than fusing indices of regular tensors.

\subsection{Multiplication of two SU(2)-invariant matrices\label{sec:tnsu2:multiply}}

Let $\hat{M}$ and $\hat{N}$ be two SU(2)-invariant matrices given in the canonical form
\begin{align}
\hat{M} = \bigoplus_j (\hat{M}_j \otimes \hat{I}_{2j+1}), ~~~ \hat{N} = \bigoplus_j (\hat{N}_j \otimes \hat{I}_{2j+1}). \label{eq:matmult11}
\end{align}
Then the SU(2)-invariant matrix $\hat{T} = \hat{M} \hat{N}$ obtained by multiplying together matrices $\hat{M}$ and $\hat{N}$ has the canonical form
\begin{equation}
	\hat{T} = \bigoplus_j (\hat{T}_j \otimes \hat{I}_{2j+1}),\label{eq:matmult4}
\end{equation}
where $\hat{T}_j$ is obtained by multiplying matrices $\hat{M}_j$ and $\hat{N}_j$,
\begin{equation}
\hat{T}_j = \hat{M}_j \hat{N}_j.\label{eq:blockmult}
\end{equation}
Clearly, computational gain is obtained as a result of performing the multiplication $\hat{T} = \hat{R}\hat{S}$ block-wise. This is illustrated by the following example.

\begin{figure}[t]
  \includegraphics[width=8cm]{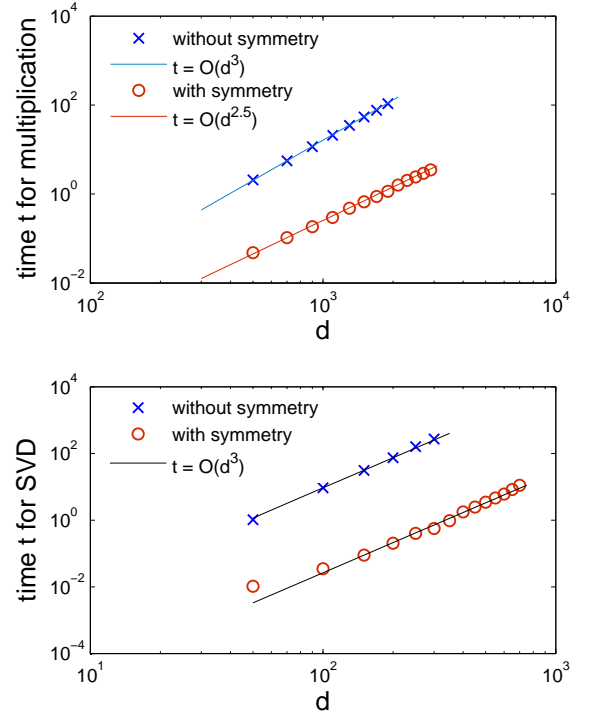}
  \caption{(Color online) Computation times (in seconds) required to multiply two matrices (upper panel) and to perform a singular value decomposition (lower panel) as a function of the size of the indices. Matrices of size $9d \times 9d$ are considered. The figures compare the time required to perform these operations using regular matrices and SU(2)-invariant matrices, where for the SU(2) matrices each index contains three different values of the spin $j=0,1,2$, each with degeneracy $d$, and the canonical form of \eref{eq:struct2} is used. That is, each matrix decomposes into three blocks of size $d \times d$. For large $d$ exploiting the block diagonal form of SU(2)-invariant matrices results in shorter computation time for both multiplication and singular value decomposition. The asymptotic cost scales with $d$ as $O(d^3)$ while the size of the matrices grows as $O(d^2)$.(For matrix multiplication a tighter bound of $O(d^{2.5})$ for the scaling of computational time with $d$ is seen in this example.) We note that the fixed-cost overheads associated with symmetric manipulations could potentially vary substantially with choice of programming language, compiler, and machine architecture. The results given here show the performance of our MATLAB implementation of SU(2) symmetry.
  \label{fig:multsvdcompare}}
\end{figure}

\textit{Example 5.} Consider vector space $\mathbb{V}$ that decomposes as $\mathbb{V} \cong d_j\mathbb{V}_j$ where $j$ assumes values $1,\cdots,q$ and let $d_j=d, \forall j$. The dimension of the space $\mathbb{V}$ is
\begin{equation}
\dim(\mathbb{V}) = d\sum_{j=1}^{q}\Delta_j=dq(q+2).\nonumber
\end{equation}
An SU(2)-invariant matrix $\hat{T}: \mathbb{V} \rightarrow \mathbb{V}$ decomposes into $q$ blocks $\hat{T}_j$ where each block has size $d\times d$. Therefore, the SU(2)-invariant matrix $\hat{T}$ contains $d^2q$ coefficients. For comparison, a regular matrix of the same size contains $d^2q^2(q+2)^2$ coefficients, a number greater by a factor of $O(q^3)$. Let us now consider multiplying two such matrices. We use an algorithm that requires $O(l^3)$ computational time to multiply two matrices of size $l\times l$. The cost of performing $q$ multiplications of $d\times d$ blocks in \eref{eq:blockmult} scales as $O(d^3q)$. In contrast, the cost of multiplying two regular matrices of the same size scales as $O(d^3q^3(q+2)^3)$, requiring $O(q^5)$ times more computational time. Figure \ref{fig:multsvdcompare} shows a comparison of the computation times when multiplying two matrices for both SU(2)-invariant and regular matrices.\markend


\subsection{Factorization of an SU(2)-invariant matrix\label{sec:tnsu2:factorize}}

The factorization of an SU(2)-invariant matrix $\hat{T}$ can also benefit from the block-diagonal structure. Consider, for instance, the singular value decomposition (SVD), $\hat{T} = \hat{U}\hat{S}\hat{V}$, where $\hat{U}$ and $\hat{V}$ are unitary matrices and $\hat{S}$ is a diagonal matrix with non-negative components. If $\hat{T}$ has the canonical form
\begin{equation}
\hat{T} = \bigoplus_j (\hat{T}_j \otimes \hat{I}_{2j+1}),
\end{equation}
we can obtain the SU(2)-invariant matrices
\begin{align}
	\hat{U} = \bigoplus_j (\hat{U}_j \otimes \hat{I}_{2j+1}),\nonumber \\
	~\hat{S} = \bigoplus_j (\hat{S}_j \otimes \hat{I}_{2j+1}),\nonumber \\
	~\hat{V} = \bigoplus_j (\hat{V}_j \otimes \hat{I}_{2j+1}),\nonumber
	\label{eq:svd0}
\end{align}
by performing SVD of each degeneracy matrix $\hat{T}_j$ independently,
\begin{equation}
	\hat{T}_j = \hat{U}_j \hat{S}_j \hat{V}_j.
	\label{eq:svd1}
\end{equation}
A different factorization of $\hat{T}$, such as spectral decomposition or polar decomposition, can be obtained by the analogous factorization of the blocks $\hat{T}_j$. The computational savings are analogous to those described in Example 5 for the multiplication of matrices. Figure \ref{fig:multsvdcompare} shows a comparison of computation times required to perform a singular value decomposition on SU(2)-invariant and regular matrices using MATLAB.


\subsection{Discussion}
In this section we have seen that SU(2)-invariant tensors can be written in the canonical form of \eref{eq:Tcanon} and that this canonical form offers a compact description in terms of components that are not constrained by the symmetry (degeneracy components). Moreover reversal, permutation and reshape of indices can be implemented only at the level of the degeneracy tensors while the intertwiners, which are completely constrained by the symmetry, only contribute numerical factors in the manipulation. In particular, these factors depend only on the $j$'s and not on the $m$'s [Eqs.~(\ref{eq:cupk}),(\ref{eq:kperm}),(\ref{eq:kfuse})]. The canonical decomposition therefore allows for a \textit{manifestly} SU(2)-invariant treatment of tensors - that is, an SU(2)-invariant tensor is completely specified by the degeneracy tensors and the underlying fusion-splitting tree. Computationally, this implies a reduction both in memory cost, since the Clebsch-Gordan coefficients are not required to be stored in memory (instead we only store the $\hat{F}$ and $\braider$ coefficients), and in computational times since the Clebsch-Gordan coefficients are not required to be manipulated explicitly.

We have also seen that maintaining the canonical form during tensor manipulations adds some computational overhead when reversing or reshaping (fusing or splitting) indices but reduces computation time when permuting indices (for sufficiently large tensors) and when multiplying or factorizing matrices (for sufficiently large matrix sizes). The cost of reversing, reshaping and permuting indices is proportional to the size $|\hat{T}|$ of the tensors, whereas the cost of multiplying and factorizing matrices is a larger power of the matrix size, for example, $|\hat{T}|^{3/2}$. The use of the canonical form when manipulating large tensors therefore frequently results in an overall reduction in computation time, making it a very attractive option in the context of tensor network algorithms. This is exemplified in the next section, where we apply the MERA to study the ground state of a quantum spin chain which has an SU(2) symmetry.

On the other hand, however, the cost of maintaining the invariant tensors in the canonical form becomes more relevant when dealing with smaller tensors. In the next section we will also see that in some situations, this additional cost may significantly reduce, or even offset, the benefits of using the canonical form. In this event, and in the specific context of algorithms where the same tensor manipulations are iterated many times, it is possible to significantly decrease the additional cost by precomputing the parts of the tensor manipulations that are repeated on each iteration (see App.~\ref{sec:tree:precom}). The performance of precomputing is illustrated in the next section.


\section{Tensor network algorithms with SU(2) symmetry: A practical example\label{sec:mera}}

In this section we demonstrate the implementation of SU(2) symmetry in tensor network algorithms with practical examples. We do so in the context of the Multi-scale Entanglement Renormalization Ansatz, or MERA, and present numerical results from our reference implementation of SU(2) symmetry in MATLAB.


\subsection{Multi-scale entanglement renormalization ansatz\label{sec:mera:ansatz}}

\begin{figure}[t]
  \includegraphics[width=8cm]{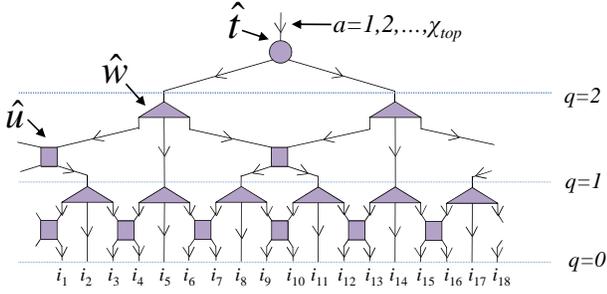}
\caption{(Color online) MERA for a system of $L=2\times 3^{2}= 18$ sites made of two layers of disentanglers $\hat{u}$ and isometries $\hat{w}$ and a top tensor $\hat{t}$.}
\label{fig:mera}
\end{figure}

Figure \ref{fig:mera} shows a MERA that represent states $\ket{\Psi} \in \mathbb{V}^{(\mathcal{L})}$ of a lattice $\mathcal{L}$ made of $L=18$ sites. Recall that the MERA is made of layers of isometric tensors, known as disentanglers $\hat{u}$ and isometries $\hat{w}$, that implement a coarse-graining transformation. In this particular scheme, isometries map three sites into one and the coarse-graining transformation reduces the $L=18$ sites of $\mathcal{L}$ into two sites using two layers of tensors. A collection of states on these two sites is then encoded in a top tensor $\hat{t}$, whose upper index $a=1,2,\cdots, \chi_{\tiny \mbox{top}}$ is used to label $\chi_{\tiny \mbox{top}}$ states $\ket{\Psi_a} \in \mathbb{V}^{(\mathcal{L})}$. This particular arrangement of tensors corresponds to the 3:1 MERA described in Ref.~\onlinecite{Evenbly09}. We will consider a MERA analogous to that of \fref{fig:mera} but with $Q$ layers of disentanglers and isometries, which we will use to describe states on a lattice $\mathcal{L}$ made of $2\times 3^{Q}$ sites.

We will use the MERA as a variational ansatz for ground states and excited states of quantum spin models described by a local Hamiltonian $\hat{H}$. In order to find an approximation to the ground state of $\hat{H}$, we set $\chi_{\tiny\mbox{top}}=1$ and optimize the tensors in the MERA so as to minimize the expectation value
\begin{equation}
	\bra{\Psi} \hat{H} \ket{\Psi}
\end{equation}
where $\ket{\Psi}\in \mathbb{V}^{(\mathcal{L})}$ is the pure state represented by the MERA. In order to find an approximation to the $\chi_{\tiny\mbox{top}}>1$ eigenstates of $\hat{H}$ with lowest energies, we optimize the tensors in the MERA so as to minimize the expectation value
\begin{equation}
	\sum_{a=1}^{\chi_{\tiny\mbox{top}}}\bra{\Psi_a} \hat{H} \ket{\Psi_a},~~~~\braket{\Psi_a}{\Psi_{a'}} = \delta_{aa'}.
\end{equation}
The optimization is carried out using the MERA algorithm described in Ref.~\onlinecite{Evenbly09}, which requires contracting tensor networks (by sequentially multiplying pairs of tensors) and performing singular value decompositions.


\subsection{MERA with SU(2) symmetry\label{sec:mera:ansatz}}

An SU(2)-invariant version of the MERA, or SU(2) MERA for short, is obtained by simply considering SU(2)-invariant versions of all of the isometric tensors, namely the disentanglers $\hat{u}$, isometries $\hat{w}$, and the top tensor $\hat{t}$. This requires assigning a spin operator to each index of the MERA. We can characterize the spin operator by two vectors, $\vec{j}$ and $\vec{d}$: a list of the different values the spin takes and the degeneracy associated with each such spin, respectively. For instance, an index characterized by $\vec{j} = \{0, 1\}$ and $\vec{d} = \{2, 1\}$ is associated to a vector space $\mathbb{V}$ that decomposes as $\mathbb{V} \cong d_0\mathbb{V}_0 \oplus d_1 \mathbb{V}_1$ with $d_0 = 2$ and $d_1 = 1$.

Let us explain how a spin operator is assigned to each link of the MERA. Each open index of the first layer of disentanglers corresponds to one site of $\mathcal{L}$. The spin operator on any such index is therefore given by the quantum spin model under consideration. For example, a lattice with a spin-$\frac{1}{2}$ associated to each site corresponds to assigning spin-$\frac{1}{2}$ operators [\eref{eq:eg2c1}] to each of the open indices.

For the open index of the tensor $\hat{t}$ at the very top the MERA, the assignment of spins will depend on spin sector $J$ that one is interested in. For instance, in order to find an approximation to the $8^{\tiny \mbox{th}}$ lowest [($2J+1$)-fold degenerate] energy level of the quantum spin model within the spin sector $J$, we choose $\vec{j} = \{J\}$ and $\vec{d} = \{8\}$.

For each of the remaining indices of the MERA, the assignment of the pair $(\vec{j}, \vec{d})$ needs careful consideration and a final choice may only be possible after numerically testing several options and selecting the one which produces the lowest expectation value of the energy.

\begin{figure}[t]
  \includegraphics[width=7.5cm]{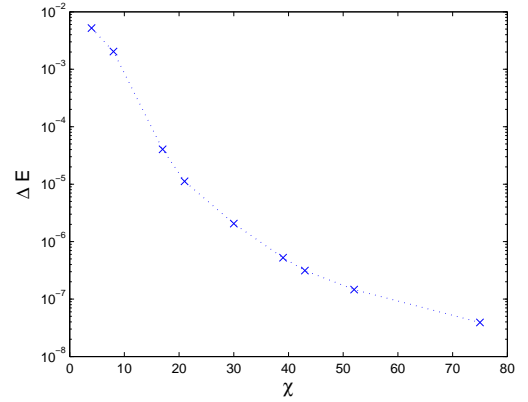}
\caption{(Color online) Error in ground state energy $\Delta E$ (in the singlet sector $J=0$) as a function of $\chi$ for the Heisenberg model with $2L=108$ spins and periodic boundary conditions. The error is calculated with respect to the exact solutions and is seen to decay polynomially with $\chi$ for the particular choice of spins listed in Table \ref{table:degdist}. \label{fig:gserror}}
\end{figure}

\begin{table}
\centering 
\begin{tabular}{c| c| c} 
\hline\hline 
Total bond dimension, $\chi$ &  Spins $\vec{j}$ & Degeneracies $\vec{d}$ \\ [0.5ex] 
\hline
4 & $\{0,1\}$ & $\{1,1\}$ \\
8 & $\left\{0,1\right\}$ & $\left\{2,2\right\}$  \\
17 & $\left\{0,1,2\right\}$ & $\left\{3,3,1\right\}$  \\
21 & $\left\{0,1,2\right\}$ & $\left\{4,4,1\right\}$  \\
30 & $\left\{0,1,2\right\}$ & $\left\{5,5,2\right\}$  \\
39 & $\left\{0, 1, 2\right\}$ & $\left\{6,6,3\right\}$  \\ 
43 & $\left\{0, 1, 2\right\}$ & $\left\{7,7,3\right\}$  \\
52 & $\left\{0, 1, 2\right\}$ & $\left\{8,8,4\right\}$  \\
75 & $\left\{0, 1, 2, 3\right\}$ & $\left\{9,9,5,2\right\}$  \\
[1ex]
\hline 
\hline 
\end{tabular}
\caption{Example of spin assignment in an SU(2) MERA for the anti-ferromagnetic spin chain with $L = 54$ sites (or $108$ spins).\label{table:degdist} 
}
\end{table}

For demonstrative purposes, we will use the SU(2) MERA as a variational ansatz to obtain the ground state and excited states of the spin-$\frac{1}{2}$ antiferromagnetic quantum Heisenberg chain that is given by,
\begin{align}
\hat{H} = \sum_{s=1}^L \hat{h}^{(s, s+1)}, \label{eq:heisenberg}
\end{align}
where
\begin{align}
\hat{h}^{(s, s+1)} &= 4\left(\hat{J}_x^{(s)}\hat{J}_x^{(s+1)} + \hat{J}_y^{(s)}\hat{J}_y^{(s+1)} + \hat{J}_z^{(s)}\hat{J}_z^{(s+1)}\right),
\end{align}
$\hat{J}_{x}, \hat{J}_y$ and $\hat{J}_z$ are the spin-$\frac{1}{2}$ operators [\eref{eq:eg2c1}]. The hamiltonian $\hat{H}$ commutes with the group SU(2) which is readily verified by noticing that
\begin{equation}
[\hat{h}^{(s, s+1)}, \hat{J}^{(s)}_{\alpha} + \hat{J}^{(s+1)}_{\alpha}] = 0,~~~\alpha={x,y,x}.
\end{equation}
Each spin-$\frac{1}{2}$ degree of freedom of the Heisenberg chain is described by a vector space $\mathbb{V} \cong \mathbb{V}_{\half}$ that is spanned by two orthonormal states [\eref{eq:basiseg1}],
\begin{equation}
\ket{j=\half, m=-\half} \mbox{ and } \ket{j=\half, m=\half}. \nonumber
\end{equation}
For computational convenience, we will consider a lattice $\mathcal{L}$ where each site contains two spins. Therefore each site of $\mathcal{L}$ is described by a space $\mathbb{V} \cong \mathbb{V}_0 \oplus \mathbb{V}_1$, where $d_0=1$ and $d_1=1$, also discussed in Example B6. This corresponds to the assignment $\vec{j} = \{0, 1\}$ and $\vec{d} = \{1, 1\}$ at the open legs at the bottom of the MERA. Thus, a lattice $\mathcal{L}$ made of $L$ sites corresponds to a chain of $2L$ spins.

Table \ref{table:degdist} lists some of the spin and degeneracy dimensions assignment (for the internal links of the MERA) that we have used in the numerical computations for $L=54$ (or 108 spins). For a given value of $\vec{j}$ and $\vec{d}$ the corresponding dimension $\chi$ can be obtained as,
\begin{equation}
\chi = \sum_{j \in \vec{j}} (2j+1)\times d_j.
\end{equation}
Figure~\ref{fig:gserror} shows the error in the ground state energy of the Heisenberg chain as a function of the bond dimension $\chi$, for the assignments of $\vec{j}$ and $\vec{d}$ that are listed in Table \ref{table:degdist}. For the choice of spin assignments listed in the table the error is seen to decay polynomially with $\chi$, indicating increasingly accurate approximations to the ground state.

\begin{figure}[t]
  \includegraphics[width=7cm]{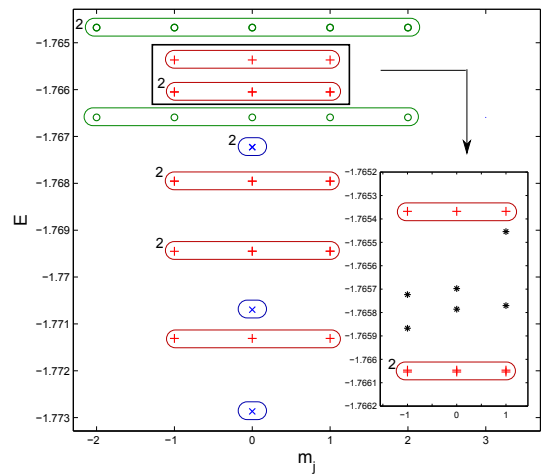}
\caption{(Color online)  Low energy spectrum of $\hat{H}$ with $L=54$ sites (=108 spins). Depicted states have spin $J$ of zero ($\times$, blue loops), one (+, red loops), or two ($\circ$, green loop). The superscript $^2$ close to the boundary of a loop indicates that the loop encloses two-fold degenerate states e.g., the second, third and fourth spin-1 triplets are twofold degenerate. The inset shows a zoom in of the region enclosed within the box. It compares the energies of the two-fold degenerate spin-one states within the box with those obtained using the regular MERA (black asterix points). Since the symmetry is not protected the states obtained with the regular MERA corresponding to different $m_J$ do not have the same energies.
\label{fig:spec}}
\end{figure}


\subsection{Advantages of exploiting the symmetry}

We now discuss some of the advantages of using the SU(2) MERA.


\subsubsection{Selection of spin sector}

An important advantage of the SU(2) MERA is that it exactly preserves the SU(2) symmetry. In other words, the states resulting from a numerical optimization are exact eigenvectors of the total spin operator $\textbf{J}^2 : \mathbb{V}^{(\mathcal{L})} \rightarrow \mathbb{V}^{(\mathcal{L})}$. In addition, the total spin $J$ can be pre-selected at the onset of optimization by specifying it in the open index of the top tensor $\hat{t}$.

Figure~\ref{fig:spec} shows the low energy spectrum of the Heisenberg model $\hat{H}$ for a periodic system of $L=54$ sites (or $108$ spins), including the ground state and several excited states in the spin sectors $J=0, 1, 2$. The states have been organized according to spin projection $m_J$. We see that states with different spin projections $m_J$ (for a given $J$) are obtained to be exactly degenerate, as implied by the symmetry.

\begin{figure}[t]
  \includegraphics[width=8cm]{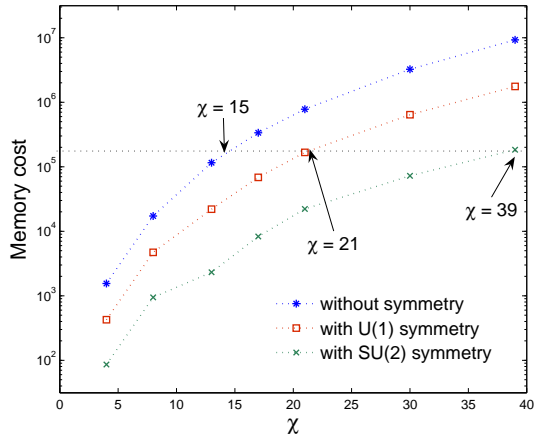}
\caption{(Color online)  Memory cost (in number of components) for storing the MERA as a function of the bond dimension $\chi$. The horizontal line on this graph shows that this reduction in memory cost equates to the ability to store MERAs with a higher bond dimension $\chi$: For the same amount of memory required to store a MERA with bond dimension $\chi=15$ one may choose instead to store a U(1)-symmetric MERA with $\chi=26$ or an SU(2)-symmetric MERA with $\chi=39$. \label{fig:memcompare}}
\end{figure}

\begin{figure}[t]
  \includegraphics[width=8cm]{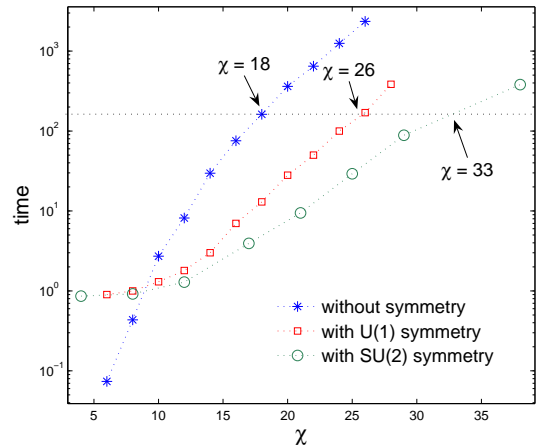}
\caption{(Color online) Computation time (in seconds) for one iteration of the MERA energy minimization algorithm as a function of the bond dimension $\chi$. For sufficiently large $\chi$ exploiting the SU(2) symmetry leads to reductions in computation time. The horizontal line on this graph shows that this reduction in computation time equates to the ability to evaluate MERAs with a higher bond dimension $\chi$: For the same cost per iteration incurred when optimizing a regular MERA in MATLAB with bond dimension $\chi=18$ one may choose instead to optimize a U(1)-symmetric MERA with $\chi=26$ or an SU(2)-symmetric MERA $\chi=33$.\label{fig:meracompare}}
\end{figure}

Similar computations can be performed with the regular MERA. However, the regular MERA cannot guarantee that the states obtained in this way are exact eigenvectors of $\textbf{J}^2$. Instead the resulting states are likely to have total spin fluctuations. This is shown in inset of \fref{fig:spec}, which corresponds to the zoom in of the region in the plot that is enclosed within the box. The inset shows (black asterix points) the corresponding energies obtained for the enclosed two-fold degenerate $J=1$ states using the regular MERA. We see that the states corresponding to different values of $m_J$ are obtained with different energies.

Also note that by using the SU(2) MERA, the three sectors $J=0,1$ and $2$ can be addressed with independent computations. This implies, for instance, that finding the gap between the first singlet ($J=0$) and the first $J=2$ state, can be addressed with two independent computations by respectively setting $(J=0, \chi_{\tiny\mbox{top}}=1)$ and $(J=2, \chi_{\tiny\mbox{top}}=1)$ on the open index of the top tensor $\hat{t}$. However, in order to capture the first $J=2$ state using the regular MERA, we would need to consider at least $\chi_{\tiny\mbox{top}} = 20$ (at a larger computational cost and possibly lower accuracy), since this state has only the $20$\textsuperscript{th} lowest energy overall.


\subsubsection{Reduction in computational costs}

The use of SU(2)-invariant tensors in the MERA also results in a reduction of computational costs. We compared the computational costs (memory and CPU) associated with using the regular MERA and the SU(2) MERA. We also found it instructive to compare the analogous costs associated with a MERA that is made of tensors that remain invariant under only a subgroup U(1) of the symmetry group. This entails introducing the spin projection operators $\hat{J}_z$ on the links of the MERA and imposing the invariance of constituent tensors under the action of these operators. For such a U(1) MERA, imposing such constraints corresponds to conservation of the total spin projection $m_J$, while the total spin may fluctuate. (The explicit construction of the U(1) MERA was discussed by the authors' in Ref.~\onlinecite{Singh11}.)

Figure \ref{fig:memcompare} shows a comparison of the total number of complex coefficients that are required to be stored for $L=54$ sites (corresponding to 108 spins) in the three cases: regular MERA, U(1) MERA and the SU(2) MERA. U(1)-invariant tensors\cite{Singh11} have a block structure in the eigenbasis of $\hat{J}_z$ operators on each index of the tensor, and therefore they incur a smaller memory cost in comparison to regular tensors. For example, it can be seen that for the same memory required to store a regular MERA with $\chi=15$, one can instead consider storing a U(1) MERA with $\chi=21$. On the other hand, SU(2)-invariant tensors are substantially more sparse. When written in the canonical form, SU(2)-invariant tensors are not only block-sparse but each block, in turn, decomposes into a degeneracy part and a structural part such that the structural part need not be stored in memory. With the same amount of memory that is required to store, for example, a $\chi=15$ regular MERA, one can already store a $\chi=39$ SU(2) MERA.

In \fref{fig:meracompare} we show an analogous comparison of the computational performance in the three cases. We plot the computational time required for one iteration of the energy minimization algorithm of Ref.~\onlinecite{Evenbly09} (during which all tensors in the MERA are updated once), as a function of the total bond dimension $\chi$ for the cases of regular MERA, U(1) MERA  and SU(2) MERA. We see that for sufficiently large $\chi$, using SU(2)-invariant tensors leads to a shorter time per iteration of the optimization algorithm. In the symmetric versions of the algorithm we considered \textit{precomputation} of repeated operations, see App. C.


\section{Summary and Outlook\label{sec:outlook}}

In this paper we have addressed the theoretical and implementation aspects of incorporating a global SU(2) symmetry into tensor network algorithms. On the theoretical side we described how SU(2)-invariant tensors decompose in a compact canonical form that is made of two terms - degeneracy tensors that are unconstrained by the symmetry and intertwiners of SU(2) that are completely determined by the group. We explained how a set of primitive tensor manipulations (reversal, permutation and reshape of indices and matrix multiplication and matrix factorizations) are adapted to the canonical form of SU(2)-invariant tensors. To this end we introduced certain transformations (listed in Table \Rmnum{1}), determined completely by the symmetry group, that play an instrumental role in the manipulation of SU(2)-invariant tensors.

On the implementation side we described (see App. C) a practical scheme to implement SU(2) symmetry into tensor network algorithms. This scheme is based on organizing the non-trivial components of an SU(2)-invariant tensor into an SU(2)-invariant vector. A highlight of this approach is that the reversal, permutation and reshape of indices of an SU(2)-invariant reduce to matrix operations, specifically, the multiplication of an SU(2)-invariant matrix and vector.

Finally, we described the SU(2)-invariant MERA and used it to demonstrate how incorporating the symmetry allows for the selection of total spin and also the significant reduction of computational costs (by a factor of between forty and fifty). These gains may be used either to reduce overall computation time or to permit substantial increases in the MERA bond dimension $\chi$, and consequently in the accuracy of the results obtained.

Though we have focused on SU(2) symmetry, the formalism presented here may equally well be applied to any non-Abelian group that is compact, completely reducible and multiplicity free. In particular, one can consider composite symmetries such as SU(2)$\times$U(1), corresponding to spin isotropy and particle number conservation, and SU(2)$\times$SU(2), corresponding to conservation of spin and isospin, etc. Such a composite symmetry is characterized by a set of charges $(a_1, a_2, a_3,\ldots)$. When fusing two such sets of charges $(a_1,a_2,a_3,\ldots)$ and $(a'_1,a'_2,a'_3,\ldots)$, each charge $a_i$ is combined with its counterpart $a'_i$ according to the relevant fusion rule. Once again, this behaviour may be encoded into a single fuse tensor $\fuser$.

On the other hand, our formalism can be extended to non-Abelian groups e.g. SU(3) where \textit{inner} and/or \textit{outer} multiplicity appears in the representations of the group. In the present formalism, inner multiplicity corresponds to the occurrence of multiple states with the same value of $m_j$ for a given $j$. This can be accounted in a straightforward way by replacing the existing label $m_j$ with the pair $(m_j, p_j)$ where the additional label $p_j \in \{1,2,\ldots\}$  allows distinction between states with inner multiplicity. The label $p_j$ does not appear on the degeneracy tensors, nor do the $\hat{F}$ and $\hat{R}$ coefficients depend on $p_j$. However, this is not the case with outer multiplicity. Outer multiplicity corresponds to the occurrence of multiple copies of the same irrep $j$ in the tensor product of two irreps (fusion rules). In this case \eref{eq:tensorirrep} is replaced with
\begin{equation}
\mathbb{V}^{(A)}_{j_A} \otimes \mathbb{V}^{(B)}_{j_B} \cong \bigoplus_{j} \mathbb{M}^{(AB)}_{j} \otimes \mathbb{V}^{(AB)}_{j},
\label{eq:outermult}
\end{equation}
where $\mathbb{M}^{(AB)}_{j}$ is the (outer) multiplicity space of irrep $j$. In order to account for this multiplicity we replace $j$ with the pair $(j, q_j)$ throughout the discussion, where $q_j \in \{1,2,\ldots,\dim(\mathbb{M}^{(AB)}_{j})\}$ labels different copies of $j$ in the decompositions (\ref{eq:outermult}).

Our formalism can also be extended to incorporate more general symmetry constraints such as those associated with conservation of total fermionic and anyonic charge. We proceed by defining the transformations listed in Table \Rmnum{1} for the relevant charges. As an example, consider fermionic constraints where the relevant charge, $p$, is the parity of fermion particle number. Charge $p$ takes two values, $p=0$ and $p=1$ corresponding to even or odd number of fermions. The fuse tensor $\fuser$ encodes the fusion rules that specify how charges $p$ and $p'$ fuse together to obtain a charge $p''$. These correspond to the fusion rules for the group $Z_2$, given as,

\begin{center}
\begin{tabular}{| C{1cm} | C{1cm} | C{1cm} |} \hline
 & $p'=0$ & $p'=1$\tabularnewline \hline
$p=0$ & $p''=0$ & $p''=1$ \tabularnewline \hline
$p=1$ & $p''=1$ & $p''=0$ \tabularnewline \hline
\end{tabular}
\end{center}

The recoupling coefficients $\hat{F}_{p_1 p_2 p_3 p_4}^{p_{12} p_{23}}$, associated with the fusion of three charges $p_1, p_2$ and $p_3$ are simple in this case owing to the Abelian fusion rules. They take value $\hat{F}_{p_1 p_2 p_3 p_4}^{p_{12} p_{23}} = 1$ for all values of intermediate charges $p_{12}$ and $p_{23}$ that appear when fusing the three charges one way or the other. The final ingredient is the tensor $\braider$ with components $\braider_{pp'\rightarrow p''}$, which in this case is defined as,
\begin{align}
\braider_{0, 0 \rightarrow 0} = 1,~~ \braider_{0, 1 \rightarrow 1} = 1,~~ \braider_{1, 0 \rightarrow 1} = 1,~~ \braider_{1, 1 \rightarrow 0} = -1.\nonumber
\end{align}

In a similar way, one can encode the corresponding fusion rules for anyonic charges into the fuse tensor $\fuser$. For anyonic charges, the recoupling coefficients $\hat{F}$ are obtained as solutions to the \textit{pentagon equations} whereas the tensors $\braider$ are replaced with the anyonic braid generators that are obtained as solutions to the \textit{hexagon equations}, see Refs.~\onlinecite{Trebst08, Feiguin07, Pfeifer10}. Thus, having defined these tensors for the relevant charges, the formalism and the implementation framework presented in this paper can be readily adapted to incorporate the constraints corresponding to the presence of fermionic or anyonic charges.


\textbf{Acknowledgments.---} The authors thank Robert Pfeifer for insightful discussions. A significant part of this work was completed when the authors were employed at the University of Queensland, Australia.

\appendix

\section{Tensor network formalism \label{sec:tensor}}

In this appendix we review the basic formalism of tensors and tensor networks. Even though we do not make any explicit reference to symmetry here, our formalism is directed towards SU(2)-invariant tensors.

\subsection{Tensors \label{sec:tensor:tensors}}

A \textit{tensor} $\hat{T}$ is a multi-dimensional array of complex numbers $\hat{T}_{i_1i_2 \ldots i_k}$. The \textit{rank} of a tensor is the number $k$ of indices. For instance, a rank-0 tensor $(k=0)$ is a complex number. Similarly, rank-1 $(k=1)$ and rank-2 $(k=2)$ tensors corresponds to vectors and matrices, respectively. The \textit{size} of an index $i$, denoted $|i|$, is the number of values that the index takes, $i \in {1, 2, \ldots, |i|}$. Each index $i_l,~l=1,2,\ldots,k$, of the tensor is also equipped with a direction: `in' or `out', that is, either incoming into the tensor or outgoing from the tensor respectively.
The \textit{size} of a tensor $\hat{T}$, denoted $|\hat{T}|$, is the number of complex numbers it contains, namely, $|\hat{T}|~=~|i_1|~\times~|i_2|~\times~\ldots~\times~|i_k|$.

It is convenient to use a graphical representation of tensors, as illustrated in Fig. 1, where a tensor is depicted as a ``blob'' (or by a shape e.g., circle, square etc.) and each of its indices is represented by a line emerging perpendicular to the boundary of the blob. Each line carries an arrow that indicates the direction of the corresponding index. By convention, all arrows in a diagram point downward in the page. Therefore arrows are redundant. Nonetheless we draw them explicitly to emphasize the direction. In order to specify which index corresponds to which emerging line, we follow the prescription that the lines corresponding to indices $\{i_1, i_2, \ldots , i_k\}$ emerge in counterclockwise order. The first index corresponds to the line emerging closest to a mark (black dot) inside the boundary of the blob (or to the first line encountered while proceeding counterclockwise from nine o'clock in case the tensor is depicted as a circle without a mark).

\begin{figure}[t]
\includegraphics[width=5.5cm]{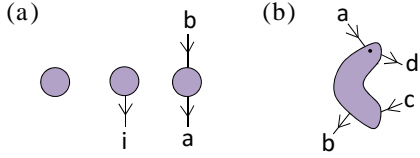}
\caption{(Color online) Graphical representation of tensors by means of a shape e.g. circle or a ``blob''. Indices of the tensor correspond to lines emerging perpendicular to the boundary of the shape. Indices may be incoming or outgoing as indicated by arrows. (a) Graphical representation of tensors with rank $0,1$ and $2$, corresponding to a complex number $c \in \mathbb{C}$, a vector $\ket{v} \in \mathbb{C}^{[i]}$ and a matrix $\hat{M} \in \mathbb{C}^{|a|\times|b|}$. (b) Graphical representation of a tensor $\hat{T}$ with components $\hat{T}_{abcd}$ and directions $\vec{D} = \{\mbox{`in', `out', `in', `out'}\}$. Indices emerge in a counterclockwise order, the first index is the one emerging closest to a mark (dot) inside the blob. By convention all arrows in a tensor diagram point downward in the page. \label{fig:tensor}}
\end{figure}


\subsection{Elementary manipulations of a tensor\label{sec:tensor:manipulations}}

A tensor can be transformed into another tensor in several elementary ways. These include, \textit{reversing} the direction of one or several of its indices, \textit{permuting} its indices, and/or \textit{reshaping} its indices.

\begin{figure}[t]
  \includegraphics[width=7cm]{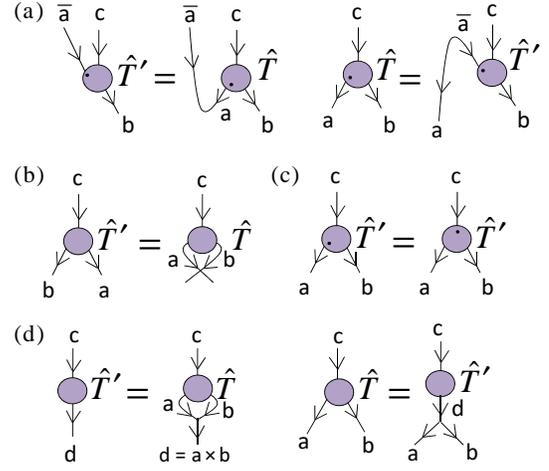}
\caption{(Color online) Transformations of a tensor: (a) Reversing the direction of an index, \eref{eq:bend}, is depicted by bending lines. (b) Swapping two outgoing indices, \eref{eq:permute}. (c) Cyclic permutation of indices, e.g. $(\hat{T}')_{cab} = \hat{T}_{abc}$, is depicted by simply shifting the starting mark (black dot) close to the first index of the resulting tensor. (d) Fusion of indices $a$ and $b$ into $d = a \times b$, \eref{eq:fuse}; splitting of index $d=a \times b$ into $a$ and $b$, \eref{eq:split}.\label{fig:tensorman}}
\end{figure}

\textit{Reversing} an index corresponds to creating a new tensor $\hat{T}'$ from $\hat{T}$ by flipping the direction of the index e.g.
\begin{equation}
	(\hat{T}')_{\overline{a}bc} = \hat{T}_{abc},
	\label{eq:bend}
\end{equation}
where $\overline{a}$ denotes the index that is obtained by reversing the direction of $a$, and $\hat{T}'$ is component-wise equal to tensor $\hat{T}$. Reversal of an index is depicted [\fref{fig:tensorman}(a)] by ``bending'' the line corresponding to the index upward if the index is outgoing or downward if it is incoming (since we allow arrows to point only downward). In this paper, we will use the terminology ``reversing an index'' and ``bending an index'' interchangeably.

A \textit{permutation} of indices corresponds to creating a new tensor $\hat{T}'$ from $\hat{T}$ by simply changing the order in which the indices appear, e.g.
\begin{equation}
	(\hat{T}')_{bac} = \hat{T}_{abc}.
	\label{eq:permute}
\end{equation}
Permutation of indices is depicted by intercrossing the lines corresponding to the indices, as illustrated in \fref{fig:tensorman}(b). A cyclic permutation of indices e.g. $(\hat{T}')_{cab} = \hat{T}_{abc}$ can also be depicted by simply shifting the starting mark (black dot) to a new location within the blob instead of intercrossing lines, \fref{fig:tensorman}(c).

A tensor $\hat{T}$ can be \textit{reshaped} into a new tensor $\hat{T}'$ by ``fusing'' and/or ``splitting'' some of its indices. For instance, in
\begin{eqnarray}
	(\hat{T}')_{dc} = \hat{T}_{abc},~~~~~~~d = a\times b,
	\label{eq:fuse}
\end{eqnarray}
tensor $\hat{T}'$ is obtained from tensor $\hat{T}$ by fusing indices $a \in \left\{1, \cdots, |a|\right\}$ and $b \in \left\{1, \cdots, |b|\right\}$ together into a single index $d$ of size $|d| = |a| \cdot |b|$ that runs over all pairs of values of $a$ and $b$, i.e.  $ d \in \left\{ (1,1), (1,2), \cdots, (|a|, |b|-1), (|a|,|b|) \right\}$, whereas in
\begin{eqnarray}
	\hat{T}_{abc} = (\hat{T}')_{dc},~~~~~~~d = a\times b,
	\label{eq:split}
\end{eqnarray}
tensor $\hat{T}$ is recovered from $\hat{T}'$ by splitting index $d$ of $\hat{T}'$ back into outgoing indices $a$ and $b$, see \fref{fig:tensorman}(d)-(e).

\begin{figure}[b]
  \includegraphics[width=7.5cm]{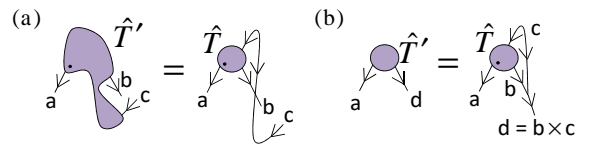}
\caption{(Color online) (a) Swapping an incoming and outgoing index as decomposed into two reversals and a swap. (b) Fusion of an incoming and an outgoing index as decomposed into a reversal and a fusion of outgoing indices.\label{fig:tensorman1}}
\end{figure}

Though the direction of an index appears to be of limited relevance here, \eref{eq:bend}, it will play an important role when we consider SU(2)-invariant tensors where it specifies how the group acts on that index. In particular, directions have to be carefully considered when permuting and reshaping indices that have different directions.

For example, consider a rank-$3$ tensor $\hat{T}$ with outgoing indices $a$ and $b$ and incoming index $c$ and let $\hat{T}'$ denote the tensor obtained from $\hat{T}$ by permuting its indices as $(\hat{T}')_{acb} = \hat{T}_{abc}$. Notice that, in order to intercross the lines corresponding to $b$ and $c$ in the corresponding graphical representation we first have to bend $c$ downward, then intercross $c$ and $b$ and bend back $c$ upward, see \fref{fig:tensorman1}(a). (Alternatively, we could bend $b$ upward, intercross $b$ and $c$, and then bend back $b$ downward.) This graphical example motivates the decomposition of an arbitrary permutation of indices into a sequence of reversal of indices and \textit{swaps}, where a swap corresponds to interchanging the position of two adjacent indices with the same direction [\fref{fig:tensorman}(b)].

Analogously, the fusion of an incoming and an outgoing index corresponds to first bending, say, the incoming index downward and then fusing it with the outgoing index, as illustrated in \fref{fig:tensorman1}(b). In this case, the original tensor is recovered by splitting back the fused index and then bending the index that was initially incoming. These decompositions of permutations and reshapes into more basic steps involving bending of lines will appear more relevant and more useful in the context of SU(2)-invariant tensors where bending lines corresponds to transforming the tensor in a non-trivial way.

\begin{figure}[t]
  \includegraphics[width=8cm]{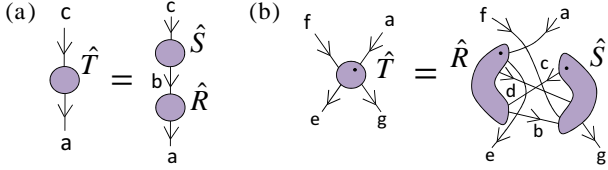}
\caption{(Color online) (a) Graphical representation of the matrix multiplication of two matrices $\hat{R}$ and $\hat{S}$ into a new matrix $\hat{T}$, \eref{eq:Mmultiply}. (b) Graphical representation of an example of the multiplication of two tensors $\hat{R}$ and $\hat{S}$ into a new tensor $\hat{T}$, \eref{eq:tensormult}. \label{fig:multiply1}}
\end{figure}


\subsection{Multiplication of two tensors\label{sec:tensor:multiply}}

Given two matrices $\hat{R}$ and $\hat{S}$ with components $\hat{R}_{ab}$ and $\hat{S}_{bc}$, we can multiply them together to obtain a new matrix $\hat{T}$, $\hat{T} = \hat{R}\cdot \hat{S}$, with components
\begin{equation}
	\hat{T}_{ac} = \sum_{b} \hat{R}_{ab}\hat{S}_{bc},
	\label{eq:Mmultiply}
\end{equation}
by summing over or \textit{contracting} index $b$. The multiplication of matrices $\hat{R}$ and $\hat{S}$ is represented graphically by connecting together the emerging lines of $\hat{R}$ and $\hat{S}$ corresponding to the contracted index, as shown in \fref{fig:multiply1}(a).

Matrix multiplication can be generalized to tensors, such that, an incoming index of one tensor is identified and contracted with an outgoing index of the other. For instance, given tensor $\hat{R}$ with components $\hat{R}_{abcde}$ and directions $\{\mbox{`in', `out', `in', `out', `out'}\}$, and tensor $\hat{S}$ with components $\hat{S}_{cdfbg}$ and directions $\{\mbox{`out', `in', `in', `in', `out'}\}$, we can define a tensor $\hat{T}$ with components $\hat{T}_{afeg}$ that are given by
\begin{equation}
	\hat{T}_{afeg} = \sum_{bcd} \hat{R}_{abcde}\hat{S}_{cdfbg}.
\label{eq:tensormult}
\end{equation}
Note that each of the indices $b, c$ and $d$ that are contracted is incoming into one tensor and outgoing from the other. The multiplication is represented graphically by connecting together the lines emerging from $\hat{R}$ and $\hat{S}$ corresponding to each of these indices, as shown in \fref{fig:multiply1}(b).

\begin{figure}[t]
  \includegraphics[width=8cm]{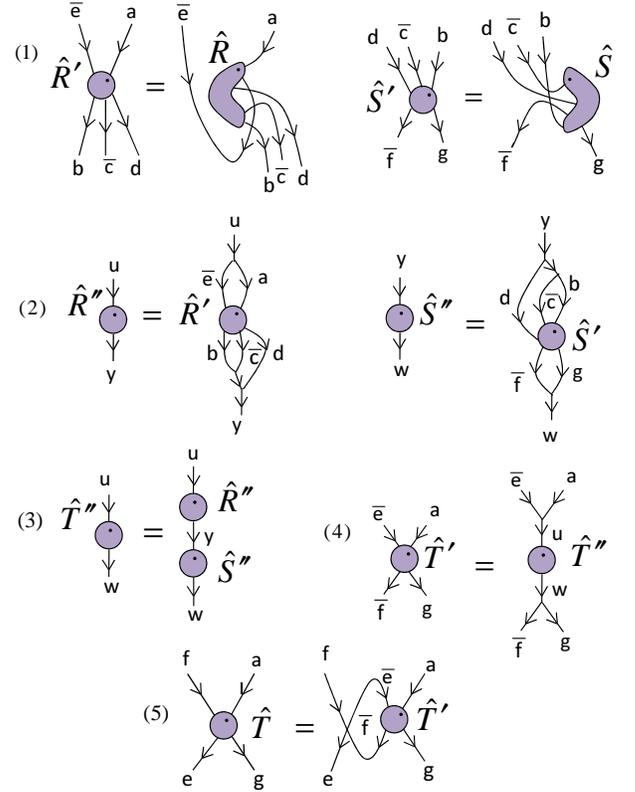}
\caption{(Color online) Graphical representations of the five elementary steps 1-5 into which one can decompose the multiplication of the tensors of \eref{eq:tensormult}.\label{fig:multiply2}}
\end{figure}

Multiplication of two tensors can be broken down into a sequence of elementary steps by transforming the tensors into matrices, multiplying the matrices together, and then transforming the resulting matrix back into a tensor. Next we describe these steps for the contraction given in \eref{eq:tensormult}. They are illustrated in \fref{fig:multiply2}.

\begin{enumerate}

	\item \textit{Reverse} and \textit{Permute} the indices of tensor $\hat{R}$ in such a way that the indices $b, c$ and $d$ that are contracted appear in the last positions as \textit{outgoing} indices and in a given order, e.g. $bcd$, and the remaining indices $a$ and $e$ appear in the first positions as \textit{incoming} indices; similarly reverse and permute the indices of $\hat{S}$ so that the indices $b, c$ and $d$ appear in the first positions as \textit{incoming} indices and in the same order, $bcd$, and the remaining indices $f$ and $g$ appear in the last positions as \textit{outgoing} indices,
	\begin{align}
	(\hat{R}')_{a\overline{e} ~b\overline{c}d} &= (\hat{R})_{abcde},   \nonumber \\
	(\hat{S}')_{b\overline{c}d\overline{f}g} &= (\hat{S})_{cdfbg}. \label{eq:multi1}
	\end{align}
		
	\item \textit{Reshape} tensor $\hat{R}'$ into a matrix $\hat{R}''$ by fusing into a single index $u$ all the indices that are not contracted, $u = a\times \overline{e}$, and into a single index $y$ all indices that are contracted, $y = b \times \overline{c} \times d$; similarly reshape tensor $\hat{S}'$ into a matrix $\hat{S}''$ with indices $y$ and $w = \overline{f}\times g$ (in order to obtain the same index $y$, the three indices $b, \overline{c}$ and $d$ are fused according to the same sequence of pairwise fusions for both the tensors as shown in the figure),
		\begin{align}
		(\hat{R}'')_{uy} &= (\hat{R}')_{a\overline{e}b\overline{c}d}, \nonumber \\
		(\hat{S}'')_{yw} &= (\hat{S}')_{b\overline{c}d\overline{f}g}. \label{eq:multi2}
	\end{align}
	
	\item \textit{Multiply} matrices $\hat{R}''$ and $\hat{S}''$ to obtain a matrix $\hat{T}'$ with components
	\begin{equation}
	(\hat{T}'')_{uw} = \sum_{y} (\hat{R}'')_{uy} ~~(\hat{S}'')_{y w}. \label{eq:multi3}
	\end{equation}
	
	\item \textit{Reshape} matrix $\hat{T}''$ into a tensor $\hat{T}'$ by splitting indices $u = a\times \overline{e}$ and $w = \overline{f}\times g$,
		\begin{equation}
	(\hat{T'})_{a\overline{e}\overline{f}g} = (\hat{T}'')_{uw}.  \label{eq:multi4}
	\end{equation}

	\item \textit{Reverse} and \textit{Permute} indices of tensor $\hat{T}'$ in the order in which they appear in $\hat{T}$,
	\begin{equation}
	\hat{T}_{afeg} = (\hat{T}')_{a\overline{e}\overline{f}g}.  \label{eq:multi5}
	\end{equation}
\end{enumerate}
The contraction of \eref{eq:tensormult} can be implemented at once, without breaking the multiplication down into elementary steps. However, it is often more convenient to compose the above elementary steps since, for instance, in this way one can use existing linear algebra libraries for matrix multiplication. In addition, it can be seen that the leading computational cost in multiplying two large tensors is not changed when decomposing the contraction in the above steps.


\subsection{Factorization of a tensor\label{sec:tensor:factorize}}

\begin{figure}[t]
  \includegraphics[width=6.5cm]{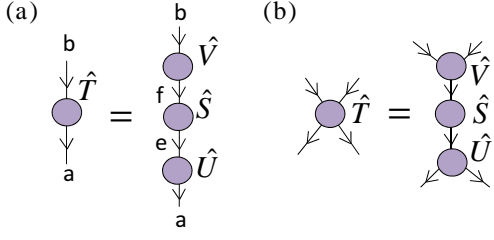}
\caption{(Color online) (a) Factorization of a matrix $\hat{T}$ according to a singular value decomposition, \eref{eq:singular}. (b) Factorization of a rank-4 tensor $\hat{T}$ according to one of several possible singular value decompositions. \label{fig:decompose}}
\end{figure}

A matrix $\hat{T}$ can be factorized into the product of two (or more) matrices in one of several canonical forms. For instance, the \textit{singular value decomposition}
\begin{equation}
	\hat{T}_{ab} = \sum_{c,d} \hat{U}_{ac}\hat{S}_{cd}\hat{V}_{db}
	= \sum_{c} \hat{U}_{ac}s_{c}\hat{V}_{cb}
	\label{eq:singular}
\end{equation}
factorizes $\hat{T}$ into the product of two unitary matrices $\hat{U}$ and $\hat{V}$, and a diagonal matrix $\hat{S}$ with non-negative diagonal elements $s_c = \hat{S}_{cc}$ known as the \textit{singular values} of $\hat{T}$ [\fref{fig:decompose}(a)].
 On the other hand, the \textit{eigenvalue} or \textit{spectral decomposition} of a square matrix $\hat{T}$ is of the form
\begin{equation}
	\hat{T}_{ab} = \sum_{c,d} \hat{M}_{ac}D_{cd}(\hat{M}^{-1})_{db}
	= \sum_{c} \hat{M}_{ac}\lambda_{c}(\hat{M}^{-1})_{cb}
	\label{eq:spectral}
\end{equation}
where $\hat{M}$ is an invertible matrix whose columns encode the eigenvectors $\ket{\lambda_c}$ of $\hat{T}$,
\begin{equation}
	\hat{T} \ket{\lambda_{c}} = \lambda_c \ket{\lambda_c},
\end{equation}
$\hat{M}^{-1}$ is the inverse of $\hat{M}$, and $\hat{D}$ is a diagonal matrix, with the eigenvalues $\lambda_c=\hat{D}_{cc}$ on its diagonal. Other useful factorizations include the LU decomposition, the QR decomposition, etc. We refer to any such decomposition generically as a \textit{matrix factorization}.

A tensor $\hat{T}$ with more than two indices can be converted into a matrix in several ways by specifying how to join its indices into two subsets. After specifying how tensor $\hat{T}$ is to be regarded as a matrix, we can factorize $\hat{T}$ according to any of the above matrix factorizations, as illustrated in \fref{fig:decompose}(b) for a singular value decomposition. Generally, this requires first reversing directions, permuting and reshaping the indices of $\hat{T}$ to form a matrix, then decomposing the latter, and finally restoring the open indices of the resulting matrices into their original form by undoing the reshapes, permutations and reversal of directions.


\subsection{Tensor networks and their manipulation\label{sec:tensor:TN}}

\begin{figure}[t]
  \includegraphics[width=8.5cm]{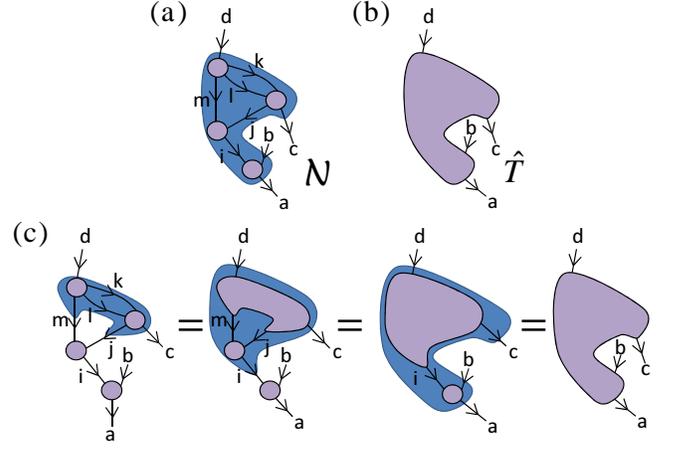}
\caption{(Color online) (a) Example of a tensor network $\mathcal{N}$. (b) Tensor $\hat{T}$ of which the tensor network $\mathcal{N}$ could be a representation. (c) Tensor $\hat{T}$ can be obtained from $\mathcal{N}$ through a sequence of contractions of pairs of tensors. Shading indicates the two tensors to be multiplied together at each step. The product tensor at each step is conveniently depicted by the blob that covers the two tensors that are multiplied.\label{fig:TN}}
\end{figure}

A \textit{tensor network} $\mathcal{N}$ is a set of tensors whose indices are connected according to a network pattern, e.g. \fref{fig:TN}.

Given a tensor network $\mathcal{N}$, a single tensor $\hat{T}$ can be obtained by contracting all the indices that connect the tensors in $\mathcal{N}$ [\fref{fig:TN}(b)]. Here, the indices of tensor $\hat{T}$ correspond to the open indices of the tensor network $\mathcal{N}$. We then say that the tensor network $\mathcal{N}$ is a tensor network decomposition of $\hat{T}$. One way to obtain $\hat{T}$ from $\mathcal{N}$ is through a sequence of contractions involving two tensors at a time [\fref{fig:TN}(c)]. Notice how a tensor that is obtained by contracting a region of a tensor network is conveniently depicted by a blob or shape that covers that region.

From a tensor network decomposition $\mathcal{N}$ for a tensor $\hat{T}$, another tensor network decomposition for the same tensor $\hat{T}$ can be obtained in many ways. One possibility is to replace two tensors in $\mathcal{N}$ with the tensor resulting from contracting them together, as is done in each step of \fref{fig:TN}(c). Another way is to replace a tensor in $\mathcal{N}$ with a decomposition of that tensor (e.g. with a singular value decomposition). In this paper, we will be concerned with manipulations of a tensor network that, as in the case of multiplying two tensors or decomposing a tensor, can be broken down into a sequence of operations from the following list:
\begin{enumerate}
	\item Reversal of direction of indices of a tensor, \eref{eq:bend}.
	\item Permutation of the indices of a tensor, \eref{eq:permute}.
	\item Reshape of the indices of a tensor, Eqs.~(\ref{eq:fuse})-(\ref{eq:split}).
	\item Multiplication of two matrices, \eref{eq:Mmultiply}.
	\item Factorization of a matrix e.g. singular value decomposition \eref{eq:singular} or spectral decomposition \eref{eq:spectral}.
\end{enumerate}
These operations constitute a set $\mathcal{P}$ of \textit{primitive} operations for tensor network manipulations (or, at least, for the type of manipulations we will be concerned with). In Sec.~\ref{sec:tnsu2} (and also in App.~\ref{sec:tree} we discuss how this set $\mathcal{P}$ of primitive operations can be generalized to tensors that are invariant under the action of the group SU(2).


\subsection{Tensor network states for quantum many-body systems\label{sec:tensor:TNstates}}

Tensor networks are used as a means to represent the wave-function of certain quantum many-body systems on a lattice. Let us consider a lattice $\mathcal{L}$ made of $L$ sites, each described by a complex vector space $\mathbb{V}$ of dimension $d$. A generic pure state $\ket{\Psi} \in \mathbb{V}^{\otimes L}$ of $\mathcal{L}$ can always be expanded as
\begin{equation}
\label{eq:purePsi}
\ket{\Psi} = \sum_{i_{1}, i_{2},\cdots, i_{L}} \hat{\Psi}_{i_{1} i_{2}\cdots i_{L}} \ket{i_{1}}\ket{ i_{2}} \cdots \ket{i_{L}},
\end{equation}
where $i_{s} = 1, \cdots, d$ labels a basis $\ket{i_s}$ of $\mathbb{V}$ for site $s \in \mathcal{L}$. Tensor $\hat{\Psi}$, with components $\hat\Psi_{i_{1} i_{2}\cdots i_{L}}$, contains $d^L$ complex coefficients. This is a number that grows exponentially with the size $L$ of the lattice. Thus, the representation of a \textit{generic} pure state $\ket{\Psi} \in \mathbb{V}^{\otimes L}$ is \textit{inefficient}. However, it turns out that an \textit{efficient} representation of \textit{certain} pure states can be obtained by expressing tensor $\hat{\Psi}$ in terms of a tensor network. Popular tensor networks such as the MPS, PEPS, TTN and the MERA correspond to decomposition of tensor $\hat{\Psi}$ into a set of tensors that are interconnected according to a given network pattern. The open indices of each of these tensor networks correspond to the indices $i_1, i_2, \cdots, i_L$ of tensor $\hat{\Psi}$. All these tensor networks contain $O(L)$ tensors. If $p$ is the rank of the tensors in one of these tensor networks, and $\chi$ is the size of their indices, then the tensor network depends on $O(L\chi^p)$ complex coefficients. For a fixed value of $\chi$ this number grows linearly in $L$, and not exponentially. It therefore does indeed offer an efficient description of the pure state $\ket{\Psi} \in \mathbb{V}^{\otimes L}$ that it represents. Of course only a subset of pure states can be decomposed in this way. Such states, often referred to as tensor network states, are used as variational ans\"atze, with the $O(L\chi^p)$ complex coefficients as the variational parameters.

Given a tensor network state, a variety of algorithms (see e.g. Refs.~\onlinecite{Fannes92}-\onlinecite{Koenig09}) are used for tasks such as: ($i$) computation of the expectation value $\bra{\Psi}\hat o\ket{\Psi}$ of a local observable $\hat o$, ($ii$) optimization of the variational parameters so as to minimize the expectation value of the energy $\bra{\Psi}\hat{H}\ket{\Psi}$, or ($iii$) simulation of time evolution, e.g. $e^{-\rmi\hat H t}\ket{\Psi}$. These tasks are accomplished by manipulating tensor networks.

On most occasions, all required manipulations can be reduced to a sequence of primitive operations in the set $\mathcal{P}$ introduced in Sec.~\ref{sec:tensor:TN}. Thus, in order to adapt the tensor network algorithms of e.g. Refs.~\onlinecite{Fannes92}-\onlinecite{Koenig09} to the presence of a symmetry, we only need to modify the set $\mathcal{P}$ of primitive tensor network operations. This will be done in Sec.~\ref{sec:tnsu2}.


\subsection{Tensors as linear maps} \label{sec:tensor:linearmap}
A tensor can be used to define a linear map between a tensor product of vector spaces and complex numbers, $\mathbb{C}$, in the following way. Let us use index $i$ of the tensor to label a basis $\ket{i}$ of a complex vector space $\mathbb{V}^{(i)} \cong \mathbb{C}^{|i|}$ of dimension $|i|$. Then a rank-one  tensor with an outgoing index $i$ represents a vector in $\mathbb{V}^{(i)}$ or alternatively a linear map from $\mathbb{V}^{(i)}$ to $\mathbb{C}$. Analogously, a rank-two tensor $\hat{T}$ with an incoming index $a$ and an outgoing index $b$ represents a matrix or equivalently a linear map $\hat{T}: (\mathbb{V}^{(a)})^* \otimes \mathbb{V}^{(b)} \rightarrow \mathbb{C}$ where $(\mathbb{V}^{(a)})^*$ is the dual of vector space $\mathbb{V}^{(a)}$.

More generally, we can use a rank-$k$ tensor $\hat{T}$ to define a linear map from the tensor product of $k$ vector spaces to $\mathbb{C}$ in the following way. Define a set $\mathbb{W}^{(i_l)},~l=1,2,\ldots,k$, of $k$ spaces,
\begin{equation}
 \mathbb{W}^{(i_l)} = \left\{
	\begin{array}{cc} \mathbb{V}^{(i_l)} &\mbox{ if } \vec{D}(l)=\mbox{`out'},\\
	 									(\mathbb{V}^{(i_l)})^* &\mbox{if } \vec{D}(l)=\mbox{`in'},
	\end{array} \right.
\end{equation}
where the $(\mathbb{V}^{(i_l)})^*$ is the dual of vector space $\mathbb{V}^{(i_l)}$ and $\vec{D}$ denotes the directions associated with the indices of tensor $\hat{T}$, namely, $\vec{D}(l) =$ `in' if $i_l$ is an incoming index and $\vec{D}(l) =$ `out' if $i_l$ is outgoing. Then tensor $\hat{T}$ can be regarded as the linear map
\begin{equation}
\hat{T} : \bigotimes_l  \mathbb{W}^{(i_l)} \rightarrow \mathbb{C}. \label{eq:tensormap}
\end{equation}
In this view, a tensor network $\mathcal{N}$ can be regarded as a composition of linear maps - namely, one linear map for each tensor in $\mathcal{N}$. Manipulations of a tensor network, namely, reversal, permutation and reshaping of indices of tensors can also be interpreted as linear maps. For instance, reversal of an index of a tensor corresponds to mapping the vector space that is associated with the index to its dual - e.g. in \eref{eq:bend}, if index $a$ is associated to a vector space $\mathbb{V}^{(a)}$, then index $\overline{a}$ that is obtained by reversing the direction of $a$ is associated with the dual space $(\mathbb{V}^{(a)})^*$.


\section{Representation theory of the group SU(2) \label{sec:symmetry}}

In this appendix we review basic background material concerning the representation theory of the group SU(2). We first consider the action of SU(2) on a vector space that is an irreducible representation of the group and then more generally on a vector space which decomposes as a direct sum of (possibly degenerate) irreducible representations. We then consider vectors in such a space that are invariant under the action of SU(2) as well as linear operators that are SU(2)-invariant. Then we consider the action of SU(2) on the tensor product of two irreducible representations and also on two reducible representations, and its generalization to the tensor product of an arbitrary number of representations. (In this appendix we also introduce the transformations that play an instrumental role in adapting the set $\mathcal{P}$ of primitive tensor network manipulations to the presence of the symmetry [Sec. \ref{sec:tnsu2}], see Table \Rmnum{1}.)


\subsection{Irreducible representations\label{sec:symmetry:irreps}}
Let $\mathbb{V}$ be a finite dimensional vector space on which the group SU(2) acts by means of unitary transformations $\hat{W}_{\textbf{r}}$,
\begin{align}
\hat{W}_{\textbf{r}}^{\dagger}\hat{W}_{\textbf{r}} &= \hat{W}_{\textbf{r}}\hat{W}_{\textbf{r}}^{\dagger} = \hat{I},~~~\forall \textbf{r}\in \mathbb{R}^3\\
\hat{W}_{\textbf{r}_1}\hat{W}_{\textbf{r}_2} &= \hat{W}_{\textbf{r}_1 + \textbf{r}_2},~~~~~~~~\forall \textbf{r}_1,\textbf{r}_2 \in \mathbb{R}^3,
\end{align}
where $\textbf{r} \equiv (r_x,r_y,r_z) \in \mathbb{R}^3$ parameterizes the elements of SU(2). The transformations $\hat{W}_{\textbf{r}}$ are a representation of the group SU(2), which is \textit{generated} by
hermitian operators $\hat{J}_x, \hat{J}_y$ and $\hat{J}_z$ that close the lie algebra su(2), namely,
\begin{equation}
[\hat{J}_{\alpha}, \hat{J}_{\beta}] = i\sum_{\gamma=x,y,z}\epsilon_{\alpha\beta\gamma}\hat{J}_{\gamma}, ~~~~~~~ \alpha,\beta = x,y,z,
\label{eq:algebra}
\end{equation}
where $\epsilon_{\alpha\beta\gamma}$ is the Levi-Civita symbol. In terms of the operators $\hat{J}_{\alpha}$ the transformations $\hat{W}_{\textbf{r}}$ read
\begin{equation}
\hat{W}_{\textbf{r}} = e^{i(r_x\hat{J}_x+r_y\hat{J}_y+r_z\hat{J}_z)}.
\label{eq:exp}
\end{equation}

If $\mathbb{V}$ transforms as an irreducible representation (or irrep) of SU(2) with charge $j \in \{0,\frac{1}{2},1,\frac{3}{2},2,\ldots\}$ it has dimension $\Delta_j = 2j+1$. For concreteness, in this paper we identify the charge $j$ as labeling the angular momentum or spin and the operators $\hat{J}_x, \hat{J}_y$ and $\hat{J}_z$ with the projection of spin along the three spatial directions $x, y$ and $z$. We denote by $\ket{jm_j}$ vectors that form an orthonormal basis of $\mathbb{V}$ and which are the simultaneous eigenvectors of the operator $\textbf{J}^2= \hat{J}_{x}^2+\hat{J}_{y}^2+\hat{J}_{z}^2$ and $\hat{J}_{z}$. That is,
\begin{align}
\textbf{J}^{2}\ket{j m_j} &= j(j+1)\ket{j m_j},\\
\hat{J}_{z} \ket{jm_j} &= m_j \ket{jm_j},
\label{eq:irrepz}
\end{align}
where $m_j \in \{-j,-j+1,\ldots,j\}$ is the spin projection along the $z$ direction. In the basis $\ket{jm_j}$ the action of the operators $\hat{J}_{x}$ and $\hat{J}_{y}$ is conveniently described in terms of the raising operator $\hat{J}_{+} = \hat{J}_{x} + i\hat{J}_{y}$ and the lowering operator $\hat{J}_{-} = \hat{J}_{x} - i\hat{J}_{y}$ as
\begin{align}
\hat{J}_{\pm} \ket{jm_j} &= \sqrt{j(j+1) - m_j (m_j \pm 1)} \ket{j, (m_j \pm 1)}.
\label{eq:irrepladder}
\end{align}

\textit{Example B1.} Vector space $\mathbb{V}$ that is a spin $j=0$ irrep of SU(2) has dimension one i.e. $\mathbb{V}~\cong~\mathbb{C}$ and the operators $\hat{J}_{\alpha}$ are trivial, $\hat{J}_x~=~\hat{J}_y~=~\hat{J}_z~=~(0)$.\markend

\textit{Example B2.} Consider a two-dimensional vector space $\mathbb{V}$ that transforms as an irrep $j=\half$. Then the orthogonal vectors (in column vector notation)
\begin{align}
\begin{pmatrix} 0 \\ 1 \end{pmatrix}  \equiv \;  \ket{j\!=\!\half, m_{\half}\!=\!-\half},
\begin{pmatrix} 1 \\ 0 \end{pmatrix}  \equiv \;  \ket{j\!=\!\half, m_{\half}\!=\!\half},
\label{eq:basiseg1}
\end{align}
form a basis of $\mathbb{V}$. In this basis operators $\hat{J}_{\alpha}$ and $\textbf{J}^2$ read as
\begin{align}
\hat{J}_{x}  &\equiv \; \begin{pmatrix} 0 & \half \\ \half & 0 \end{pmatrix}, ~~~~
\hat{J}_{y}  \equiv \;   \begin{pmatrix} 0 & -\frac{i}{2}  \\ \frac{i}{2} & 0 \end{pmatrix}, \nonumber \\
\hat{J}_{z}  &\equiv \; \begin{pmatrix} \half & 0 \\ 0 & -\half \end{pmatrix}, ~~ \textbf{J}^2 \equiv \;  \begin{pmatrix} \frac{3}{4} & 0 \\ 0 & \frac{3}{4} \end{pmatrix}. \label{eq:eg2c1}
\end{align}
Or in terms of Pauli matrices $\hat{\sigma}_{\alpha}$
\begin{equation}
\hat{J}_{\alpha} = \frac{\hat{\sigma}_{\alpha}}{2}, ~~~ \alpha = x,y,z.\markend
\end{equation}

\textit{Example B3.} Consider a three-dimensional vector space $\mathbb{V}$ that transforms as an irrep $j=1$. The orthogonal vectors
\begin{align}
\begin{pmatrix} 0 \\ 0 \\ 1\end{pmatrix}  &\equiv \;  \ket{j\!=\!1, m_{1}\!=\!-1}, ~~
\begin{pmatrix} 0 \\ 1 \\ 0 \end{pmatrix}  &\equiv \;  \ket{j\!=\!1, m_{1}\!=\!0}, \nonumber \\
\begin{pmatrix} 1 \\ 0 \\ 0 \end{pmatrix}  &\equiv \;  \ket{j\!=\!1, m_{1}\!=\!1},
\label{eq:basiseg2}
\end{align}
form a basis of $\mathbb{V}$. In this basis operators $\hat{J}_{\alpha}$ and $\textbf{J}^2$ read as
\begin{align}
\hat{J}_x  &\equiv \;  \begin{pmatrix} 0 & \frac{1}{\sqrt{2}} & 0\\\frac{1}{\sqrt{2}} & 0 & \frac{1}{\sqrt{2}}\\0 & \frac{1}{\sqrt{2}} & 0 \end{pmatrix},
\hat{J}_y  \equiv \;  \begin{pmatrix} 0 & -\frac{i}{\sqrt{2}} & 0\\\frac{i}{\sqrt{2}} & 0 & -\frac{i}{\sqrt{2}}\\0 & \frac{i}{\sqrt{2}} & 0 \end{pmatrix}, \nonumber \\
\hat{J}_z  &\equiv \; \begin{pmatrix} 1 & 0 & 0\\0 & 0 & 0\\ 0 & 0 & -1 \end{pmatrix}, ~~~~~\textbf{J}^2 \equiv \begin{pmatrix} 2 & 0 & 0\\0 & 2 & 0\\ 0 & 0 & 2 \end{pmatrix}.\markend
\label{eq:eg2c2}
\end{align}

\subsection{Reducible representations \label{sec:symmetry:rreps}}

More generally, SU(2) can act on a vector space $\mathbb{V}$ reducibly, in that, $\mathbb{V}$ may decompose as the direct sum of irreps of SU(2),
\begin{equation}
\mathbb{V} \cong \bigoplus_j d_j\mathbb{V}_{j}.
\end{equation}
Here space $\mathbb{V}_{j}$ accommodates a spin $j$ irrep of SU(2) and $d_j$ is the number of times $\mathbb{V}_{j}$ appears in the decomposition. The decomposition can also be written in terms of a $d_j$-dimensional space $\mathbb{D}_j$,
\begin{equation}
\boxed{
\mathbb{V} \cong \bigoplus_j\left(\mathbb{D}_j \otimes \mathbb{V}_j\right).
}
\label{eq:decoV}
\end{equation}
We say that irrep $j$ is $d_j$-fold degenerate and that $\mathbb{D}_j$ is the degeneracy space. The total dimension of space $\mathbb{V}$ is
\begin{equation}
\dim(\mathbb{V}) = \sum_j d_j \Delta_j,~~~\Delta_j = 2j+1.
\end{equation}
Let $t_j=1,2,\ldots,d_j$ label an orthonormal basis $\ket{jt_j}$ in the degeneracy space $\mathbb{D}_j$. Then a natural choice for a basis of $\mathbb{V}$ is the set of orthonormal vectors $\ket{jt_jm_j} \equiv \ket{jt_j}\otimes\ket{jm_j}$, where $j$ assumes various values that occur in the direct sum decomposition \eref{eq:decoV}. In this basis the action of SU(2) on $\mathbb{V}$ is given by
\begin{equation}
\hat{W}_{\textbf{r}} \equiv \bigoplus_j \left(\hat{I}_{d_j} \otimes \hat{W}_{\textbf{r},j}\right),
\label{eq:decow}
\end{equation}
as generated by the operators
\begin{equation}
\hat{J}_{\alpha} \equiv \bigoplus_j \left(\hat{I}_{d_j} \otimes \hat{J}_{\alpha, j}\right),~~~\alpha=x,y,z,
\label{eq:decoS}
\end{equation}
where $\hat{J}_{\alpha, j}$ now denote (with an explicit subscript $j$) the generators of the spin $j$ irreducible representation $\hat{W}_{\textbf{r},j}:\mathbb{V}_j \rightarrow \mathbb{V}_j$ and $\hat{I}_{d_j}$ is the $d_j \times d_j$ Identity matrix.

\textit{Example B4.} Consider a vector space $\mathbb{V}$ of dimension six that transforms as an irrep $j=\half$ with a finite degeneracy $d_{\half} = 3$. It decomposes as $\mathbb{V}\cong(\mathbb{D}_{\half}\otimes\mathbb{V}_{\half})$, where $\mathbb{D}_{\half}$ is a three-dimensional degeneracy space and $\mathbb{V}_{\half}$ corresponds to the space of Example B2. The vectors $\ket{jt_j}$,
\begin{align}
\begin{pmatrix} 0 \\ 0 \\ 1 \end{pmatrix}  &\equiv \ket{j=\!\half, t_0\!= 1},~~~
\begin{pmatrix} 0 \\ 1 \\ 0 \end{pmatrix}  \equiv   \ket{j=\!\half, t_0\!= 2},\nonumber \\
\begin{pmatrix} 1 \\ 0 \\ 0 \end{pmatrix}  &\equiv \ket{j=\!\half, t_1\!= 3},
\label{eq:basiseg21}
\end{align}
form a basis of $\mathbb{D}_{\half}$ and the vectors $\ket{jm_j}$ of (\ref{eq:basiseg1}) form a basis of $\mathbb{V}_{\half}$. Then in the basis $\ket{jt_j}\otimes\ket{jm_j}$ of $\mathbb{V}$ the operators $\hat{J}_{\alpha}:\mathbb{V}\rightarrow\mathbb{V}$ take the form of \eref{eq:decoS}, that is,
\begin{equation}
\hat{J}_x  \equiv \begin{pmatrix} 1 & 0 & 0 \\ 0 & 1 & 0 \\ 0 & 0 & 1 \end{pmatrix} \otimes \begin{pmatrix} 0 & \half \\ \half & 0 \end{pmatrix} =
\begin{pmatrix} 0 & \half & 0 & 0 & 0 & 0 \\ \half & 0 & 0 & 0 & 0 & 0 \\ 0 & 0 & 0 & \half & 0 & 0 \\ 0 & 0 & \half & 0 & 0 & 0 \\ 0 & 0 & 0 & 0 & 0 & \half \\ 0 & 0 & 0 & 0 & \half & 0 \end{pmatrix}.
\label{eq:eg2}
\end{equation}
Similarly we have
\begin{align}
\hat{J}_y  &\equiv
\begin{pmatrix} 0 & -\frac{i}{2} & 0 & 0 & 0 & 0 \\ \frac{i}{2} & 0 & 0 & 0 & 0 & 0 \\ 0 & 0 & 0 & -\frac{i}{2} & 0 & 0 \\ 0 & 0 & \frac{i}{2} & 0 & 0 & 0 \\ 0 & 0 & 0 & 0 & 0 & -\frac{i}{2} \\ 0 & 0 & 0 & 0 & \frac{i}{2} & 0 \end{pmatrix}, \nonumber \\
\hat{J}_z  &\equiv
\begin{pmatrix} \half  & 0 & 0 & 0 & 0 & 0 \\ 0 & -\half  & 0 & 0 & 0 & 0 \\ 0 & 0 & \half  & 0 & 0 & 0 \\ 0 & 0 & 0 & -\half  & 0 & 0 \\ 0 & 0 & 0 & 0 & \half  & 0 \\ 0 & 0 & 0 & 0 & 0 & -\half \end{pmatrix}.
\end{align}
The operator $\textbf{J}^2$ reads
\begin{equation}
\textbf{J}^2  \equiv 
 \begin{pmatrix} \frac{3}{4} & 0 & 0 & 0 & 0 & 0 \\ 0 & \frac{3}{4} & 0 & 0 & 0 & 0 \\ 0 & 0 & \frac{3}{4} & 0 & 0 & 0 \\ 0 & 0 & 0 & \frac{3}{4} & 0 & 0 \\ 0 & 0 & 0 & 0 & \frac{3}{4} & 0 \\ 0 & 0 & 0 & 0 & 0 & \frac{3}{4} \end{pmatrix}.\markend
 \label{eq:eg2s2}
\end{equation}

\textit{Example B5.} Consider a five-dimensional Hilbert space $\mathbb{V}$ that decomposes into two different irreps $j=0$ and $j=1$ with degeneracy $d_0=2$ and $d_1=1$ respectively so that irrep $j=0$ is two-fold degenerate. The space $\mathbb{V}$ decomposes as $\mathbb{V}~\cong~(\mathbb{D}_0~\otimes~ \mathbb{V}_{0})~\oplus~(\mathbb{D}_1~\otimes~\mathbb{V}_{1})$, where $\mathbb{D}_0$ is the two-dimensional degeneracy space of irrep $j=0$ and $\mathbb{D}_1$ is the one-dimensional degeneracy space of irrep $j=1$. The orthogonal vectors
\begin{align}
\begin{pmatrix} 0 \\ 0 \\ 0 \\ 0 \\ 1 \end{pmatrix}  &\equiv \;  \ket{j=0, t_0 = 1, m_0=0},\nonumber \\
\begin{pmatrix} 0 \\ 0 \\ 0 \\ 1 \\ 0 \end{pmatrix}  &\equiv \;  \ket{j=0, t_0 = 2, m_0=0}, \nonumber \\
\begin{pmatrix} 0 \\ 0 \\ 1 \\ 0 \\ 0 \end{pmatrix}  &\equiv \;  \ket{j=1, t_1 = 1, m_1=-1},\nonumber \\
\begin{pmatrix} 0 \\ 1 \\ 0 \\ 0 \\ 0 \end{pmatrix}  &\equiv \;  \ket{j=1, t_1 = 1, m_1=0},\nonumber
\end{align}
\begin{align}
\begin{pmatrix} 1 \\ 0 \\ 0 \\ 0 \\ 0 \end{pmatrix}  &\equiv \;  \ket{j=1, t_1 = 1, m_1=1},
\label{eq:basiseg3}
\end{align}
form a basis of $\mathbb{V}$. In this basis the operators $\hat{J}_{\alpha}$ take the form
\begin{equation}
\hat{J}_{\alpha} = (\hat{I}_{d_0} \otimes \hat{J}_{\alpha, 0}) \oplus (\hat{I}_{d_1} \otimes \hat{J}_{\alpha, 1}),~~~\alpha=x,y,z,
\end{equation}
where $\hat{J}_{\alpha, 0}$ and $\hat{J}_{\alpha, 1}$ are operators that generate irrep $j=0$ (Example B1) and irrep $j=1$ (Example B3) respectively. Operators $\hat{J}_{\alpha}$ and $\textbf{J}^2$ read as
\begin{align}
\hat{J}_x  &\equiv \; \begin{pmatrix} 0 & 0 & 0 & 0 & 0 \\ 0 & 0 & 0 & 0 & 0 \\ 0 & 0 & 0 & \frac{1}{\sqrt{2}} & 0 \\0 & 0 & \frac{1}{\sqrt{2}} & 0 & \frac{1}{\sqrt{2}} \\ 0 & 0 & 0 & \frac{1}{\sqrt{2}} & 0 \end{pmatrix},\nonumber \\
\hat{J}_y  &\equiv \;  \begin{pmatrix} 0 & 0 & 0 & 0 & 0 \\ 0 & 0 & 0 & 0 & 0 \\ 0 & 0 & 0 & -\frac{i}{\sqrt{2}} & 0 \\0 & 0 & \frac{i}{\sqrt{2}} & 0 & -\frac{i}{\sqrt{2}} \\0 & 0 & 0 & \frac{i}{\sqrt{2}} & 0 \end{pmatrix}, \nonumber \\
\hat{J}_z  &\equiv \; \begin{pmatrix} 0 & 0 & 0 & 0 & 0 \\ 0 & 0 & 0 & 0 & 0 \\ 0 & 0 & 1 & 0 & 0 \\ 0 & 0 & 0 & 0 & 0 \\0 & 0 & 0 & 0 & -1 \end{pmatrix},\nonumber \\
\textbf{J}^2 &\equiv \; \begin{pmatrix} 0 & 0 & 0 & 0 & 0 \\ 0 & 0 & 0 & 0 & 0 \\ 0 & 0 & 2 & 0 & 0 \\0 & 0 & 0 & 2 & 0 \\0 & 0 & 0 & 0 & 2 \end{pmatrix}.\markend
\label{eq:eg3}
\end{align}


\subsection{Invariant states and operators \label{sec:symmetry:states}}

In this paper we are interested in states and operators that are invariant under the action of SU(2).

A pure state $\ket{\Psi} \in \mathbb{V}$ is \textit{invariant} if it transforms trivially under the action of SU(2),
\begin{equation}
\hat{W}_{\textbf{r}}\ket{\Psi} = \ket{\Psi},~~~\forall \textbf{r}\in\mathbb{R}^3.
\label{eq:invPsi1}
\end{equation}
Equivalently, the state $\ket{\Psi}$ is annihilated by the action of the generators [\eref{eq:exp}]
\begin{equation}
\hat{J}_{\alpha}\ket{\Psi} = 0, ~~~~~~~~~~~\alpha=x,y,z,
\label{eq:invPsi2}
\end{equation}
and therefore also by the operator $\textbf{J}^2$,
\begin{equation}
\textbf{J}^2 \ket{\Psi} = 0.
\label{eq:invPsi3}
\end{equation}
This implies that $\ket{\Psi}$ corresponds to a pure state with $j=0$ and $m=0$. Thus, it can be expanded in the basis $\{\ket{j=0,t_0,m_0=0}\}$ of the subspace $(\mathbb{D}_{0} \otimes \mathbb{V}_0) \subseteq \mathbb{V}$,
\begin{equation}
	\boxed{\ket{\Psi} = \sum_{t_0=1}^{d_0} (\Psi_0)_{t_0} \ket{j=0,t_0,m_0=0},}
	\label{eq:nPsi3}
\end{equation}
where we have used $(\Psi_0)_{t_0}$ as a shorthand for $(\Psi_{j=0})_{t_0,m_0=0}$ and $d_0$ is the dimension of the degeneracy space $\mathbb{D}_0$.

A linear operator $\hat{T}: \mathbb{V} \rightarrow \mathbb{V}$ is SU(2)-invariant if it commutes with the action of the group,
\begin{equation}
	[\hat{T}, \hat{W}_{\textbf{r}}] = 0, ~~~~~~~~~\forall \textbf{r} \in \mathbb{R}^3,
	\label{eq:invOp2}
\end{equation}
or equivalently, if it commutes with the generators $\hat{J}_{\alpha}$,
\begin{equation}
	[\hat{T}, \hat{J}_{\alpha}] = 0, ~~~~~~~~~~~\alpha=x,y,z.
	\label{eq:invOp1}
\end{equation}
According to Schur's lemma, an SU(2)-invariant operator $\hat{T}$ decomposes as
\begin{align}
&\boxed{\hat{T} = \bigoplus_{j} \left(\hat{T}_{j} \otimes \hat{I}_{2j+1}\right),}\label{eq:Schur}
\end{align}
where $\hat{T}_{j}$ is a $d_j\times d_j$ matrix that acts on the degeneracy space $\mathbb{D}_j$ and $\hat{I}_{2j+1}$ is the $(2j+1)\times(2j+1)$ Identity matrix acting on the irrep $\mathbb{V}_j$. This decomposition implies, for instance, that operator $\hat{T}$ transforms pure states belonging to the spin $j$ subspace to pure states within the same subspace. Thus, SU(2)-invariant operators \textit{conserve} spin($j$).

\textit{Example B6.} Using \eref{eq:algebra} it follows that
\begin{equation}
[\textbf{J}^2, \hat{J}_{\alpha}] = 0, ~~~ \alpha=x,y,z, \nonumber
\end{equation}
that is, the operator $\textbf{J}^2 = \sum_{\alpha}\hat{J}^2_{\alpha}$ is SU(2)-invariant and has the form (\ref{eq:Schur}), as can be verified in Examples B2-B5. In particular, for an irreducible representation $j$ we have
\begin{equation}
\textbf{J}^2 = j(j+1) \hat{I}_{2j+1}.
\end{equation}

\textit{Example B7.} An SU(2)-invariant \textit{mixed} state in a vector space $\mathbb{V}$ is described by a density matrix $\hat{\rho}: \mathbb{V} \rightarrow \mathbb{V}$ that is an SU(2)-invariant operator. That is,
\begin{equation}
[\hat{\rho}, \hat{J}_{\alpha}] = 0, ~~~ \alpha=x,y,z. \label{eq:su2rho}
\end{equation}
Note that in the decomposition (\ref{eq:Schur}) of $\hat{\rho}$, $j$ may also take values different from zero (while an SU(2)-invariant pure state corresponds to only $j=0$).

\textit{Example B8.} A generic state $\ket{\Psi}$ in the vector space $\mathbb{V} \cong 2\mathbb{V}_0 \oplus \mathbb{V}_1$ of Example B5 has the form
\begin{equation}
	\ket{\Psi} = \begin{pmatrix} (\Psi_0)_{1,0} \\ (\Psi_0)_{2,0} \\ ~~(\Psi_1)_{1,\small{-1}} \\ (\Psi_1)_{1,0} \\ (\Psi_1)_{1,1} \end{pmatrix},
	\label{eq:ex2rev}
\end{equation}
where we have used $(\Psi_0)_{1,0}$, for instance, as shorthand notation for $(\Psi_{j=0})_{t_0=1,m_0=0}$ and so on. Clearly, $\ket{\Psi}$ is generally not a state with $j=0$, and thus not SU(2)-invariant. An SU(2)-invariant vector $\ket{\Psi_0}$ has the form
\begin{equation}
	\ket{\Psi_0} = \begin{pmatrix} (\Psi_0)_{1,0} \\ (\Psi_0)_{2,0} \\ 0 \\ 0 \\ 0 \end{pmatrix}, ~~~~~~~~~~~~~
	\label{eq:ex3rev}
\end{equation}
with non-trivial components only in the spin $j=0$ subspace. Notice that this state is annihilated by the action of the operators $\hat{J}_{\alpha}$ of \eref{eq:eg3} in accordance with \eref{eq:invPsi2}. Analogously a state with a well defined spin $j=1$ must be of the form
\begin{equation}
	\ket{\Psi_1} = \begin{pmatrix} 0 \\ 0 \\ ~~(\Psi_1)_{1,-1} \\ (\Psi_1)_{1,0} \\ (\Psi_1)_{1,1} \end{pmatrix},
	\label{eq:ex2rev1}
\end{equation}
with non-trivial components only in the spin $j=1$ subspace.

An SU(2)-invariant operator $\hat{T}:\mathbb{V} \rightarrow \mathbb{V}$ has the form
\begin{align}
	\hat{T} &= \; \begin{pmatrix} \left(T_0\right)_{11} & \left(T_0\right)_{12}\\ \left(T_0\right)_{21} &  \left(T_0\right)_{22}\end{pmatrix} \otimes (1) \oplus \begin{pmatrix} \left(T_1\right)_{11} \end{pmatrix} \otimes \begin{pmatrix} 1 & 0 & 0\\ 0 & 1 & 0 \\ 0 & 0 & 1 \end{pmatrix} \nonumber \\
	&= \begin{pmatrix} \left(T_0\right)_{11} & \left(T_0\right)_{12} & 0 & 0 & 0\\ \left(T_0\right)_{21} & \left(T_0\right)_{22} & 0 & 0 & 0 \\ 0 & 0 & \left(T_1\right)_{11} & 0 & 0 \\ 0 & 0 & 0 & \left(T_1\right)_{11} & 0 \\ 0 & 0 & 0 & 0 & \left(T_1\right)_{11} \end{pmatrix},
	\label{eq:sparseT3}
\end{align}
where $\left(T_0\right)_{11},~\left(T_0\right)_{12},~\left(T_0\right)_{21},~\left(T_0\right)_{22},~\left(T_1\right)_{11}~\in~\mathbb{C}$.\markend
\\
\\
Notice that an SU(2)-invariant vector e.g. $\ket{\Psi_0}$ of \eref{eq:ex3rev} and an SU(2)-invariant matrix e.g. $\hat{T}$ of \eref{eq:sparseT3} have a \textit{sparse} structure, that is, several components are identically zero. In particular, the non-trivial components of the SU(2)-invariant matrix $\hat{T}$ are organized into blocks $\hat{T}_j$. This structure can be exploited to store $\hat{T}$ compactly in memory by only storing the blocks $\hat{T}_j$. Moreover, multiplication and factorizations of an SU(2)-invariant matrix can be performed block-wise (as described in Sec. \ref{sec:tnsu2}) resulting in a significant computational speedup (see \fref{fig:multsvdcompare}) for these operations. Our strategy for exploiting the symmetry in the context of tensor network algorithms is based on identifying the analogous sparse block structure for generic SU(2)-invariant \textit{tensors}, as is described in Section~\ref{sec:tnsu2}.


\subsection{Tensor product of two irreducible representations\label{sec:symmetry:tpIrrep}}
Let $\mathbb{V}^{(A)}$ and $\mathbb{V}^{(B)}$ be two vector spaces which carry irreps $j_A$ and $j_B$ of SU(2) as generated by spin operators $\hat{J}^{(A)}_{\alpha}$ and $\hat{J}^{(B)}_{\alpha}$. Also consider the action of SU(2) on the tensor product space $\mathbb{V}^{(AB)} \cong \mathbb{V}^{(A)} \otimes \mathbb{V}^{(B)}$ that is generated by the \textit{total} spin operators $\hat{J}^{(AB)}_{\alpha}$,
\begin{equation}
\hat{J}^{(AB)}_{\alpha} \equiv \hat{J}^{(A)}_{\alpha}\otimes \hat{I} + \hat{I} \otimes \hat{J}^{(B)}_{\alpha},~~~\alpha=x,y,z,
\end{equation}
and which corresponds to the unitary transformations,
\begin{equation}
\hat{W}^{(AB)}_{\textbf{r}} \equiv e^{i(r_x\hat{J}^{(AB)}_x + r_y\hat{J}^{(AB)}_y + r_z\hat{J}^{(AB)}_z)}.
\end{equation}
The space $\mathbb{V}^{(AB)}$ is in general reducible and decomposes as
\begin{equation}
\mathbb{V}^{(AB)} \cong \bigoplus_{j_{AB}} \mathbb{V}_{j_{AB}}^{(AB)},
\label{eq:tensorirrep}
\end{equation}
where the total spin $j_{AB}$ assumes values $|j_A-j_B|,~|j_A-j_B|+1, \ldots,~j_A+j_B$.

Introduce a \textit{coupled} basis $\ket{j_{AB}m_{AB}}$ in $\mathbb{V}^{(AB)}$ given by
\begin{align}
	{\textbf{J}^2}^{(AB)}\ket{j_{AB} m_{j_{AB}}} &= j_{AB}(j_{AB}+1) \ket{j_{AB} m_{j_{AB}}}, \nonumber \\
	\hat{J}_z^{(AB)}~\ket{j_{AB} m_{j_{AB}}}
	&= m_{j_{AB}}~\ket{j_{AB} m_{j_{AB}}},
\end{align}
where ${\textbf{J}^2}^{(AB)} = \sum_{\alpha}\hat{J}^{2(AB)}_{\alpha}$. If $\ket{j_Am_A}$ and $\ket{j_Bm_B}$ denote the basis of spaces $\mathbb{V}^{(A)}$ and $\mathbb{V}^{(B)}$ then the coupled basis $\ket{j_{AB}m_{AB}}$ is related to the product basis $\ket{j_{A}m_{j_A};j_B m_{j_B}} \equiv \ket{j_{A}m_{j_A}} \otimes \ket{j_B m_{j_B}}$ by means of the transformation
\begin{equation}
\boxed{
\begin{split}
\ket{j_{AB} m_{j_{AB}}} = \sum_{m_{j_{A}} m_{j_{B}}} \cfusespin{A}{B}{AB}& \\
\ket{j_{A} m_{j_{A}} ;j_{B} m_{j_{B}}}&,
\end{split}
\label{eq:cg}
}
\end{equation}
where $\cfusespin{A}{B}{AB}$ are the \textit{Clebsch-Gordan coefficients} of SU(2). These coefficients are real and are identically zero unless the $j's$ and the $m's$ fulfill
\begin{align}
|j_A-j_B| \leq j_{AB} \leq j_A+j_B, \\
m_{j_{AB}} = m_{j_A} + m_{j_B}.\label{eq:compatiblej}
\end{align}
We say that $j_A, j_B$ and $j_{AB}$ are \textit{compatible} if they satisfy the above inequality.

The product basis can in turn be expressed in terms of the coupled basis as
\begin{equation}
\boxed{
\begin{split}
\ket{j_{A} m_{j_{A}};j_{B} m_{j_{B}}} = \sum_{m_{j_{AB}}} \csplitspin{AB}{A}{B}& \\
\ket{j_{AB}m_{j_{AB}}}&,\label{eq:revcg}
\end{split}
}
\end{equation}
where
\begin{equation}
\csplitspin{AB}{A}{B} \equiv \cfusespin{A}{B}{AB}.
\label{eq:splitcg}
\end{equation}

The basis $\ket{j_{AB}m_{j_{AB}}}$ is orthonormal and complete. Then it follows that
\begin{align}
\sum_{j_{AB}m_{j_{AB}}}&\!\!\!\!\!\cfuse{j_{A}m_{j_{A}}}{j_{B}m_{j_{B}}}{j_{AB}m_{j_{AB}}}\cdot\csplitt{j_{AB}m_{j_{AB}}}{j'_{A}m_{j'_{A}}}{j'_{B}m_{j'_{B}}} \nonumber\\
&~~~~~~~~~~~~~=\delta_{j_{A}j'_{A}}\delta_{m_{j_{A}}m_{j'_{A}}} \delta_{j_{B}j'_{B}}\delta_{m_{j_{B}}m_{j'_{B}}}, \label{eq:cgunitary0}\\
\sum_{m_{j_A}m_{j_B}}&\!\!\!\!\!\csplitt{j_{AB}m_{j_{AB}}}{j_{A}m_{j_{A}}}{j_{B}m_{j_{B}}}\cdot\cfuse{j_{A}m_{j_{A}}}{j_{B}m_{j_{B}}}{j'_{AB}m_{j'_{AB}}} \nonumber \\
&~~~~~~~~~~~~~=\delta_{j_{AB}j'_{AB}}\delta_{m_{j_{AB}}m_{j'_{AB}}}.
\label{eq:cgunitary}
\end{align}

The change of basis (\ref{eq:revcg}) [and also (\ref{eq:cg})] is related in a simple way to the corresponding change of basis with spaces $\mathbb{V}^{(A)}$ and $\mathbb{V}^{(B)}$ swapped as
\begin{equation}
\boxed{
\begin{split}
\csplitspin{AB}{B}{A} = &\braid{A}{B}{AB}\times \\
&~~~~~~\csplitspin{AB}{A}{B},\label{eq:su2swap}
\end{split}
}
\end{equation}
where the factor $\braid{A}{B}{AB}$ depends only on the value of the $j$'s,
\begin{equation}
\braid{A}{B}{AB} \equiv (-1)^{j_A+j_B-j_{AB}}. \label{eq:braid3}
\end{equation}

\begin{figure}[t]
  \includegraphics[width=8.5 cm]{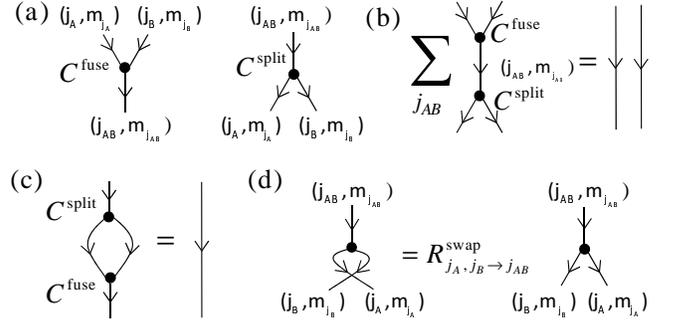}
\caption{(a) The graphical representation of the transformations $\cfuser$ and $\cspliter$. (b) Depiction of \eref{eq:cgunitary0}. The sum is over all values of $j_{AB}$ (and $m_{j_{AB}}$) that are compatible with $j_A$ and $j_B$. (c) Depiction of \eref{eq:cgunitary}. The identity holds for all compatible values of $j_A, j_B,$ and $j_{AB}$. Any contraction involving $\cfuser$ and $\cspliter$ corresponds to summing the $m$'s and, if required, also the $j$'s. We explicitly indicate a summation only when the $j$'s are summed. (d) The tensor obtained by swapping indices $(j_A,m_{j_A})$ and $(j_B,m_{j_B})$ of tensor $\cspliter$ is proportional to another $\cspliter$ tensor, \eref{eq:su2swap}. \label{fig:cg}}
\end{figure}

The graphical representation of the transformations $\cfuser$ and $\cspliter$ is shown in \fref{fig:cg}(a). Note the arrangement of arrows and the order in which indices $(j_A, m_{j_{A}}), (j_B, m_{j_{B}})$ and $(j_{AB}, m_{j_{AB}})$ are assigned to the three lines in the graphical representations of $\cfuser$ and $\cspliter$. This graphical representation allows for an intuitive depiction of Eqs.(\ref{eq:cgunitary0})-(\ref{eq:su2swap}), as shown in \fref{fig:cg}(b)-(d).

\textit{Example B9.} Consider two vector spaces $\mathbb{V}^{(A)}$ and $\mathbb{V}^{(B)}$, both transforming as the spin $\frac{1}{2}$ irrep, and let $\ket{j_{A}=\frac{1}{2}, m_{j_{A}}}$ and $\ket{j_{B}=\frac{1}{2}, m_{j_{B}}}$ denote the spin basis in the respective spaces. The space $\mathbb{V}^{(AB)} \cong \mathbb{V}^{(A)} \otimes \mathbb{V}^{(B)} $ decomposes as $\mathbb{V}^{(AB)} \cong \mathbb{V}^{(AB)}_{0} \oplus \mathbb{V}^{(AB)}_{1}$. The coupled basis $\ket{j_{AB}, m_{j_{AB}}}$ of $\mathbb{V}^{(AB)}$,
\begin{align}
&\ket{j_{AB}=0, m_{j_{AB}}=0},~\ket{j_{AB}=1, m_{j_{AB}}=-1},\nonumber \\
&\ket{j_{AB}=1, m_{j_{AB}}=0},~\ket{j_{AB}=1, m_{j_{AB}}=1}, \nonumber
\end{align}
is related to the product basis,
\begin{equation}
\ket{j_{A}=\frac{1}{2}, m_{j_{A}};j_{B}=\frac{1}{2}, m_{j_{B}}},\nonumber
\end{equation}
by means of the Clebsch-Gordan coefficients, for instance,
\begin{align}
&\ket{j_{AB}=0,m_{AB}=0} = \nonumber \\
&\cfuse{\half\mhalf}{\half\half}{00} \ket{j_A\!\!=\!\!\half, m_A\!\!=\!\!\mhalf;j_B\!\!=\!\!\half, m_B\!\!=\!\! \half}\nonumber \\
&+\cfuse{\half\half}{\half\mhalf}{00}\ket{j_A\!\!=\!\!\half, m_A\!\!=\!\!\half;j_B\!\!=\!\!\half, m_B\!\!=\!\!\mhalf}, \nonumber
\end{align}
where the numerical values of the Clebsch-Gordan coefficients can be read off from standard tables\cite{cgtable}. Here we have
\begin{align}
\cfuse{\half\mhalf}{\half\half}{00} &= -\frac{1}{\sqrt{2}}, \cfuse{\half\half}{\half\mhalf}{00} = \frac{1}{\sqrt{2}} \nonumber \\
\cfuse{\half\mhalf}{\half\mhalf}{1-1} &= \cfuse{\half\half}{\half\half}{11} ~~= 1, \nonumber \\
\cfuse{\half\half}{\half\mhalf}{10} &= \cfuse{\half\mhalf}{\half\half}{10} = \frac{1}{\sqrt{2}}. \nonumber
\end{align}
It can also be readily verified, for instance, that
\begin{align}
\left(\cfuse{\half\mhalf}{\half\half}{00}\right)^2 + \left(\cfuse{\half\half}{\half\mhalf}{00}\right)^2 &= 1,\nonumber \\
\cfuse{\half\mhalf}{\half\half}{00} &= -~\cfuse{\half\half}{\half\mhalf}{00}.\nonumber
\end{align}
in accordance with Eqs.(\ref{eq:cgunitary})-(\ref{eq:su2swap}).\markend


\subsection{Tensor product of two reducible representations\label{sec:symmetry:tpGeneral}}

More generally, consider vector spaces $\mathbb{V}^{(A)}$ and $\mathbb{V}^{(B)}$ that transform reducibly under the action of SU(2). That is, they decompose as
\begin{equation}
\mathbb{V}^{(A)} \cong \bigoplus_{j_{A}}\left(\mathbb{D}^{(A)}_{j_A} \otimes \mathbb{V}^{(A)}_{j_A}\right), ~~~\mathbb{V}^{(B)} \cong \bigoplus_{j_{B}}\left(\mathbb{D}^{(B)}_{j_B} \otimes \mathbb{V}^{(B)}_{j_B}\right).
\label{eq:AandB}
\end{equation}
The product space $\mathbb{V}^{(AB)} = \mathbb{V}_{(A)} \otimes \mathbb{V}_{(B)}$ decomposes as
\begin{equation}
\mathbb{V}^{(AB)} \cong \bigoplus_{j_{AB}} d_{j_{AB}} \mathbb{V}^{(AB)}_{j_{AB}} \cong \bigoplus_{j_{AB}}\left(\mathbb{D}^{(AB)}_{j_{AB}} \otimes \mathbb{V}^{(AB)}_{j_{AB}}\right),
\label{eq:decoVAB}
\end{equation}
where the degeneracy $d_{j_{AB}}$ of a total spin $j_{AB}$ has a contribution from all pairs of irreps $j_A$ and $j_B$ that are compatible with it, that is,
\begin{equation}
\mathbb{D}^{(AB)}_{j_{AB}} \cong \bigoplus \left(\mathbb{D}^{(A)}_{j_{A}} \otimes \mathbb{D}^{(B)}_{j_{B}} \right).
\label{eq:decoDAB}
\end{equation}
Let $\ket{j_At_{j_A}m_{j_A}}$ and $\ket{j_Bt_{j_B}m_{j_B}}$ denote the spin basis of spaces $\mathbb{V}^{(A)}$ and $\mathbb{V}^{(B)}$ respectively. We can then introduce a coupled basis $\ket{j_{AB} t_{j_{AB}} m_{j_{AB}}}$ in $\mathbb{V}^{(AB)}$ that fulfills
\begin{align}
	{\textbf{J}^2}^{(AB)}\ket{j_{AB} t_{j_{AB}} m_{j_{AB}}} &= j_{AB}(j_{AB}+1) \ket{j_{AB} t_{j_{AB}} m_{j_{AB}}}, \nonumber \\
	\hat{J}_z^{(AB)}~\ket{j_{AB} t_{j_{AB}} m_{j_{AB}}}
	&= m_{j_{AB}}~\ket{j_{AB} t_{j_{AB}} m_{j_{AB}}}.
\end{align}
and is related to the product basis
\begin{equation}
\ket{j_{A}t_{j_{A}}m_{j_A};j_B t_{j_{B}} m_{j_B}} \equiv \ket{j_{A}t_{j_{A}}m_{j_A}} \otimes \ket{j_B t_{j_{B}}m_{j_B}}, \nonumber
\end{equation}
by means of the transformation
\begin{equation}
\boxed{
\begin{split}
&\ket{j_{AB} t_{j_{AB}} m_{j_{AB}}} = \\
&~~\sum_{t_{j_{A}} t_{j_{B}}}\sum_{m_{j_{A}} m_{j_{B}}} \fusespin{A}{B}{AB} \\
&~~~~~~~~~~~~~~~~~~~~~~~~~~~~~~~~~~\ket{j_{A} t_{j_{A}}m_{j_{A}};j_{B} t_{j_{B}} m_{j_{B}}}.
\end{split}
\label{eq:basischange}
}
\end{equation}
The coefficients $\fusespin{A}{B}{AB}$ can be expressed in terms of the Clebsch-Gordan coefficients as
\begin{equation}
\boxed{
\begin{split}
&\fusespin{A}{B}{AB} = \\
&~~~\tfusespin{A}{B}{AB} \cfusespin{A}{B}{AB}.
\end{split}
\label{eq:basischange1}
}
\end{equation}
Let us explain how this expression is obtained. By definition we have
\begin{align}
&\fusespin{A}{B}{AB}\equiv \nonumber\\
&~~~~~~~~~~~\braket{j_{AB} t_{j_{AB}} m_{j_{AB}}}{j_{A} t_{j_{A}}m_{j_{A}};j_{B} t_{j_{B}} m_{j_{B}}}.\label{eq:fusedef}
\end{align}
According to the direct sum decomposition (\ref{eq:decoVAB}) each vector $\ket{j_{AB} t_{j_{AB}} m_{j_{AB}}}$ belongs to the subspace $(\mathbb{D}^{(AB)}_{j_{AB}} \otimes \mathbb{V}^{(AB)}_{j_{AB}})$ where it factorizes as
\begin{equation}
\ket{j_{AB} t_{j_{AB}} m_{j_{AB}}} = \ket{j_{AB} t_{j_{AB}}} \otimes \ket{j_{AB} m_{j_{AB}}}.
\end{equation}
Similarly, we can factorize vectors $\ket{j_{A} t_{j_{A}} m_{j_{A}}}$ and $\ket{j_{B} t_{j_{B}} m_{j_{B}}}$. Substituting these factorizations into \eref{eq:fusedef} and re-arranging terms we obtain
\begin{align}
&\fusespin{A}{B}{AB} \nonumber \\
&~~~~~ =\braket{j_{AB}t_{j_{AB}}}{j_{A}t_{j_A};j_{B}t_{j_B}} \braket{j_{AB}m_{j_{AB}}}{j_{A}m_{j_A};j_{B}m_{j_B}},\nonumber \\
&~~~~~ = \tfusespin{A}{B}{AB} \cfusespin{A}{B}{AB},\label{eq:fusedecompose}
\end{align}
where $\ket{j_{A}t_{j_A};j_{B}t_{j_B}} \equiv \ket{j_{A}t_{j_A}} \otimes \ket{j_{B}t_{j_B}}$. Here coefficients $\tfusespin{A}{B}{AB}$ can all be chosen to be either zero or one corresponding to the choice of a change of basis that maps vectors $\ket{j_{A}t_{j_A};j_B t_{j_B}}$ to vectors $\ket{j_{AB} t_{j_{AB}}}$ in a one-to-one way (see Examples B9 and B10).
\begin{figure}[t]
  \includegraphics[width=7.5cm]{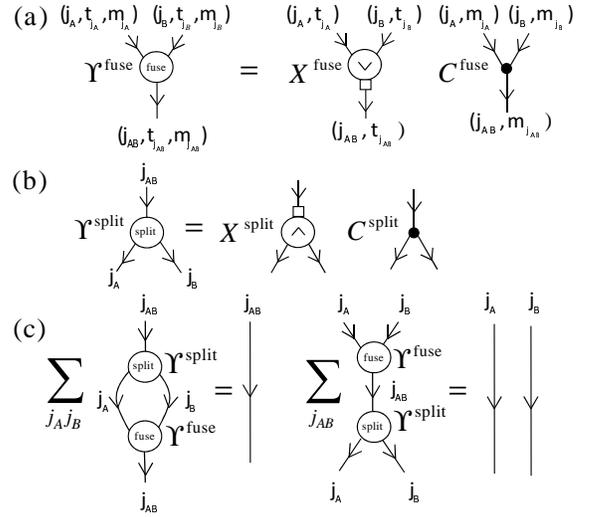}
\caption{The graphical representation of (a) the fuse tensor $\fuser$, \eref{eq:basischange} and (b) the split tensor, \eref{eq:revbasischange}, and their decomposition into $X$ and $C$ tensors. (c) Tensors $\fuser$ and $\spliter$ are unitary and thus yield the Identity when contracted pairwise as shown. The sum is over all the $j$'s (and implicitly over the corresponding $t$'s and $m$'s) on the contracted indices that are compatible with the given $j$'s on the open indices. \label{fig:su2fuse}}
\end{figure}
The product basis can be expressed in terms of the coupled basis as
\begin{equation}
\boxed{
\begin{split}
&\ket{j_{A} t_{j_{A}}m_{j_{A}};j_{B} t_{j_{B}} m_{j_{B}}} = \\
&\sum_{j_{AB}}\sum_{t_{j_{AB}}}\sum_{m_{j_{AB}}}
\splitspin{AB}{A}{B} \\
&~~~~~~~~~~~~~~~~~~~~~~~~~~~~~~~~~~~~~~~\ket{j_{AB} t_{j_{AB}} m_{j_{AB}}},\label{eq:revbasischange}
\end{split}
}
\end{equation}
where
\begin{equation}
\boxed{
\begin{split}
&\splitspin{AB}{A}{B} = \\
&\tsplitspin{AB}{A}{B}\!\!\cdot~\csplitspin{AB}{A}{B},
\label{eq:u1split1}
\end{split}
}
\end{equation}
and
\begin{equation}
\tsplitspin{AB}{A}{B} \equiv \tfusespin{A}{B}{AB}.
\label{eq:xsplit}
\end{equation}

The graphical representation\cite{fusetensor} of the transformations $\fuser$ and $\spliter$ and their decomposition into $X$ and $C$ terms is shown in \fref{fig:su2fuse}(a)-(b). By construction, $\fuser$ and $\spliter$ fulfill the equalities that are depicted in \fref{fig:su2fuse}(c).

\textit{Example B10.} Consider vector spaces $\mathbb{V}^{(A)}$ and $\mathbb{V}^{(B)}$ that both correspond to the vector space of Example B4,
\begin{align}
\mathbb{V}^{(A)} \cong (\mathbb{D}_{\half}^{(A)}\otimes\mathbb{V}^{(A)}_{\half}),~~\mathbb{V}^{(B)} \cong (\mathbb{D}_{\half}^{(B)} \otimes \mathbb{V}^{(B)}_{\half}).\nonumber
\end{align}
The product space $\mathbb{V}^{(AB)} \cong \mathbb{V}^{(A)} \otimes \mathbb{V}^{(B)}$ decomposes as
\begin{align}
\mathbb{V}^{(AB)} \cong (\mathbb{D}^{(AB)}_{0} \otimes \mathbb{V}^{(AB)}_{0}) \oplus (\mathbb{D}^{(AB)}_{1} \otimes \mathbb{V}^{(AB)}_{1}) \nonumber
\end{align}
where
\begin{align}
\mathbb{V}_{\half}^{(A)} \otimes \mathbb{V}_{\half}^{(B)} \cong (\mathbb{V}^{(AB)}_{0} \oplus \mathbb{V}^{(AB)}_{1}),  \label{eq:eg71}
\end{align}
and
\begin{align}
\mathbb{D}^{(AB)}_{0} &\cong (\mathbb{D}_{\half}^{(A)} \otimes \mathbb{D}_{\half}^{(B)}),\nonumber\\
\mathbb{D}^{(AB)}_{1} &\cong (\mathbb{D}_{\half}^{(A)} \otimes \mathbb{D}_{\half}^{(B)}).\label{eq:eg72}
\end{align}
A coupled basis $\ket{j_{AB}t_{j_{AB}}m_{j_{AB}}}$ can be introduced in $\mathbb{V}^{(AB)}$ by performing a change of basis $\fuser$ from the product basis for the spaces $\mathbb{V}^{(A)}$ and $\mathbb{V}^{(B)}$. For fixed values of $j_A, j_B$ and $j_{AB}$ the transformation $\fuser$ decomposes into $\tfuser$ and $\cfuser$ parts. The $\cfuser$ part relates the coupled basis $\ket{j_{AB}m_{j_{AB}}}$ of $\mathbb{V}^{(AB)}_{j_{AB}}$ to the basis $\ket{j_{A}=\frac{1}{2},m_{j_{A}};j_B=\frac{1}{2},m_{j_B}}$ of $(\mathbb{V}^{(A)}_{\frac{1}{2}} \otimes \mathbb{V}^{(B)}_{\frac{1}{2}})$, as described in Example B9. Analogously, the $\tfuser$ part relates the coupled basis $\ket{j_{AB}t_{j_{AB}}}$ to the product basis $\ket{j_{A}=\frac{1}{2},t_{j_{A}};j_B=\frac{1}{2},t_{j_B}}$ in the degeneracy spaces $\mathbb{D}_{j_{AB}}^{(AB)}, j_{AB}=0,1$. The coefficients $\tfusespin{A}{B}{AB}$ that correspond to the change of basis in $\mathbb{D}^{(AB)}_{0}$ can all be chosen to be zero except
\begin{align}
\tfuse{\half 1}{\half 1}{01} &=~~ \tfuse{\half 1}{\half 2}{02} =~~ \tfuse{\half 1}{\half 3}{03}= \nonumber \\
\tfuse{\half 2}{\half 1}{01} &=~~ \tfuse{\half 2}{\half 2}{02} =~~ \tfuse{\half 2}{\half 3}{03}= \nonumber\\
\tfuse{\half 3}{\half 1}{01} &=~~ \tfuse{\half 3}{\half 2}{02} =~~ \tfuse{\half 3}{\half 3}{03} = 1.\nonumber
\end{align}
Similarly, the non-zero coefficients $\tfusespin{A}{B}{AB}$ that correspond to the change of basis in $\mathbb{D}^{(AB)}_{1}$ are
\begin{align}
\tfuse{\half 1}{\half 1}{11} &=~~ \tfuse{\half 1}{\half 2}{12} =~~ \tfuse{\half 1}{\half 3}{13}= \nonumber\\
\tfuse{\half 2}{\half 1}{11} &=~~ \tfuse{\half 2}{\half 2}{12} =~~ \tfuse{\half 2}{\half 3}{13}= \nonumber \\
\tfuse{\half 3}{\half 1}{11} &=~~ \tfuse{\half 3}{\half 2}{12} =~~ \tfuse{\half 3}{\half 3}{13} = 1.\nonumber
\end{align}
Note that the change of basis in a degeneracy space, as described by $\tfuser$, is not constrained by symmetry and therefore we can fix the value of coefficients $\tfusespin{A}{B}{AB}$ to be 0 or 1 in this simple way.\markend

\textit{Example B11.} Let $\mathbb{V}^{(A)}$ and $\mathbb{V}^{(B)}$ correspond to the vector spaces of Example B3 and Example B5 respectively, that is,
\begin{align}
\mathbb{V}^{(A)} &\cong (\mathbb{D}^{(A)}_{1} \otimes \mathbb{V}^{(A)}_{1}), \nonumber \\
\mathbb{V}^{(B)} &\cong (\mathbb{D}_{0}^{(B)}\otimes\mathbb{V}^{(B)}_{0}) \oplus (\mathbb{D}_{1}^{(B)}\otimes\mathbb{V}^{(B)}_{1}).\nonumber
\end{align}
The space $\mathbb{V}^{(AB)} \cong \mathbb{V}^{(A)} \otimes \mathbb{V}^{(B)}$ decomposes as
\begin{align}
\mathbb{V}^{(AB)} \cong (\mathbb{D}_{0}^{(AB)} \otimes \mathbb{V}^{(AB)}_{0}) &\oplus(\mathbb{D}_{1}^{(AB)} \otimes \mathbb{V}^{(AB)}_{1}) \nonumber \\
&\oplus (\mathbb{D}_{2}^{(AB)} \otimes\mathbb{V}^{(AB)}_{2}),\nonumber
\end{align}
where
\begin{align}
\mathbb{D}_{0}^{(AB)} &\cong \mathbb{D}_{1}^{(A)} \otimes \mathbb{D}_{1}^{(B)} \nonumber \\
\mathbb{D}_{1}^{(AB)} &\cong (\mathbb{D}_{1}^{(A)} \otimes \mathbb{D}_{0}^{(B)}) \oplus (\mathbb{D}_{1}^{(A)} \otimes \mathbb{D}_{1}^{(B)}) \nonumber \\
\mathbb{D}_{2}^{(AB)} &\cong \mathbb{D}_{1}^{(A)} \otimes \mathbb{D}_{1}^{(B)}.\nonumber
\end{align}
Here the non-zero coefficients $\tfusespin{A}{B}{AB}$ that correspond to the change from the product basis to the coupled basis in the three degeneracy spaces $\mathbb{D}_{0}^{(AB)}, \mathbb{D}_{1}^{(AB)}$ and $\mathbb{D}_{2}^{(AB)}$ are
\begin{align}
&\tfuse{1 1}{1 1}{0 1}=\nonumber \\
&\tfuse{1 1}{0 1}{1 1}=~\tfuse{1 1}{0 2}{1 2}=~\tfuse{1 1}{1 1}{1 3}= \nonumber \\
&\tfuse{1 1}{1 1}{2 1}=1.\nonumber\markend
\end{align}


\subsection{Intertwiners and F-moves\label{sec:symmetry:fmove}}
Now consider the action of SU(2) on a space $\mathbb{V}$ that is a tensor product of $L$ vector spaces,
\begin{equation}
\mathbb{V} \equiv \bigotimes_{l=1}^{L} \mathbb{V}^{(l)}, \label{eq:latticespace}
\end{equation}
where each vector space $\mathbb{V}^{(l)}, l=1,2,\ldots,L,$ transforms as a finite dimensional representation of SU(2) as generated by spin operators $\hat{J}^{(l)}_{\alpha}, \alpha=x,y,z$. We consider the action of SU(2) on the space $\mathbb{V}$ that is generated by the total spin operators,
\begin{equation}
\boxed{\hat{J}_{\alpha} \equiv \sum_{l=1}^{L} \hat{J}^{(l)}_{\alpha}, ~~~ \alpha = x,y,z,} \label{eq:totspinops}
\end{equation}
(each term in the sum acts as $\hat{J}^{(l)}_{\alpha}$ on site $l$ and the Identity on the remaining sites) and which corresponds to the unitary transformations
\begin{equation}
	\hat{W}_{\textbf{r}} \equiv e^{\rmi\textbf{r}\cdot \textbf{J}} = \bigotimes_{l=1}^L e^{\rmi\textbf{r}\cdot \textbf{J}^{(l)}} = \bigotimes_{l=1}^L \hat{W}_{\textbf{r}}^{(l)}.\label{eq:latticerep}
\end{equation}
In the tensor product of $L$ representations one can consider different coupled spin bases corresponding to the existence of different ways of decomposing the tensor product space into the factor spaces. For example, the tensor product $\mathbb{V}^{(ABC)} \cong \mathbb{V}^{(A)} \otimes \mathbb{V}^{(B)} \otimes \mathbb{V}^{(C)}$ of $L=3$ representations can be decomposed as
\begin{align}
\mathbb{V}^{(ABC)} &\cong \mathbb{V}^{(AB)} \otimes \mathbb{V}^{(C)}, \nonumber \\
\mathbb{V}^{(AB)} &\cong \mathbb{V}^{(A)} \otimes \mathbb{V}^{(B)},  \label{eq:order1}
\end{align}
or as
\begin{align}
\mathbb{V}^{(ABC)} &\cong \mathbb{V}^{(A)} \otimes \mathbb{V}^{(BC)},\nonumber \\
\mathbb{V}^{(BC)} &\cong \mathbb{V}^{(B)} \otimes \mathbb{V}^{(C)}. \label{eq:order2}
\end{align}
For simplicity, let us consider that $\mathbb{V}^{(A)}, \mathbb{V}^{(B)}$ and $\mathbb{V}^{(C)}$ transform as irreps $j_A, j_B$ and $j_C$ respectively. The space $\mathbb{V}_{AB}$ in (\ref{eq:order1}) then generally decomposes as
\begin{equation}
\mathbb{V}^{(AB)} \cong \bigoplus_{j_{AB}} \mathbb{V}^{(AB)}_{j_{AB}}.\label{eq:intermediate1}
\end{equation}
The space $\mathbb{V}^{(ABC)}$ is also reducible, and may contain several copies of an irrep $j_{ABC}$. It decomposes as
\begin{equation}
\mathbb{V}^{(ABC)} \cong \bigoplus_{j_{ABC}, j_{AB}} \mathbb{V}_{j_{ABC}, j_{AB}}^{(ABC)}, \label{eq:tensorirrep3}
\end{equation}
where we have used $j_{AB}$ to label different copies of irrep $j_{ABC}$. Let $(\hat{Q}^{j_{AB}}_{j_A j_B j_C j_{ABC}})_{m_{j_A} m_{j_B} m_{j_C} m_{j_{ABC}}}$ denote the change of basis to the corresponding coupled basis $\ket{j_{ABC}m_{j_{ABC}},j_{AB}}$. In terms of Clebsch-Gordan coefficients we have
\begin{align}
&(\hat{Q}^{j_{AB}}_{j_A j_B j_C j_{ABC}})_{m_{j_A} m_{j_B} m_{j_C} m_{j_{ABC}}}\equiv \nonumber \\
&\sum_{m_{j_{AB}}}\cfusespin{A}{B}{AB} \cdot \cfusespin{AB}{C}{ABC},
\label{eq:cob1}
\end{align}
where the coefficients $\cfusespin{A}{B}{AB}$ describe the change of basis to the coupled basis $\ket{j_{AB}m_{j_{AB}}}$ in the intermediate space $\mathbb{V}^{(AB)}$ and the coefficients $\cfusespin{AB}{C}{ABC}$ describe the change to the coupled basis $\ket{j_{ABC}m_{j_{ABC}},j_{AB}}$ in $\mathbb{V}^{(ABC)}$.
\begin{figure}[t]
  \includegraphics[width=5.5cm]{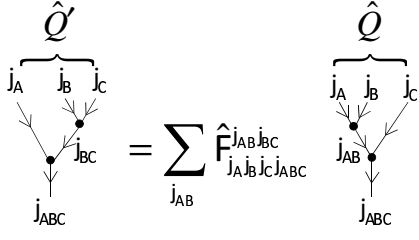}
\caption{The F-move that relates two different ways of fusing three spins $j_A, j_B, j_C$ into a total spin $j_{ABC}$, \eref{eq:fmove}.
\label{fig:fmove}}
\end{figure}

Alternatively, we can consider the decomposition (\ref{eq:order2}) of $\mathbb{V}^{(ABC)}$ into factor spaces where
\begin{equation}
\mathbb{V}^{(BC)} \cong \bigoplus_{j_{BC}} \mathbb{V}^{(BC)}_{j_{BC}}, \label{eq:intermediate2}
\end{equation}
and use $j_{BC}$ to label another coupled basis $\ket{j_{ABC}m_{j_{ABC}},j_{BC}}$ of $\mathbb{V}^{(ABC)}$. Denote by $(\hat{Q}'^{j_{BC}}_{j_A j_B j_Cj_{ABC}})_{m_{j_A} m_{j_B} m_{j_C} m_{j_{ABC}}}$ the corresponding change of basis,
\begin{align}
&(\hat{Q}'^{j_{BC}}_{j_A j_B j_C j_{ABC}})_{m_{j_A} m_{j_B} m_{j_C} m_{j_{ABC}}} \equiv \nonumber \\
&\sum_{m_{j_{BC}}}\cfusespin{B}{C}{BC}\cdot\cfusespin{A}{BC}{ABC}.
\label{eq:cob2}
\end{align}
Here $\hat{Q}^{j_{AB}}_{j_A j_B j_C j_{ABC}}$ and $\hat{Q}'^{j_{BC}}_{j_A j_B j_C j_{ABC}}$ are rank-$4$ \textit{intertwiners} or generalized Clebsch-Gordan coefficients of the group SU(2).

The two coupled bases $\ket{j_{ABC}m_{j_{ABC}},j_{AB}}$ and $\ket{j_{ABC}m_{j_{ABC}},j_{BC}}$ are related by an \textit{F-move} (see \fref{fig:fmove})
\begin{equation}
\boxed{\hat{Q}'^{j_{BC}}_{j_A j_B j_C j_{ABC}} = \sum_{j_{AB}}\hat{F}^{j_{AB}j_{BC}}_{j_Aj_Bj_Cj_{ABC}} \hat{Q}^{j_{AB}}_{j_A j_B j_C j_{ABC}},}
\label{eq:fmove}
\end{equation}
where $\hat{F}^{j_{AB}j_{BC}}_{j_Aj_Bj_Cj_{ABC}}$ are the \textit{recoupling coefficients} of SU(2). By using Eqs.~(\ref{eq:cob1}) and (\ref{eq:cob2}) the recoupling coefficients can be expressed in terms of Clebsch-Gordan coefficients as
\begin{align}
&\hat{F}^{j_{AB}j_{BC}}_{j_Aj_Bj_Cj_{ABC}} \equiv (2j_{ABC}+1)^{-1} \times \nonumber \\
&\sum \left(\cfusespin{A}{B}{AB} \cfusespin{AB}{C}{ABC}\right. \nonumber\\
&~~~\left.\cfusespin{B}{C}{BC} \cfusespin{A}{BC}{ABC}\right),\label{eq:f}
\end{align}
where the summation is over $m_{j_A},m_{j_B},m_{j_C},m_{j_{AB}},m_{j_{BC}}$ and $m_{j_{ABC}}$. Since all the $m$'s are summed over, the recoupling coefficients depend only on the $j$'s. A manifestly $m$ independent definition is given in terms of the 6-j symbols of SU(2),
\begin{align}
\hat{F}^{j_{AB}j_{BC}}_{j_Aj_Bj_Cj_{ABC}} = \kappa\left\{\begin{array}{ccc} j_A&j_B&j_{AB}\\j_C&j_{ABC}&j_{BC} \end{array}\right\},
\label{eq:sixj}
\end{align}
where
\begin{equation}
\kappa \equiv (-1)^{(j_A+j_B+j_C+j_{ABC})}\sqrt{(2j_{AB}+1)(2j_{BC}+1)}.
\end{equation}

\begin{figure}[t]
  \includegraphics[width=9cm]{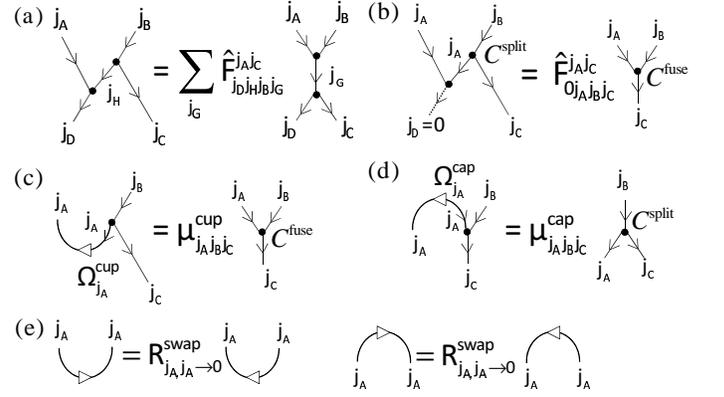}
\caption{(a) An F-move that relates two different ways of fusing two incoming and two outgoing spins. (b) A special instance of $(a)$ corresponding to setting $j_D=0$ and the only compatible value $j_G=j_C$ and $j_H=j_A$, \eref{eq:cup}. (c) Depicting the F-move $(b)$ in terms of a left-directed \textit{cup}, \eref{eq:cup}, to explicitly indicate the upward bending of index $(j_A,m_{j_A})$. (d) The inverse of $(c)$ i.e. the downward bending of index $(j_A,m_{j_A})$, as explicitly depicted by a left-directed \textit{cap}, \eref{eq:cap}. Contrast the bending of indices of the Clebsch-Gordan tensors $\cfuser$ and $\cspliter$ in $(c)-(d)$ with \fref{fig:tensorman}(a). (e) Right directed cups and caps are related to their left counterparts by a factor $\braider_{j_A,j_A\rightarrow0}$.
\label{fig:fmove1}}
\end{figure}

Other F-moves are possible corresponding to a different arrangement of arrows, as illustrated in \fref{fig:fmove1}(a). Notice that the F-move $(b)$ is a special instance of $(a)$ corresponding to the choice $j_D=0$ and only compatible values $j_{G}=j_C$ and $j_{H}=j_A$. Also notice that the index $(j_A, m_{j_A})$ appears as outgoing on the split tensor (l.h.s.) and as incoming on the fuse tensor (r.h.s.). Thus, this F-move corresponds to \textit{bending} the index $(j_A, m_{j_A})$ of the split tensor upward from the left. This is depicted more explicitly $(c)$ where we have deleted the $j_D = 0$ edge and replaced its parent tensor with a left directed ``cup''. Here we have defined
\begin{equation}
\boxed{
\begin{split}
(\mycup_{j_A})_{m_{j_A},m'_{j_A}} &\equiv \Delta^{1/2}_{j_A}\cfuse{j_Am_{j_A}}{j_A~m'_{j_A}}{00},\\
\cupf{A}{B}{C} &\equiv \Delta_{j_A}^{1/2}\hat{F}_{0j_Aj_Bj_C}^{j_Aj_C},\label{eq:cup}
\end{split}
}
\end{equation}
where $\Delta_{j_A} = 2j_A+1$ is the dimension of irrep $j_A$. The F-move $(d)$ is the inverse of $(c)$ where
\begin{equation}
\boxed{
\begin{split}
(\mycap_{j_A})_{m_{j_A},m'_{j_A}} &\equiv (-1)^{2j_A}\Delta^{1/2}_{j_A}\csplitt{00}{j_Am_{j_A}}{j_A~m'_{j_A}},\\
\capf{A}{B}{C} &\equiv (-1)^{2j_A}\Delta_{j_A}^{1/2}\hat{F}_{j_Cj_A0j_B}^{j_Bj_A}.\label{eq:cap}
\end{split}
}
\end{equation}
(Notice the additional factor $(-1)^{2j_A}$). That is, the F-move $(d)$ describes the \textit{downward} bending of index $(j_A, m_{j_A})$, as explicitly depicted by a left directed ``cap''.

More generally, the cup-cap\cite{tensorcategory} transformations and the F-moves $(c)-(d)$ play an instrumental role in bending indices of an SU(2)-invariant tensor, as described in Sec.~\ref{sec:tnsu2}. By construction the cup and cap fulfill
\begin{equation}
\mycup_j \mycap_j = \mycap_j\mycup_j = \hat{I}_{2j+1}, \label{eq:cupcap}
\end{equation}
which ensures that indices are bent in a reversible way. \textit{Right} directed cup and caps may also appear, corresponding to bending indices from the right. These are equal to their left counterparts times a factor, see \fref{fig:fmove1}(e).

Finally, we remark that in the tensor product of $L>3$ representations two different coupled bases may be related by several F-moves.


\section{Practical implementation of SU(2)-invariant tensors\label{sec:tree}}

In this appendix we describe in detail a possible implementation of the set $\mathcal{P}$ of primitive tensor network manipulations [App.~\ref{sec:tensor:TN}] for SU(2)-invariant tensor networks. Our implementation is based on \textit{tree decompositions} of SU(2)-invariant tensors. A tree decomposition of an SU(2)-invariant tensor is a canonical decomposition [Sec.~\ref{sec:su2tensors:block}] where the underlying fusion-splitting tree is made of only splitting vertices i.e. a \textit{splitting tree}. The highlight of working with tree decompositions is that the reversal, permutation and reshape of indices of SU(2)-invariant tensors simply correspond to the multiplication of an SU(2)-invariant matrix and vector (see App.~\ref{sec:tree:precom} for other benefits of using tree decompositions).


\subsection{Tree decompositions of SU(2)-invariant tensors\label{sec:tree:def}}
\begin{figure}[t]
  \includegraphics[width=7cm]{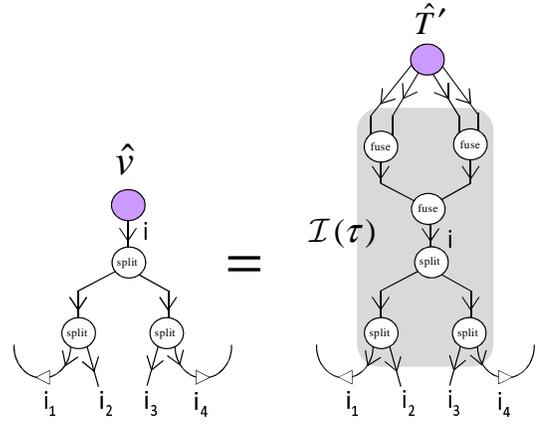}
\caption{(Color online) A tree decomposition of an SU(2)-invariant tensor $\hat{T}$ with components $(\hat{T})_{i_1i_2i_3i_4}$ and directions $\{\mbox{`in', `out', `out', `in'}\}$. It comprises of an SU(2)-invariant vector $\hat{v}$, three split tensors $\spliter$ and two cups. Tensor $\hat{T}'$ is obtained by bending the incoming indices ($i_1$ and $i_4$) of $\hat{T}$. The tree decomposition of tensor $\hat{T}$ is obtained by applying a resolution of Identity $\mathcal{I}(\tree)$ on the indices of $\hat{T}'$ where $\mathcal{I}(\tree)$ corresponds to a tensor network made of fuse and split tensors that are interconnected according to a given fusion tree $\tree$. \label{fig:treedeco1}}
\end{figure}

A tree decomposition, denoted $\mathcal{D}(\hat{T})$, of a rank-4 SU(2)-invariant tensor $\hat{T}$ with indices $i_1,i_2,i_3,i_4$ and directions `in',`out',`out', and `in' is shown in \fref{fig:treedeco1}. It consists of (i) an SU(2)-invariant vector $\hat{v}$ with components $(\hat{v})_i$, (ii) three split tensors $\spliter$, and (iii) two cups [\fref{fig:fmove1}(c),(e)]. The cups are attached to indices that are incoming in $\hat{T}$, that is, $i_1$ and $i_4$; the parity (left or right) of the cups is additionally specified. In the figure, $\hat{T}'$ denotes the SU(2)-invariant tensor with only outgoing indices that is obtained by bending the incoming indices of $\hat{T}$ (using the cap transformations, \fref{fig:fmove1}(d)). Equivalently, tensor $\hat{T}$ may be recovered by multiplying $\hat{T}'$ and the corresponding cups. 

The tree decomposition $\mathcal{D}(\hat{T})$ is obtained by inserting a resolution of Identity $\mathcal{I}(\tree)$ as shown and then multiplying together $\hat{T}'$ and all the fuse tensors in $\mathcal{I}(\tree)$ to obtain the vector $\hat{v}$. The resolution of Identity $\mathcal{I}(\tree)$ is given by a tensor network made of tensors $\fuser$ that fuse the indices of $\hat{T}'$ according to the given fusion tree $\tree$ and the corresponding tensors $\spliter$ that invert this fusion.

More generally, a tree decomposition $\mathcal{D}(\hat{T})$ of a rank-$k$ SU(2)-invariant tensor $\hat{T}$ with $k_{in}$ incoming indices consists of an SU(2)-invariant vector $\hat{v}$ with an index $i$ that is obtained by fusing indices $i_1,i_2,\ldots,i_k$ according to a fusion tree $\tree$, the $(k-1)$ split tensors that invert this fusion and $k_{in}$ cups. In practice, the tree decomposition $\mathcal{D}(\hat{T})$ of an SU(2)-invariant tensor $\hat{T}$ can be stored in memory by storing the following data:
\begin{enumerate}
	\item the indices $i_l=(j_l,t_{j_l},m_{j_l})$,
	\item the parity of the bend $\vec{B}$ on each index e.g. for the tree decomposition of \fref{fig:treedeco1} we have $\vec{B} \equiv \{\mbox{`left',`straight', `straight', `right'}\}$,
	\item the fusion tree $\tree$ according to which the indices of the tensor are fused into the single index $i$ (equivalently, we also say that the tree decomposition is based on an underlying \textit{splitting} tree $\tree$), and
	\item the vector $\hat{v}$.
\end{enumerate}

\begin{figure}[t]
  \includegraphics[width=8.5cm]{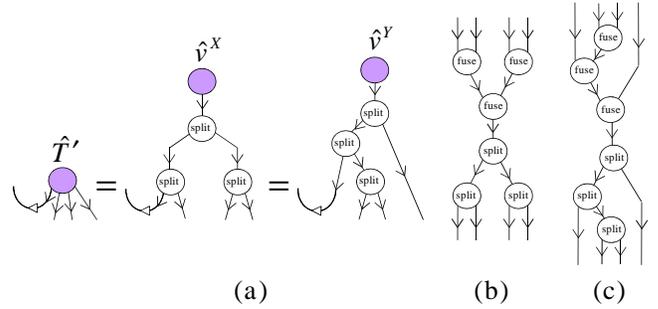}
\caption{(Color online) (a) Two different tree decompositions of a rank-$4$ SU(2)-invariant tensor corresponding to the choice of two different fusion trees $\tree^{X}$ and $\tree^{Y}$. The two tree decompositions are obtained from the tensor by means of the resolutions of Identity (b) $\mathcal{I}(\tree^{X})$ and (c) $\mathcal{I}(\tree^{Y})$ as illustrated in \fref{fig:treedeco1}.
\label{fig:treeDeco4}}
\end{figure}

\begin{figure}[t]
  \includegraphics[width=6.5cm]{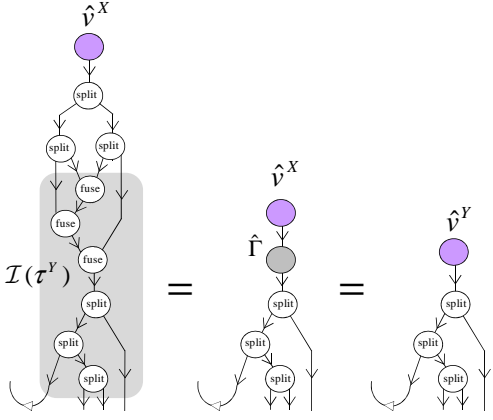}
\caption{(Color online)  Mapping between two tree decompositions [\fref{fig:treeDeco4}] $\mathcal{D}^{X}(\hat{T})$ and $\mathcal{D}^{Y}(\hat{T})$ of an SU(2)-invariant tensor $\hat{T}$. Tree decomposition $\mathcal{D}^{Y}(\hat{T})$ is obtained from $\mathcal{D}^{X}(\hat{T})$ in two steps. First the resolution of Identity $\mathcal{I}(\tree^Y)$ is applied on $\mathcal{D}^{X}(\hat{T})$ as shown and a matrix $\hat{\Gamma}$ is obtained by multiplying together the split tensors in $\mathcal{D}^{X}(\hat{T})$ and the fuse tensors in $\mathcal{I}(\tree^Y)$. Then vector $\hat{v}^Y \in \mathcal{D}^{Y}(\hat{T})$ is obtained by multiplying $\hat{\Gamma}$ with $\hat{v}^X$, \eref{eq:Gamma}. \label{fig:MtoM}}
\end{figure}


\subsection{Mapping between tree decompositions\label{sec:tree:basischange}}

The same tensor $\hat{T}$ may be expressed in different tree decompositions corresponding to different choices of the fusion tree. A different choice of left and right directed cups also generally corresponds to a different tree decomposition of the tensor (see Sec.~\ref{sec:tree:reverse}). However, here we will only consider how tree decompositions with different fusion trees are related.

Two different fusion trees $\tree^{X}$ and $\tree^{Y}$ lead to two different tree decompositions $\mathcal{D}^{X}(\hat{T})$ and $\mathcal{D}^{Y}(\hat{T})$ of the same tensor $\hat{T}$, as illustrated in \fref{fig:treeDeco4}(a). The two decompositions are obtained from the tensor $\hat{T}$ by means of the resolutions of Identity $\mathcal{I}(\tree^{X})$ and $\mathcal{I}(\tree^{Y})$ [\fref{fig:treeDeco4}(b)-(c)] as explained previously. Suppose now that we have a tensor $\hat T$ in a tree decomposition $\mathcal{D}^{X}(\hat{T})$ and we wish to transform it into another tree decomposition $\mathcal{D}^{Y}(\hat{T})$. This can be achieved by applying the resolution of Identity $\mathcal{I}(\tree^{Y})$ on the tree decomposition $\mathcal{D}^{X}(\hat{T})$ as shown in \fref{fig:MtoM}. The new vector $\hat{v}^{Y}$ is obtained from $\hat{v}^{X}$ as
\begin{equation}
	\hat{v}^{Y} = \hat{\Gamma}\hat{v}^X,
	\label{eq:Gamma}
\end{equation}
where $\hat{\Gamma}$ is a matrix that is obtained by multiplying together the split tensors $\spliter$ in $\mathcal{D}^{X}(\hat{T})$ and the fuse tensors $\fuser$ in $\mathcal{D}^{Y}(\hat{T})$. By construction, the matrix $\hat{\Gamma}$ is SU(2)-invariant and has a block-diagonal form. Notice that only the block with $j=0$ is relevant in \eref{eq:Gamma}, since $\hat{\Gamma}$ is multiplied with an SU(2)-invariant vector $\hat{v}$. (App. \ref{sec:tree:precom} describes how to obtain the matrix $\hat{\Gamma}$ in the block diagonal form in practice.)

Next we describe how the set $\mathcal{P}$ of primitive tensor network manipulations, namely, the reversal, permutation and reshape of indices, and matrix multiplication and matrix factorization, are adapted to tree decompositions of SU(2)-invariant tensors.

\begin{figure}[t]
  \includegraphics[width=8cm]{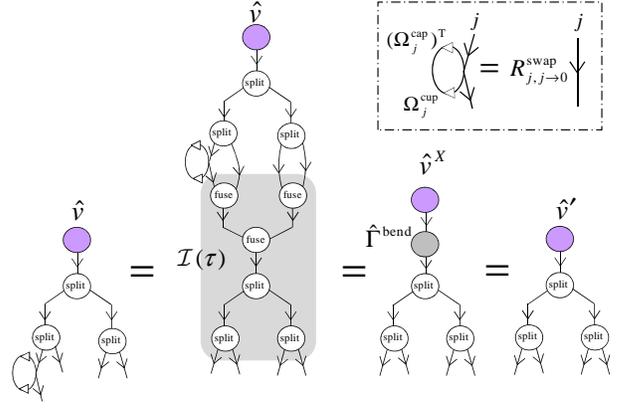}
\caption{(Color online) Bending indices of an SU(2)-invariant tensor $\hat{T}$ that is given in the tree decomposition $\mathcal{D}(\hat{T})$ to obtain another SU(2)-invariant tensor $\hat{T}'$. The only non-trivial cases are bending down a `left' index from the right (shown here) and bending a `right' index from the left, which results in a ``loop'' as shown. A tree decomposition of the tensor $\hat{T}'$ is obtained by subsuming the loop into $\mathcal{D}(\hat{T})$. This is achieved by applying the resolution of Identity $\mathcal{I}(\tree)$ as shown, straightening the loop (inset) then multiplying together the resulting swap factors, the split tensors in $\mathcal{D}(\hat{T})$ and the fuse tensors in $\mathcal{I}(\tree)$ to obtain a matrix $\hat{\Gamma}^{\tiny \mbox{bend}}$. The vector $\hat{v}'$ that comprises the tree decomposition of $\hat{T}'$ is obtained by multiplying $\hat{\Gamma}^{\tiny \mbox{bend}}$ with $\hat{v}$, \eref{eq:treebend}.}
\label{fig:treebend}
\end{figure}


\subsection{Reversal of indices\label{sec:tree:reverse}}
Consider an SU(2)-invariant tensor $\hat{T}$ that is given in a tree decomposition $\mathcal{D}(\hat{T})\equiv (\{i_1, i_2, \ldots, i_k\}, \vec{B}, \tree, \hat{v})$ and let $\hat{T}'$ denote the SU(2)-invariant tensor obtained from $\hat{T}$ by bending some of its indices. When bending a `straight' index either leftward or rightward, a tree decomposition $\mathcal{D}(\hat{T'})$ of tensor $\hat{T}'$ is obtained by simply attaching the left or right directed cup to the index in the tree decomposition $\mathcal{D}(\hat{T})$ respectively. In practice, this corresponds to only updating the parity of the index in $\vec{B}$ to `left' or `right'. In particular, the decompositions $\mathcal{D}(\hat{T})$ and $\mathcal{D}(\hat{T'})$ comprise of the same vector $\hat{v}$.

Next, when bending (downward) a `left' index from the left or a `right' index from the right, a tree decomposition of $\hat{T}'$ is obtained from $\mathcal{D}(\hat{T})$ by simply detaching the original bend (cup) from the index. Once again, in practice, this corresponds to simply updating the parity of the index in $\vec{B}$, to `straight' in this case.

However, bending a `left' index from the right or a `right' index from the left results in a ``loop'', as shown in \fref{fig:treebend}. A tree decomposition of tensor $\hat{T}'$ is obtained by subsuming the loop into $\mathcal{D}(\hat{T})$. This is achieved by first applying the resolution of Identity $\mathcal{I}(\tree)$ as illustrated in the figure. The vector $\hat{v}'$ that comprises the tree decomposition of $\hat{T}'$ is then obtained as
\begin{equation}
	\hat{v}' = \hat{\Gamma}^{\tiny \mbox{bend}}\hat{v},
	\label{eq:treebend}
\end{equation}
where $\hat{\Gamma}^{\tiny \mbox{bend}}$ is the SU(2)-invariant matrix obtained by ``straightening'' (see inset of \fref{fig:treebend}) the loop(s) and multiplying together the resulting swap factors, the split tensors $\spliter$ in $\mathcal{D}(\hat{T})$ and the fuse tensors $\fuser$ in $\mathcal{I}(\tree)$.

\begin{figure}[t]
  \includegraphics[width=8.5cm]{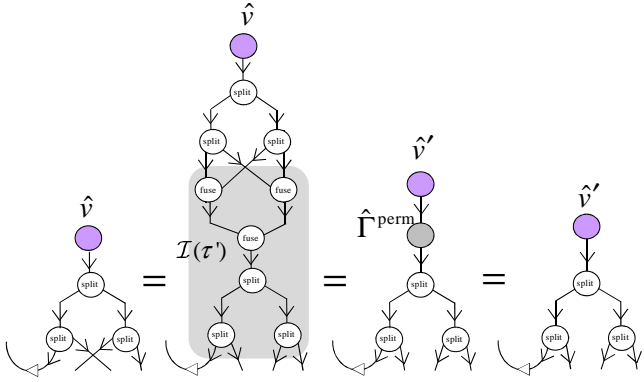}
\caption{(Color online) Permuting (or intercrossing) indices of an SU(2)-invariant tensor $\hat{T}$ that is given in a tree decomposition $\mathcal{D}(\hat{T})$ to obtain another SU(2)-invariant tensor $\hat{T}'$. A tree decomposition of the tensor $\hat{T}'$ is obtained by subsuming the intercrossings into the decomposition $\mathcal{D}(\hat{T})$. This is achieved by applying a resolution of Identity $\mathcal{I}(\tree')$ (for a specified fusion tree $\tree'$) as shown and then multiplying together the split tensors in $\mathcal{D}(\hat{T})$ and the fuse tensors in $\mathcal{I}(\tree')$ to obtain a matrix $\hat{\Gamma}^{\tiny \mbox{perm}}$. The vector $\hat{v}'$ that comprises the tree decomposition of $\hat{T}'$ is obtained by multiplying $\hat{\Gamma}^{\tiny \mbox{perm}}$ and the vector $\hat{v}$, \eref{eq:treeperm}.}
\label{fig:permute1}
\end{figure}


\subsection{Permutation of indices \label{sec:tree:permute}}
As described in App. \ref{sec:tensor:manipulations}, an arbitrary permutation of indices of a tensor can be decomposed into a sequence of reversals and pairwise swaps. For a tree decomposition this corresponds to first detaching all the cups, then intercrossing the indices and finally reattaching the cups. This is equivalent to applying the permutation of indices \textit{before} the cups (i.e. in the corresponding graphical representation the intercrossing of lines appears \textit{above} the cups). Thus, cups are irrelevant when subsuming the intercrossings (permutation of indices) into the tree decomposition.

Consider an SU(2)-invariant tensor $\hat{T}'$ that is obtained by permuting, in an arbitrary way, the indices of an SU(2)-invariant tensor $\hat{T}$ given in the tree decomposition $\mathcal{D}(\hat{T})$. A tree decomposition $\mathcal{D}(\hat{T}')$ of the tensor $\hat{T}'$ is obtained by subsuming the intercrossings into the decomposition $\mathcal{D}(\hat{T})$. This is achieved by first applying the resolution of Identity $\mathcal{I}(\tree')$, as illustrated in \fref{fig:permute1}, where $\tree'$ is the fusion tree specified for the decomposition $\mathcal{D}(\hat{T}')$. Vector $\hat{v}'$ that comprises the decomposition $\mathcal{D}(\hat{T}')$ is then obtained as
\begin{equation}
	\hat{v}' = \hat{\Gamma}^{\tiny \mbox{perm}}\hat{v},
	\label{eq:treeperm}
\end{equation}
where $\hat{\Gamma}^{\tiny \mbox{perm}}$ is the SU(2)-invariant matrix obtained by multiplying together the split tensors in $\mathcal{D}(\hat{T})$ and the fuse tensors in $\mathcal{I}(\tree')$.

\begin{figure}[t]
  \includegraphics[width=8.5cm]{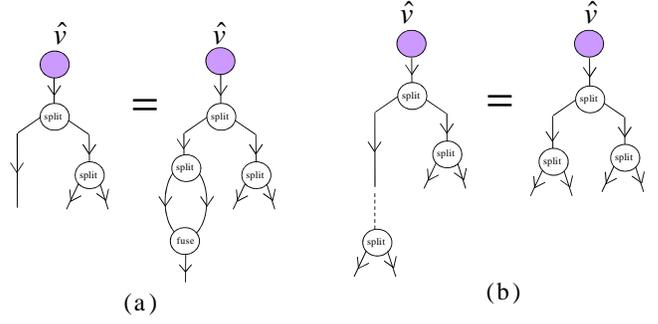}
\caption{(Color online)  Reshaping indices of an SU(2)-invariant tensor $\hat{T}$ that is given in a tree decomposition $\mathcal{D}(\hat{T})$ to obtain another SU(2)-invariant tensor $\hat{T}'$. Fusion and splitting of `straight' indices is by using transformations $\fuser$ and $\spliter$ respectively. (a) Fusion of two `straight' indices of $\hat{T}$ that belong to the same split tensor in the tree decomposition $\mathcal{D}(\hat{T})$. A tree decomposition of $\hat{T}'$ is obtained by simply deleting the split tensor, \eref{fig:su2fuse}. (b) Tensor $\hat{T}$ is recovered from $\hat{T}'$ by splitting back the fused index. That is, the tree decomposition $\mathcal{D}(\hat{T})$ is recovered by reattaching the split tensor to the tree decomposition of $\hat{T}'$.}
\label{fig:reshape}
\end{figure}


\subsection{Reshape of indices\label{sec:tree:reshape}}

Two `straight' indices of an SU(2)-invariant tensor given in a tree decomposition are fused by using the transformation $\fuser$. A `straight' index is split into two indices by using the transformation $\spliter$. Consider an SU(2)-invariant tensor $\hat{T}'$ obtained by fusing two `straight' indices $i_l$ and $i_{l+1}$ of an SU(2)-invariant tensor $\hat{T}$. Let us work in a tree decomposition $\mathcal{D}(\hat{T})$ of the tensor $\hat{T}$ in which $i_l$ and $i_{l+1}$ belong to the same split tensor. Then a tree decomposition $\mathcal{D}(\hat{T}')$ of the tensor $\hat{T}'$ is obtained by simply deleting that split tensor from $\mathcal{D}(\hat{T})$, as illustrated in \fref{fig:reshape}(a). (Since the split tensor cancels out with the applied fuse tensor, \fref{fig:su2fuse}(c)). Tensor $\hat{T}$ may be recovered from $\hat{T}'$ by splitting the fused index back into indices $i_l$ and $i_{l+1}$. That is, the tree decomposition $\mathcal{D}(\hat{T})$ is recovered by reattaching the split tensor $\spliter$ to the decomposition $\mathcal{D}(\hat{T}')$, as illustrated in \fref{fig:reshape}(b). Note that the tree decompositions $\mathcal{D}(\hat{T}')$  and $\mathcal{D}(\hat{T})$ comprise the same vector.

Finally, consider reshaping an SU(2)-invariant tensor by fusing a `left'/`right' index with a `straight' index. Once again, consider that the two indices belong to the same split node in the tree decomposition. In this case the reshape proceeds by first detaching the left/right cup and then fusing the two indices as described above.  The original tensor, and its tree decomposition, may be recovered by simply reattaching the removed $\spliter$ and the removed cup.


\subsection{Matrix multiplication and factorizations \label{sec:tree:matrixops}}
Two SU(2)-invariant matrices, each given as a tree decomposition, may be multiplied together by first obtaining the matrices in a block-diagonal form (that is, the $(\hat{P},\hat{Q})$ form described in Sec.~\ref{sec:su2tensors:block}) from the respective tree decompositions, performing block-wise multiplication (see Sec.~\ref{sec:tnsu2:multiply}) and recasting the resulting block-diagonal matrix into a tree decomposition. An SU(2)-invariant matrix may be factorized e.g. singular value decomposed in a similar way. That is, by first obtaining the matrix in a block-diagonal form, then performing block-wise factorization (see Sec.~\ref{sec:tnsu2:factorize}), and finally recasting each of the factor block-diagonal matrices into a tree decomposition.

\begin{figure}[t]
  \includegraphics[width=7.5cm]{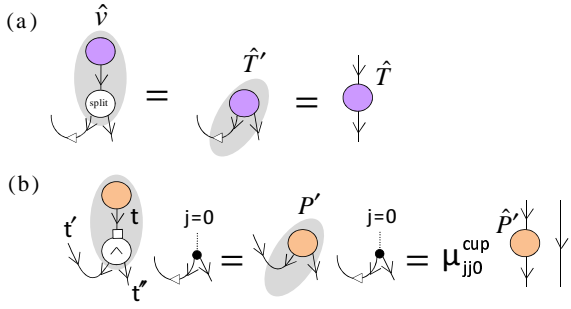}
\caption{(Color online) (a) Obtaining the block diagonal form (i.e. the $(\hat{P},\hat{Q})$ form) of an SU(2)-invariant matrix from its tree decomposition (\textit{left}) by performing two multiplications (highlighted by shading). First, the vector $\hat{v}$ is multiplied with the split tensor to obtain an intermediate SU(2)-invariant tensor $\hat{T}'$. Then $\hat{T}'$ is multiplied with the cup to obtain the block diagonal matrix $\hat{T}$. (b) The two multiplications of $(a)$ as performed in the canonical form at each step. The ``multiplication'' with the cup simply corresponds to applying the F-move of \fref{fig:fmove1}(b).}
\label{fig:recovermat}
\end{figure}

The tree decomposition of an SU(2)-invariant matrix $\hat{T}$ consists of a vector $\hat{v}$, a split tensor $\spliter$ and a cup. The block diagonal form of $\hat{T}$ can be obtained from its tree decomposition as shown in \fref{fig:recovermat}. Analogously, the tree decomposition of an SU(2)-invariant matrix $\hat{T}$ can be obtained from its block-diagonal form by reversing the depicted procedure.

\subsection{Precomputation scheme for iterative tensor network algorithms}\label{sec:tree:precom}

We conclude this appendix by describing how the SU(2)-invariant matrix $\Gamma$ of \eref{eq:Gamma}, and also the closely related SU(2)-invariant matrix $\hat{\Gamma}^{\tiny \mbox{bend}}$ of \eref{eq:treebend} and $\hat{\Gamma}^{\tiny \mbox{perm}}$ of \eref{eq:treeperm}, is obtained in the block-diagonal form.

\begin{figure}[b]
  \includegraphics[width=6.5cm]{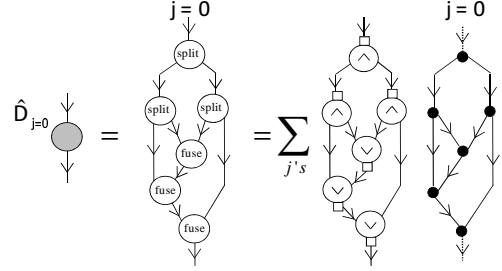}
\caption{The matrix $\hat{\Gamma}$ of \eref{eq:Gamma} is obtained by contracting a tensor network $\mathcal{M}$ made of fuse and split tensors. Only the $j=0$ block ($\hat{D}_{j=0}$ in \eref{eq:gammasplit}) of $\hat{\Gamma}$ is relevant. For fixed values of $j$'s on all the contracted indices, compatible with $j=0$ on the two open indices, each tensor $\fuser$ and $\spliter$ decomposes into $X$ and $C$ parts. Subsequently, the tensor network $\mathcal{M}$ factorizes into two terms tensors. The first one is a tensor network is made of $X$ tensors that can be contracted to obtain a matrix made of 0's and 1's. The second term is a spin network which, here, can be contracted to obtain a number (the ``value'' of the spin network) since the open indices take only one value: $j=0,m=0$, i.e. have size one. This ``evaluation'' can be achieved by applying a sequence of F-moves (instead of actually multiplying the Clebsch-Gordan tensors). Therefore, the value of the spin network is given in terms of $\hat{F}$ coefficients (and possibly also the $\braider$ coefficients when considering the contractions for obtaining the matrices $\hat{\Gamma}^{\tiny \mbox{bend}}$ or $\hat{\Gamma}^{\tiny \mbox{perm}}$).
\label{fig:gamma1}}
\end{figure}

The SU(2)-invariant matrix $\hat{\Gamma}$ of \fref{fig:MtoM} is separately shown in \fref{fig:gamma1}. It is obtained by contracting a tensor network $\mathcal{M}$ made of fuse tensors $\fuser$ and split tensors $\spliter$, and decomposes as
\begin{equation}
\hat{\Gamma} = \bigoplus_{j} (\hat{D}_{j} \otimes \hat{I}_{j}).\label{eq:gammasplit}
\end{equation}
Here we are interested only in the $j=0$ block since $\hat{\Gamma}$ is multiplied with the SU(2)-invariant vector $\hat{v}$. The (degeneracy) matrix $\hat{D}_{j=0}$ is obtained as explained by \fref{fig:gamma1}.
Note that the contraction of the $X$ tensors can be performed in a fast way by exploiting the fact that they are sparse and made of only 0's and 1's in a very specific way. We refer the reader to the appendix of Ref.~\onlinecite{Singh11} where one method for the fast multiplication of the $X$ tensors was outlined\cite{fusetensor}.

In this appendix we have described how by working in tree decompositions the reversal, permutation and reshaping of indices of SU(2)-invariant tensors corresponds to multiplying a matrix with an SU(2)-invariant vector [Eqs.~(\ref{eq:Gamma})-(\ref{eq:treeperm})]. Note that these matrices are purely \textit{structural} and do not depend on the components of tensors that are e.g. reshaped or permuted. In the specific context of algorithms where the same tensor manipulations are iterated many times the same structural matrices are computed in every iteration. In such a scenario it is possible to significantly decrease the running cost by precomputing all such matrices once and reusing them in subsequent iterations thus reducing computational times at the expense of incurring an additional memory cost. In our implementation the use of precomputation led to a significant speedup of simulations, \fref{fig:meracompare}. We also remark that by implementing tensor network algorithms in terms of only matrix operations the implementation code is readily set up for further optimization by using vectorization and parallelization techniques.



\begin{thebibliography}{74}

\bibitem{Fannes92} M. Fannes, B. Nachtergaele, and R. Werner, Commun. Math. Phys. \textbf{144}, 443 (1992).
\bibitem{Ostlund95} S. Ostlund and S. Rommer, Phys. Rev. Lett. \textbf{75}, 3537 (1995). 
\bibitem{Vidal03} G. Vidal, Phys. Rev. Lett. \textbf{91}, 147902 (2003). 
\bibitem{PerezGarcia07} D. PerezGarcia, F. Verstraete, M. M. Wolf, and J. I. Cirac, Quantum Inf. Comput. \textbf{7}, 401 (2007).

\bibitem{White92} S. R. White, Phys. Rev. Lett. \textbf{69}, 2863 (1992).
\bibitem{White93} S. R. White, Phys. Rev. B \textbf{48}, 10345 (1993).
\bibitem{Schollwoeck05} U. Schollw\"ock, Rev. Mod. Phys. \textbf{77}, 259 (2005).
\bibitem{McCulloch08} I. P. McCulloch, arXiv:0804.2509v1 [cond-mat.str-el] (2008).

\bibitem{Ramasesha96} S. Ramasesha, S. K. Pati, H. R. Krishnamurthy, Z. Shuai, and J. L. Bredas, Phys.Rev. B \textbf{54}, 7598 (1996).
\bibitem{Sierra97} G. Sierra and T. Nishino, Nucl. Phys. \textbf{B495}, 505 (1997). 
\bibitem{Tatsuaki00} W. Tatsuaki, Phys. Rev. E \textbf{61}, 3199 (2000).
\bibitem{McCulloch02} I. P. McCulloch and M. Gulacsi, Europhys. Lett. \textbf{57}, 852 (2002). 
\bibitem{Bergkvist06} S. Bergkvist, I. P. McCulloch, and A. Rosengren, Phys. Rev. A \textbf{74}, 053419 (2006).
\bibitem{Pittel06} S. Pittel and N. Sandulescu, Phys. Rev. C \textbf{73}, 014301 (2006).
\bibitem{McCulloch07} I. McCulloch, J. Stat. Mech., P10014 (2007).
\bibitem{PerezGarcia08} D. PerezGarcia, M. M. Wolf, M. Sanz, F. Verstraete, and J. I. Cirac, Phys. Rev. Lett. \textbf{100}, 167202 (2008).
\bibitem{Sanz09} M. Sanz, M. M. Wolf, D. PerezGarcia, and J. I. Cirac, Phys. Rev. A \textbf{79}, 042308 (2009).

\bibitem{Vidal04} G. Vidal, Phys. Rev. Lett. \textbf{93}, 040502 (2004). 
\bibitem{Daley04} A. J. Daley, C. Kollath, U. Schollw\"ock, and G. Vidal, J. Stat. Mech. Theor. Exp., P04005 (2004). 
\bibitem{White04} S. R. White and A. E. Feiguin, Phys. Rev. Lett. \textbf{93}, 076401 (2004).
\bibitem{Schollwoeck05b} U. Schollw\"ock, J. Phys. Soc. Jpn. \textbf{74S}, 246 (2005).
\bibitem{Daley05} A. J. Daley, S. R. Clark, D. Jaksch, and P. Zoller, Phys. Rev. A \textbf{72}, 043618(2005).
\bibitem{Vidal07} G. Vidal, Phys. Rev. Lett. \textbf{98}, 070201 (2007).
\bibitem{Danshita07} I. Danshita, J. E. Williams, C. A. R. S\'a de Melo, and C. W. Clark, Phys. Rev. A \textbf{76}, 043606(2007).
\bibitem{Muth09} D. Muth, B. Schmidt, and M. Fleischhauer, New J. Phys. \textbf{12}, 083065 (2010). 
\bibitem{Mishmash09} R. V. Mishmash, I. Danshita, C. W. Clark, and L. D. Carr, Phys. Rev. A \textbf{80}, 053612 (2009). 
\bibitem{Singh10} S. Singh, H.-Q. Zhou, and G. Vidal, New J. Phys. \textbf{12}, 033029 (2010).
\bibitem{Cai10} Z. Cai, L. Wang, X. C. Xie, and Y. Wang, Phys. Rev. A \textbf{81}, 043602 (2010).

\bibitem{Shi06} Y. Y. Shi, L.-M. Duan and G. Vidal, Phys. Rev. A \textbf{74}, 022320 (2006).

\bibitem{Vidal07b} G. Vidal, Phys. Rev. Lett. \textbf{99}, 220405 (2007).
\bibitem{Vidal08} G. Vidal, Phys. Rev. Lett. \textbf{101}, 110501 (2008).
\bibitem{Evenbly09} G. Evenbly and G. Vidal, Phys. Rev. B \textbf{79}, 144108 (2009).
\bibitem{Giovannetti08} V. Giovannetti, S. Montangero, and R. Fazio, Phys. Rev. Lett. \textbf{101},
180503 (2008).
\bibitem{Pfeifer09} R. N. C. Pfeifer, G. Evenbly, and G. Vidal, Phys. Rev. A \textbf{79},
040301(R) (2009).
\bibitem{Vidal10} G. Vidal, in \textit{Understanding Quantum Phase Transitions}, edited by L. D. Carr (Taylor \& Francis, Boca Raton, 2010).

\bibitem{Yan11} 
S. Yan, D. A. Huse, S. R. White,
Science, Vol. 332 (6034) 1173-1176 (2011).
\bibitem{Jiang12} 
H.-C. Jiang, Z. Wang, L. Balents, arXiv:1205.4289v1 [cond-mat.str-el]
\bibitem{Depenbrock12} 
S. Depenbrock, I. P. McCulloch, U. Schollwoeck,
arXiv:1205.4858v1 [cond-mat.str-el]

\bibitem{Tagliacozzo09} L. Tagliacozzo, G. Evenbly, and G. Vidal, Phys. Rev. B \textbf{80}, 235127 (2009).
\bibitem{Murg10} V. Murg, O. Legeza, R. M. Noack, and F. Verstraete, arXiv:1006.3095v1 [cond-mat.str-el] (2006).

\bibitem{Evenbly10f} G. Evenbly and G. Vidal, Phys. Rev. B \textbf{81}, 235102 (2010). 
\bibitem{Evenbly10b} G. Evenbly and G. Vidal, New J. Phys. \textbf{12}, 025007 (2010). 
\bibitem{Aguado08} M. Aguado and G. Vidal, Phys. Rev. Lett. \textbf{100}, 070404 (2008).
\bibitem{Cincio08} L. Cincio, J. Dziarmaga, and M. M. Rams, Phys. Rev. Lett. \textbf{100}, 240603 (2008).
\bibitem{Evenbly09b} G. Evenbly and G. Vidal, Phys. Rev. Lett. \textbf{102}, 180406 (2009).
\bibitem{Koenig09} R. K\"onig, B. W. Reichardt, and G. Vidal, Phys. Rev. B \textbf{79}, 195123
(2009).
\bibitem{Evenbly10} G. Evenbly and G. Vidal, Phys. Rev. Lett. \textbf{104}, 187203 (2010).

\bibitem{Verstraete04} F. Verstraete, and J. I. Cirac, arXiv:cond-mat/0407066v1 (2004).
\bibitem{Sierra98} G. Sierra and M. A. Martin-Delgado, arXiv:cond-mat/9811170v3 (1998).
\bibitem{Nishino98} T. Nishino and K. Okunishi, J. Phys. Soc. Jpn. \textbf{67}, 3066, 1998. 
\bibitem{Nishio04} Y. Nishio, N. Maeshima, A. Gendiar, and T. Nishino, arXiv:cond-mat/0401115v1 (2004).
\bibitem{Murg07} V. Murg, F. Verstraete, and J. I. Cirac, Phys. Rev. A \textbf{75}, 033605 (2007).
\bibitem{Jordan08} J. Jordan, R. Orus, G. Vidal, F. Verstraete, and J. I. Cirac, Phys. Rev. Lett. \textbf{101}, 250602 (2008).
\bibitem{Gu08} Z.-C. Gu, M. Levin, and X.-G. Wen, Phys. Rev. B \textbf{78}, 205116 (2008).
\bibitem{Jiang08} H. C. Jiang, Z. Y. Weng, and T. Xiang, Phys. Rev. Lett. \textbf{101}, 090603 (2008).
\bibitem{Xie09} Z. Y. Xie, H. C. Jiang, Q. N. Chen, Z. Y. Weng, and T. Xiang, Phys. Rev. Lett. \textbf{103}, 160601 (2009).
\bibitem{Murg09} V. Murg, F. Verstraete, and J. I. Cirac, Phys. Rev. B \textbf{79}, 195119 (2009).
\bibitem{Wang11} 
L. Wang, Z.-C. Gu, F. Verstraete, X.-G. Wen, arXiv:1112.3331v2 [cond-mat.str-el]

\bibitem{Corboz09} P. Corboz, G. Evenbly, F. Verstraete, and G. Vidal, Phys. Rev. A \textbf{81}, 010303(R) (2010). 
\bibitem{Kraus09} C. V. Kraus, N. Schuch, F. Verstraete, and J. I. Cirac, Phys. Rev. A \textbf{81}, 052338 (2010). 
\bibitem{Pineda09} C. Pineda, T. Barthel, and J. Eisert, Phys. Rev. A \textbf{81}, 050303(R) (2010).
\bibitem{Corboz09b} P. Corboz and G. Vidal, Phys. Rev. B \textbf{80}, 165129 (2009). 
\bibitem{Barthel09} T. Barthel, C. Pineda, and J. Eisert, Phys. Rev. A \textbf{80}, 042333 (2009).
\bibitem{Shi09} Q.-Q. Shi, S.-H. Li, J.-H. Zhao, and H.-Q. Zhou, arXiv:0907.5520v1 [cond-mat.str-el] (2009). S.-H. Li, Q.-Q. Shi, H.-Q. Zhou, arXiv:1001.3343v1 [cond-mat.supr-con] (2010).
\bibitem{Corboz10b}  P. Corboz, R. Orus, B. Bauer, and G. Vidal, Phys. Rev. B \textbf{81}, 165104 (2010).
\bibitem{Pizorn10} I. Pizorn and F. Verstraete, Phys. Rev. B \textbf{81}, 245110 (2010).
\bibitem{Gu10} Z.-C. Gu, F. Verstraete, and X.-G. Wen, arXiv:1004.2563v1 [cond-mat.str-el] (2010).

\bibitem{Pollmann10} F. Pollmann, A. M. Turner, E. Berg, and M. Oshikawa, Phys. Rev. B 81, 064439 (2010).
\bibitem{Chen11} X. Chen, Z.-C. Gu, and X.-G.Wen, Phys. Rev. B , 035107 (2011). 
\bibitem{Schuch11} N. Schuch, D. PerezGarcia, I. Cirac, Phys. Rev. B 84, 165139 (2011). 
\bibitem{Chen11b} X. Chen, Z.-C. Gu, X.-G. Wen, Phys. Rev. B 84, 235128 (2011) 
\bibitem{Gu12} Z.-C. Gu, X.-G. Wen, arXiv:1201.2648v1 [cond-mat.str-el]. 

\bibitem{Verstraete06} F. Verstraete, M. M. Wolf, D. Perez-Garcia, J. I. Cirac,
Phys. Rev. Lett. 96, 220601 (2006).
\bibitem{Evenbly11} G. Evenbly, G. Vidal, J Stat Phys (2011) 145:891-918.
\bibitem{Evenbly12} G. Evenbly, G. Vidal, arXiv:1205.0639v1 [quant-ph]
\bibitem{Evenbly12b} G. Evenbly, G. Vidal, \textit{Branching MERA}, in preparation.

\bibitem{Cornwell97} J. F. Cornwell, {\it Group Theory in Physics} (Academic Press, San Diego, 1997).

\bibitem{Singh09} S. Singh, R. N. C. Pfeifer, and G. Vidal, Phys. Rev. A \textbf{82}, 050301 (2010), arXiv:0907.2994v1 [cond-mat.str-el] (2009).

\bibitem{Singh11} S. Singh, R. N. C. Pfeifer and G. Vidal, Phys. Rev. B \textbf{83}, 115125 (2011), arXiv:1008.4774 [cond-mat.str-el] (2010).

\bibitem{spinnetwork} A spin network is a well-known object in mathematical physics and, especially, in loop quantum gravity, where it is used to describe states of quantum geometry, see for instance ``C. Rovelli and L. Smolin, Phys. Rev. D \textbf{53}, 5743 (1995)''. A spin network is a (directed) graph whose edges are associated with irreducible representations of a compact Lie group and whose vertices are associated with intertwiners of the edge representations adjacent to it.
%
\bibitem{Andreas12} A. Weichselbaum, arXiv:1202.5664v1.
\bibitem{multiplicity} Multiplicity may be either inner or outer. \textit{Inner multiplicity} refers to the occurrence of several states that carry the same $m$ label for a given $j$, while \textit{outer multiplicity} corresponds to the occurrence of multiple copies of the same irrep (i.e. same $j$) in the output of the fusion rules.

\bibitem{PerezGarcia10} D. PerezGarcia, M. Sanz, C. E. Gonzalez-Guillen, M. M. Wolf, and J. I. Cirac, New J. Phys. \textbf{12}, 025010 (2010).
\bibitem{Zhao10} H. H. Zhao, Z. Y. Xie, Q. N. Chen, Z. C. Wei, J. W. Cai, and T. Xiang, Phys. Rev. B \textbf{81}, 174411 (2010).

\bibitem{Pfeifer10} R.N.C. Pfeifer, P. Corboz et. al., Phys. Rev. B 82, 115126 (2010).
\bibitem{Koenig10} R. Koenig and E. Bilgin, Phys. Rev. B 82, 125118 (2010).


\bibitem{cgtable} See, for instance, en.wikipedia.org/wiki/Table of Clebsch-Gordan coefficients, and the references therein.
\bibitem{fusetensor} We notice that tensor $\tfuser$ can be decomposed into two pieces. The first piece (depicted as the circle enclosing an arrow head in \fref{fig:su2fuse}(a)) expresses a basis $\{\ket{j_{A}t_{j_{A}}; j_{B}t_{j_{B}} \equiv \ket{j_{A}t_{j_A}} \otimes \ket{j_{B}t_{j_B}}}\}$ of $\mathbb{D}^{(AB)}$ as the direct product of the basis $\{\ket{j_{A}t_{j_A}}\}$ of $\mathbb{D}^{(A)}$ and the basis $\{\ket{j_{B}t_{j_B}}\}$ of $\mathbb{D}^{(B)}$.  Note that this procedure does not always lead to the set $\{\ket{j_{A}t_{j_{A}}; j_{B}t_{j_{B}}}\}$ being ordered such that states corresponding to the same total spin $j_{AB}$ are adjacent to each other within the set. However, we require that the basis associated to an index be maintained as such (this ensures, for example, that an SU(2)-invariant matrix is block diagonal when expressed in such a basis). This ordering is achieved by means of the second piece (depicted as the small rectangle in \fref{fig:su2fuse}(a)): a permutation of basis states $\{\ket{j_{A}t_{j_{A}}; j_{B}t_{j_{B}}}\}$ that reorganizes them according to their total spin $j_{AB}$, so that they are identified in an one-to-one correspondence with the coupled states $\{\ket{j_{AB}t_{j_{AB}}}\}$. In particular, this description of the tensors $\tfuser$ and $\tspliter$ can be exploited to multiply together several such tensors (e.g. in \fref{fig:gamma1}) in a fast way. The interested reader is referred to the appendix of Ref. \onlinecite{Singh11} where details of such a sparse multiplication have been described.
\bibitem{tensorcategory} The reader who is familiar with the graphical calculus used in the context of Tensor Categories, will recognize tensors $\mycup$ and $\mycap$ as implementing the cap and cup transformation appearing therein, a formal way of bending lines in a diagram and defining transposition. For instance, see \textit{Notes on Quantum Gravity} http://www.math.ucr.edu/home/baez/qg-fall2000/QGravity/QGravity.pdf; J. D. Biamonte, S. R. Clark and D. Jaksch, arXiv:1012.0531v1.
\bibitem{fusetree} A \textit{fusion-splitting tree} is a directed trivalent tree graph made of only fusion and splitting vertices. A fusion vertex is a vertex with two incoming indices and one outgoing index, and a splitting vertex is a vertex with one incoming index and two outgoing indices. A \textit{fusion tree} is a fusion-splitting tree made of only fusion vertices. Analogously, a \textit{splitting tree} is a fusion-splitting tree made of only splitting vertices.
\bibitem{Feiguin07} A. Feiguin, S. Trebst, A.W.W. Ludwig, M. Troyer, A. Kitaev, Z. Wang, and M. H. Freedman, Phys. Rev. Lett. \textbf{98}, 160409 (2007).
\bibitem{Trebst08} S. Trebst, M. Troyer, Z. Wang, and A.W.W. Ludwig, Prog. Theor. Phys. Supp. \textbf{176}, 384 (2008).
\end{thebibliography}
\end{document}